\documentclass{jpp}  
\usepackage{graphicx}
\usepackage{natbib}
\usepackage{har2nat}
\usepackage{graphicx}  % needed for figures
\usepackage{dcolumn}   % needed for some tables
\usepackage{bm}        % for math
\usepackage{amssymb}   % for math
\usepackage{amsmath}

\usepackage{movie15}

\usepackage[latin1]{inputenc}
\usepackage{braket}
\usepackage{todonotes}
\usepackage[hang, flushmargin]{footmisc}
\usepackage{verbatim}
\usepackage{setspace}
\usepackage{indentfirst}
\usepackage{fancybox, xcolor}
\usepackage{enumitem}
\usepackage{textcomp}
\usepackage{float}
\usepackage{tikz}
\usepackage{soul}
\usepackage{pythonhighlight}

\usepackage[colorlinks,citecolor=blue,filecolor=black,linkcolor=blue,urlcolor=blue]{hyperref}
\usepackage[noabbrev]{cleveref}
\usepackage[figure]{hypcap}
\title{Radiation pressure acceleration of protons from  structured thin-foil targets}

\author{Tim Arniko Meinhold and Naveen Kumar\corresp{\email{naveen.kumar@mpi-hd.mpg.de}}}
\affiliation{Max-Planck-Institut f\"ur Kernphysik, Saupfercheckweg 1, D-69117 Heidelberg, Germany}

\date{\today}

\begin{document}
%\ams{}
\maketitle

\begin{abstract}
The process of radiation pressure acceleration (RPA) of ions is investigated with the aim of suppressing the Rayleigh-Taylor like transverse instabilities in laser-foil interaction. This is achieved by imposing surface and density modulations on the target surface. We also study the efficacy of RPA of ions from density modulated and structured targets in the radiation dominated regime where the radiation reaction effects are important. We show that the use of density modulated and structured targets and the radiation reaction effects can help in achieving the twin goals of high ion energy (in GeV range) and lower energy spread.
\end{abstract}

\section{\label{sec:Intro}Introduction}
\textcolor{black}{Radiation pressure acceleration (RPA) of ions has attracted a significant attention in the last two decades~\cite{Esirkepov:2004aa,Chen:2009aa,Palmer:2011aa,Robinson:2008aa,Chen:2009aa,Macchi:2009tx,Pegoraro:2007aa,Chen:2011aa,Dollar:2012ws,Khudik:2014aa,Eliasson:2015aa,Wan:2018aa,Wang:2021aa}. The two important characteristics of the RPA are the higher laser energy conversion to the ions and the quality of ion energy spectrum, and due to these reasons the ion beams accelerated by the RPA can have ultrashort pulse duration, and extremely high peak energy needed for applications in many areas~\cite{Daido:2012vq,Atzeni:2004tt,Roth:2001aa,Mackinnon:2006aa,Borghesi:2002aa,Li:2006aa,Honrubia:2015aa}, including  ion beam therapy~\cite{Malka:2004aa}.}

\textcolor{black}{The idea of using plasma as a medium to accelerate charged particles under electromagnetic waves and the use of the photon beams for sailing are not new and they were discussed already in 1950s~\cite{Veksler:1957aa,Garwin:1958aa,TSU:1959aa}. The proposal to use the laser for interstellar travel was} discussed \textcolor{black}{by R. Forward in 1962 ~\cite{McInnes:1999aa,Forward:1984aa} and later on reinvented by G. Marx~\cite{MARX:1966aa}, who first worked out the equation by considering a simple model of a mirror accelerated by the laser pulse~\cite{MARX:1966aa,Forward:1984aa}.} This line of thought is not in the realm of science-fiction, and the photon sail can accelerate the interstellar probe to about 20\% of light velocity within minutes~\cite{Heller:2017aa}. The radiation pressure acceleration also has genesis in the studies of Einstein when Einstein studied the reflection of the light from a mirror and deduced that the ratio $E/\omega$, where $E$ is the electric field and $\omega$ is the frequency of the light, is an invariant. This invariant constant was later found out to be the Planck constant ($E=\hbar \omega$), which Einstein subsequently used to explain the photoelectric effect~\cite{Pauli:1981aa}.   With the availability of ultra-intense lasers $I_l \sim 10^{23}$W/cm$^2$, in a near-future~\cite{eli,cilex,xcels,vulcan}, the RPA of ions has the potential to produce high-energy ion beams with higher energy conversion efficiency compared to other mechanisms of laser-driven ion acceleration~\cite{Forslund:1970aa,Silva:2004aa,Haberberger:2012aa,Liu:2016aa}. Apart from the technological requirements for the efficient RPA of ions, \emph{e.g} the need for the high-contrast, large focal spot-size of the ultra-intense laser pulse, \textcolor{black}{the issue of the transverse instabilities remains important. The onset of the transverse instabilities limits the effectiveness of the RPA of ions}. The onset and physical mechanisms of these transverse instabilities have been recently studied focusing on the intrinsic origin of the instability~\cite{Pegoraro:2007aa,Khudik:2014aa,Eliasson:2015aa,Wan:2018aa,Wan:2020aa}. Several methods \emph{e.g.} tailored electromagnetic pulses with sharp intensities~\cite{Pegoraro:2007aa}, modulation of the RPA~\cite{Bulanov:2009aa}, use of surface modulated targets~\cite{Chen:2011aa} have been proposed to alleviate the influence of the transverse instabilities on the RPA of ions. Moreover, in the ultra-relativistic regime of the laser plasma interaction, envisaged in ~\cite{eli,cilex,xcels,vulcan}, the effect of the radiation reaction (RR) force in the laser-driven electron dynamics has to be taken into account~\cite{Chen:2010vo,Macchi:2011ts}.

In this paper, we not only study the RPA of ions from the density modulated and structured targets, but we also study the influence of the RR force on development of the transverse instabilities from density modulated and structured targets. \textcolor{black}{This paper is organised as follows: in Sec.\ref{sec:Param}, we discuss the parameters of the PIC simulations, followed by the energy spectra of the accelerated ions in Sec.\ref{sec:Eprofile}. Afterwards, we analyse the Rayleigh-Taylor (RT) like transverse instability growth rate for surface modulated targets in Sec.\ref{sec:theory}  and follow-up with the Fourier analysis of the ion density fluctuations in Sec.\ref{Insta_ana}. Before, we conclude in Sec.\ref{sec:conclusion}, we briefly show the results from a simulation run having a laser pulse with spatial Gaussian profile interacting with a target consisting of both density and surface modulations.}

\section{\label{sec:PIC}  PIC simulations setup and results}

We first begin with showing the results on the RPA of protons in ultra-relativistic regimes from the density modulated and structured targets. Afterwards, we extend the results to the radiation dominated regime including the effect of the RR force on the RPA of protons from the density modulated and structured targets.

\subsection{\label{sec:Param} PIC simulations setups and shapes of structured and density modulated targets}

For PIC simulations, we use the open source PIC code SMILEI~\cite{derouillat2018smilei}.\textcolor{black}{We carry out 2D3V (two-dimensional in space and three dimensional in velocity) simulations} employing a simulation box of the size of \(L_x \times L_y = 18\lambda_L \times 10\lambda_L\), where \(\lambda_L = 0.8 \times 10^{-6}\,\)m is the laser wavelength. Thus, for $\Delta x=\Delta y=0.06\lambda_L$, it yields $1800 \times 1000$ cells in the simulation box. \textcolor{black}{This resolution is comparable or smaller than the previous studies on ion acceleration~\cite{Esirkepov:2004aa,Pegoraro:2007aa,Zigler:2013aa,Wang:2021aa,Haberberger:2012aa}}. The time step of the simulation is \(4.19\, \times 10^{-2} \tau_L,\) where \(\tau_L = 2\pi/\omega_0 = 2.67\) fs is the laser period, corresponding to a total of \(1.5\times 10^{4} \) iterations. \textcolor{black}{For the sake of computational efficiency, we take protons instead of high-Z ions.}
We use 16 particles per cells per species for PIC simulations. The plasma is fully ionized with a maximum density of  $n_e = 250\,n_c$, where $n_c = \omega_0^2 m_e/4\pi e^2= 4.36\times 10^{23}\,$cm$^{-3}$  is the non-relativistic critical density for the laser pulse of frequency $\omega_0 = 2.36 \times 10^{15}\,$s$^{-1}$ corresponding to the Ti:Sa laser pulse system. Here $e$ and $m_e$ are the electronic charge and mass, respectively. Plasma ions are assumed to be cold, however the plasma electrons have a small temperature, $T_e \sim 10^{-3}\,m_e c^2$, where  $c$ the velocity of the light in vacuum. We employ the moving window in our PIC simulations. At the onset of the moving window, the circularly polarized laser pulse injection into the simulation is turned off. Consequently, we have a finite   laser pulse duration in our simulations. \textcolor{black}{The laser pulse is turned off when the moving window starts in the simulations, effectively limiting the pulse duration. For the flat target, we have pulse duration  $t/\tau_L \approx 35$\,fs and for density and surface modulated targets, we have $t/\tau_L \approx 30$\,fs}. Since the velocity of the moving window, \textcolor{black}{$\upsilon_{\rm mov}$}, closely follows the group velocity of the laser pulse which in turn depends on the laser and plasma parameters, we have different velocities \textcolor{black}{in the range, 
$\upsilon_{\rm mov}=(0.67-0.84) c$}, of the moving window.  \textcolor{black}{The optimum target thickness for the RPA is given by $\xi \simeq a_0$, where  $\xi = \pi n_e d /n_c \lambda_L$, $d$ is the target thickness, and $a_0=e E_0/m_e \omega_0 c$ is the dimensionless amplitude of the circularly polarized laser pulse~\cite{Macchi:2009tx}.  For $a_0=150$ and $n_e = 250\,n_c$, we get $d \simeq 0.19\lambda_L$.} \textcolor{black}{Notwithstanding with technological improvements, manufacturing of the structured targets is always challenging. Thus, from the point of view of manufacturing structured targets and limiting the deleterious pre-pulse effects, thicker targets $(d \ge 1.0 \lambda_L)$ may be preferred over ultra-thin targets ($d \ll 1.0\lambda_L$) for performing experiments. }

\begin{figure}
\centering
\begin{tikzpicture}
	\node[above right] (img) at (0cm,0cm) {\includegraphics[width=0.80\textwidth]{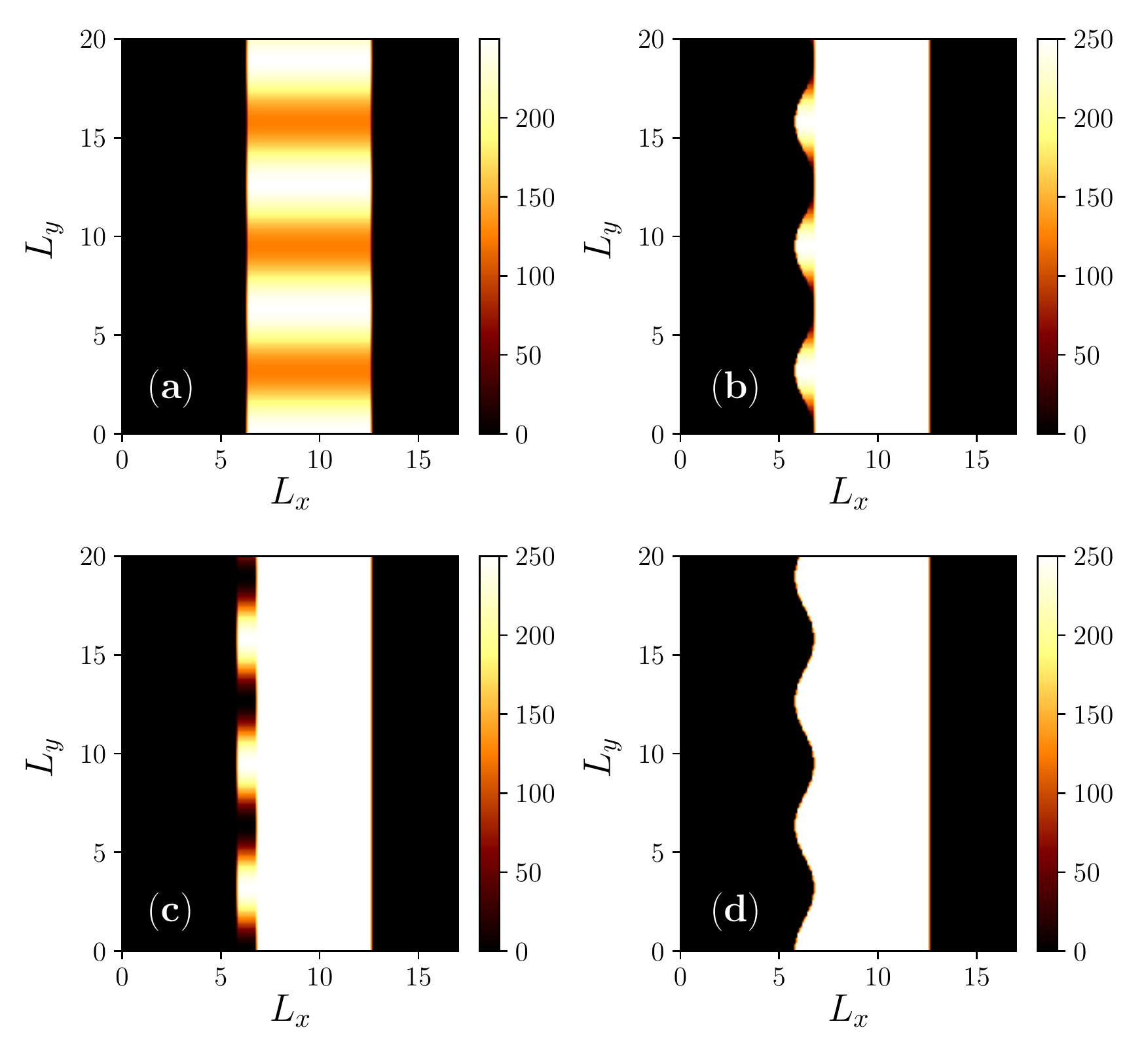}};
	\node at (8.15cm,5.05cm) {$\bm{rp}$};
	\node at (3.0cm,5.05cm) {$\bm{rec}$};
	\node at (3.0cm,9.95cm) {$\bm{dm}$};
	\node at (8.20cm,9.95cm) {$\bm{rpg}$};
	\node at (4.8cm,9.95cm) {\begin{scriptsize}$\bm{n_e/n_c}$\end{scriptsize}};
	\node at (4.8cm,5.15cm) {\begin{scriptsize}$\bm{n_e/n_c}$\end{scriptsize}};
	\node at (10.0cm,5.15cm) {\begin{scriptsize}$\bm{n_e/n_c}$\end{scriptsize}};
	\node at (10.0cm,10.0cm) {\begin{scriptsize}$\bm{n_e/n_c}$\end{scriptsize}};
\end{tikzpicture}
\caption{\label{fig:shapes} \textcolor{black}{$\bm{(a)}$} density modulated target (dm), \textcolor{black}{$\bm{(b)}$} rippled target with changing plasma density (rpg), \textcolor{black}{$\bm{(c)}$} surface modulated target with rectangular grooves (rec) and \textcolor{black}{$\bm{(d)}$} rippled target with constant plasma density (rp). \textcolor{black}{The colorbar denotes the normalised plasma density $n_e/n_c$.}}
\end{figure}
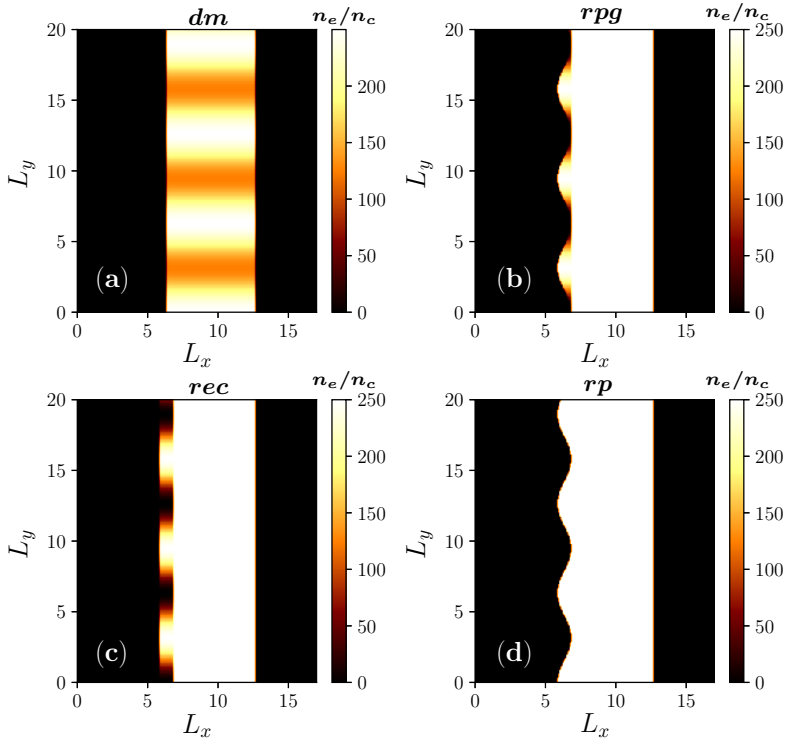

\textcolor{black}{Since we are interested in studying the physics of competitive mode feeding of the RTI-like transverse instabilities, we exclude other effects occurring due to the spatiotemporal shapes of the laser pulse}. Fig.~\ref{fig:shapes} shows the targets with different modulations. The profiles in Fig.\ref{fig:shapes} are described mathematically as follows: (a) the density modulated target with width ({$d=1.0\lambda_L$}) has a spatial density profile, $n(x,y) = n_e a_m [3+\cos (k_m y)]/2$, where $n_e=250\,n_c$. This yields minimum density \(n_{min} = a_m\,n_{e}\). Panel (b) describes the rippled plasma density target with the width ({$d=1.0\lambda_L$}), with the ripples being located in the region, $d + a_m \cos (k_m y) \leq x \leq d + a_m$, having the spatial density as, $n(x,y) = n_e [a_m\cos (k_my)-a_m]$. Panel (c) depicts the structured target with rectangular groovings, which are located in the region, \(d - a_m \leq x \leq d + a_m\), and the spatial density profile reads as, $n(x,y) = n_e [a_m\cos (k_my)-a_m]$. Finally, panel (d) depicts the  target with ripples imposed on the left side. The target has ripples localised in the region, \(d - a_m \cos (k_m y) \leq x \leq d + a_m\), with a {constant}  plasma density $n_e$ in this case. The targets with a width ({$d=1.0\lambda_L$}) are located in the region {$(1.0\lambda_L \leq x \leq 2.0\lambda_L)$} while the targets with a width ({$d=2.0\lambda_L$}) are located {\(1.0\lambda_L \leq x \leq 3.0\lambda_L \)}  in all cases. Parameter $a_m$ is a dimensionless number showing the modulations in the density while $k_m$ is normalised with the laser wavevector $k_L$ in all cases. We not only change the wavelength of the modulations, $\lambda_m$ (normalized to $\lambda_L$), but also the amplitude of the modulations $a_m$ in Figs.\ref{fig:shapes} (a), (b), (c) and (d).

\subsection{\label{sec:Eprofile}Energy spectra of ions}

\begin{figure}
\centering
\begin{tikzpicture}
	\node[above right] (img) at (0cm,0cm) {\includegraphics[width=0.65\textwidth]{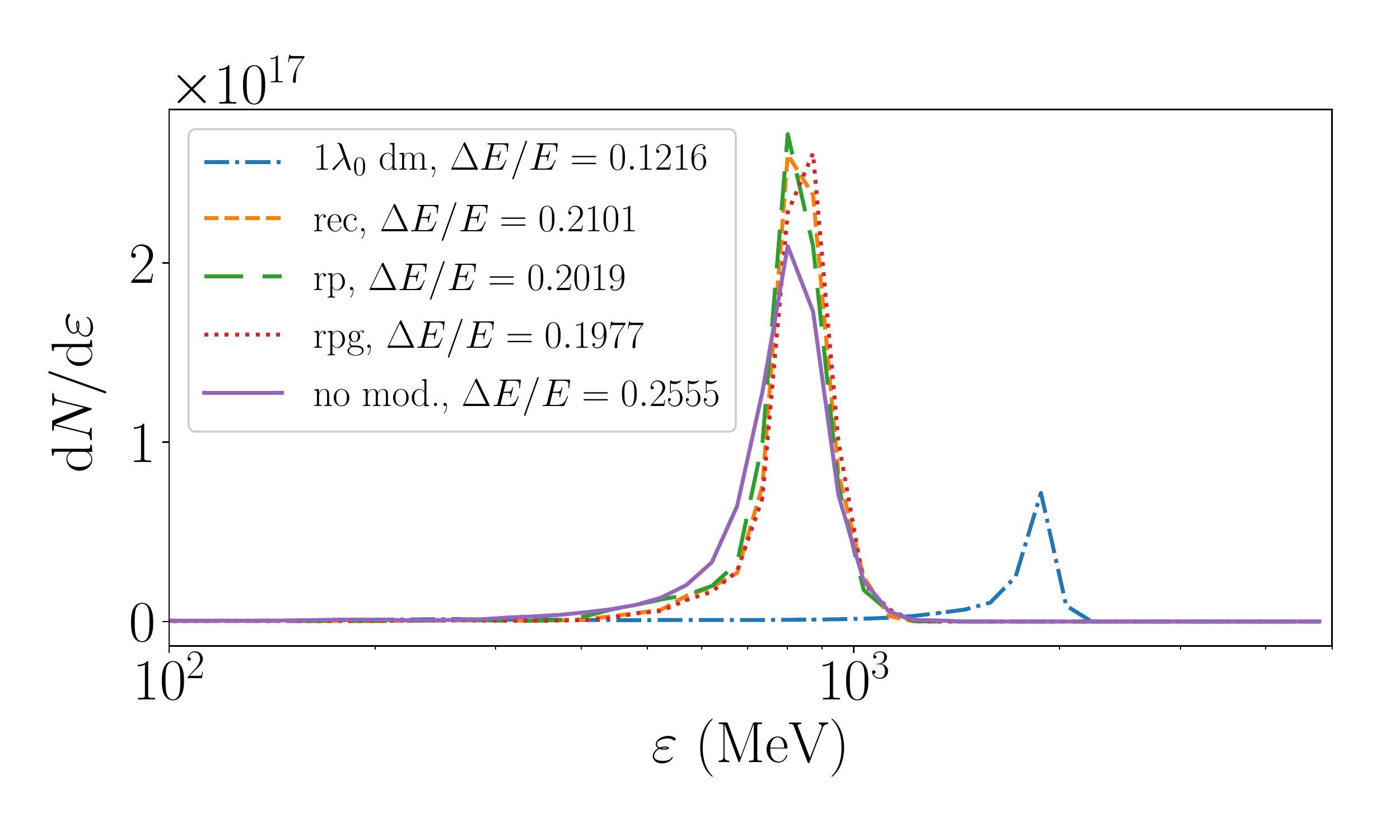}};
\end{tikzpicture}
\caption{Kinetic energy of ions for different targets from Fig.\ref{fig:shapes} with modulation parameters, $k_m=2$ and $a_m=0.25$ at $t/\tau_L = 440$. The other parameters are $a_0=150, n_e=250 n_c,$ and $d=1.0\lambda_L$ in each case. The $y$-axis represents the proton numbers, $N$ per unit length. \textcolor{black}{Moving window velocities are $\upsilon_{\rm mov} = 0.8\, c$ for dm target, and $\upsilon_{\rm mov}=0.75\,c$ for surface modulated and flat targets.} }\label{fig:Ekin}
\end{figure}

Fig.\ref{fig:Ekin} shows the energy spectra of different targets. One can immediately see that modulating the target density leads to improvement not only in quality of the energy spectra captured in the  full-width-half-maximum (FWHM) but in the case of the density modulated target [Fig.\ref{fig:shapes}(a)], it also results  in higher energy gain with {significantly smaller FWHM ($\Delta E/E\sim 12\%$) in comparison to other modulated targets and remarkably smaller compared to the flat target ($\Delta E/E\sim 26\%$) case.  The lower number of accelerated ions in the case of density modulated target can be attributed to both lower target mass in the beginning and also a slight loss ($\sim 4\%$) of the target mass in the interaction process (See Supplemental Material).
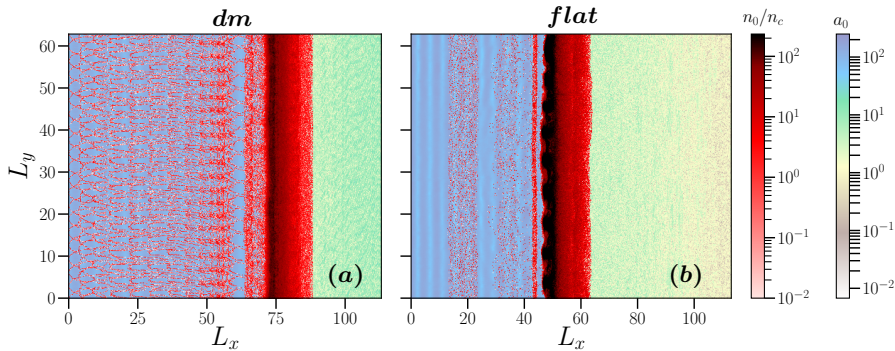
\begin{figure}
	\centering
	\begin{tikzpicture}
	\node[above right] (img) at (0cm,0cm) {\includegraphics[width=1.0\textwidth]{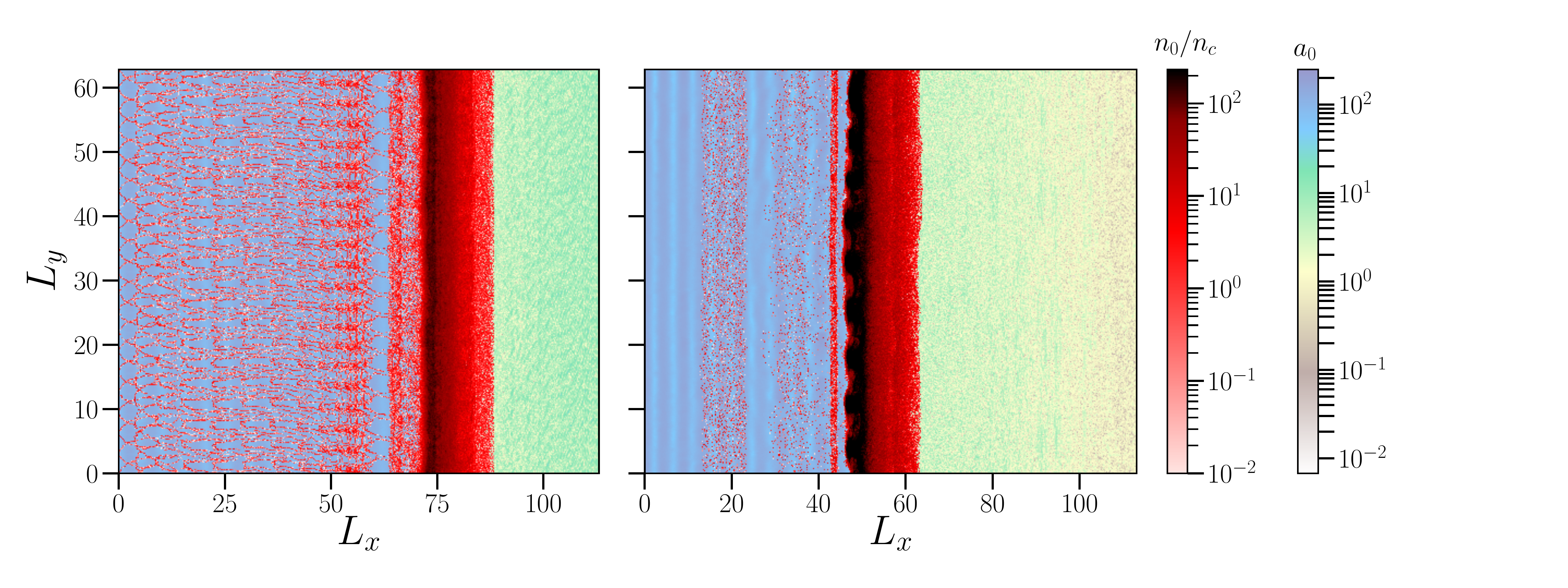}};
	%\node[above right] (img) at (4.25cm,0cm) {\includegraphics[width=0.45\textwidth]{figure3b}};
	\node at (9.3cm,1.4cm) {$\bm{(b)}$};
	\node at (7.8cm,4.8cm) {$\bm{flat}$};
	\node at (4.8cm,1.4cm) {$\bm{(a)}$};%3.7cm,1.1cm
	\node at (3.3cm,4.8cm) {$\bm{dm}$};
	\end{tikzpicture}
	\caption{\label{fig:ion_dens_1l0_2l0_RR250} Evolution of the ion density and the normalized laser electric field $a_0$ for $\bm{(a)}$ a density modulated target ($k_m=2$, $a_m=0.25$) and $\bm{(b)}$ the flat target, for $a_0=150$. Both targets have  $d=1.0\lambda_L$ width. The results are shown at {$t/\tau_L = 144$}. The other parameters are same as in Fig.\ref{fig:Ekin}.}
\end{figure}
\noindent
\textcolor{black}{The maximum energy of protons in Fig.\ref{fig:Ekin} are slightly smaller than the analytical scalings. The analytical scaling of proton energy, reproduced here again for completeness, reads as, $E_k= A m_pc^2(\gamma_f -1), \gamma_f=(1-\beta_f^2)^{-1/2}, \beta_f=((1+\mathcal{E})^2-1)/((1+\mathcal{E})^2+1), \mathcal{E}=2 \pi (Z/A)(m_e/m_p) a_0^2 \tau/ \xi$, where $m_p$ is the proton mass, $\tau$ is measured in the units of the laser period, $\tau_L$, $Z$ and $A$ are the atomic and mass numbers, respectively~\cite{Macchi:2009tx}. For the flat target ($\tau=35$) in Fig.\ref{fig:Ekin}, $E_k\sim 1.01$ GeV while for the density modulated target ($n_e^{\textrm{max}}=125 n_c, \tau=30$), it yields $E_k\sim 2.05$ GeV. These values are a bit higher than observed in  Fig.\ref{fig:Ekin}. As expected the lower plasma density for the density modulated target leads to higher proton energy gain, but it alone can not account for the lower FWHM of the proton energy spectra observed in Fig.\ref{fig:Ekin}.}
Since, the modulation in the ion beam  spectra are a good evidence of the growth of the RTI~\cite{Chen:2010vo,Palmer:2012aa,Sgattoni:2015aa}, it is apparent that the use of density modulated and structured target is efficient in suppressing the long wavelength modes of the RTI-like interchange instabilities (see movies).  We explain this in terms of the competitive feeding of different modes in the RPA of protons later in Sec.\ref{Insta_ana}. Fig.\ref{fig:ion_dens_1l0_2l0_RR250} shows a snapshot from the movies (see also Supplemental Movies) on the ion density and laser electric field evolutions. One can immediately notice the onset of the RTI like instabilities for the flat-target case [panel (b)] while the density modulated target [panel (a)] does not show the surface rippling associated with the RTI like instabilities leading to the break-up of the target at later times.  
 
The stronger impact on the proton energy spectrum in the case of density modulated target [Fig.\ref{fig:shapes}(a)] points to physical mechanism of the RTI-like interchange instability in the RPA regime of involving the coupling of both electron and ion modes~\cite{Wan:2018aa}. Not only the choice of the modulation wavelength $k_m=2$, but also the choice of modulation amplitude $a_m$ affect the late time evolution of the ion energy spectrum. Fig.~\ref{fig:DeltaE_E} captures this dependence  showing the evolution of the FWHM  density modulated (dm) target for different modulation parameters. We record  $E_{max}$ and its full width of half maximum (FWHM) at every simulation timestep. Fig.~\ref{fig:DeltaE_E} shows these discrete data-points (raw-data method) together with the Gaussian fitting (Gauss-fit method) of the simulation results. The difference between the two methods (Gauss fit and raw data methods) can be attributed to the lack of clear single peak formation in the energy spectra, especially at early and late times. We also take into account error propagation of the Gaussian fitting. On defining the energy spread as $V=\Delta E/E$, the error in energy-spread $\sigma_V$ can be estimated as $\sigma_V=\sqrt{(\sigma_{\Delta E}/E)^2 + (\sigma_E \Delta E /E^2)^2}$, where $\sigma_E$ and  $\sigma_{\Delta E}$ represent the standard deviations in the energy and the energy spread, respectively of the ion beam.  The oscillations in the energy spectra occur because the FWHM and $E_{max}$ in Fig.~\ref{fig:DeltaE_E} do not show the same temporal development. The stretching of the oscillations in the ion energy spectra is related to the speed of the target relative to the E-field of the laser and the amplitude of the energy oscillations are related to the wave vector of the density modulations. For comparison we also plot the FWHM of a flat target in  Fig.~\ref{fig:DeltaE_E}. One can notice that in the case of the flat target, the FWHM increases with time, while for the density modulated target [panel(b)], it remains lower for a longer duration. On changing the amplitude of the density modulation ($a_m=0.25$) as in Fig.~\ref{fig:DeltaE_E}(d), the FWHM remains lower and stable for longer durations but shows the disruption in the ion energy spectrum at later times. Thus, an optimisation in value of modulation parameters is required.

\subsection{\label{pot_param} Parameter maps for the optimised RPA of ions}

\begin{figure}
\centering
\begin{tikzpicture}[every node/.style={inner sep=0,outer sep=0}]
	\node[above right] (img) at (0.cm,0.0cm) {\includegraphics[width=0.37\textwidth]{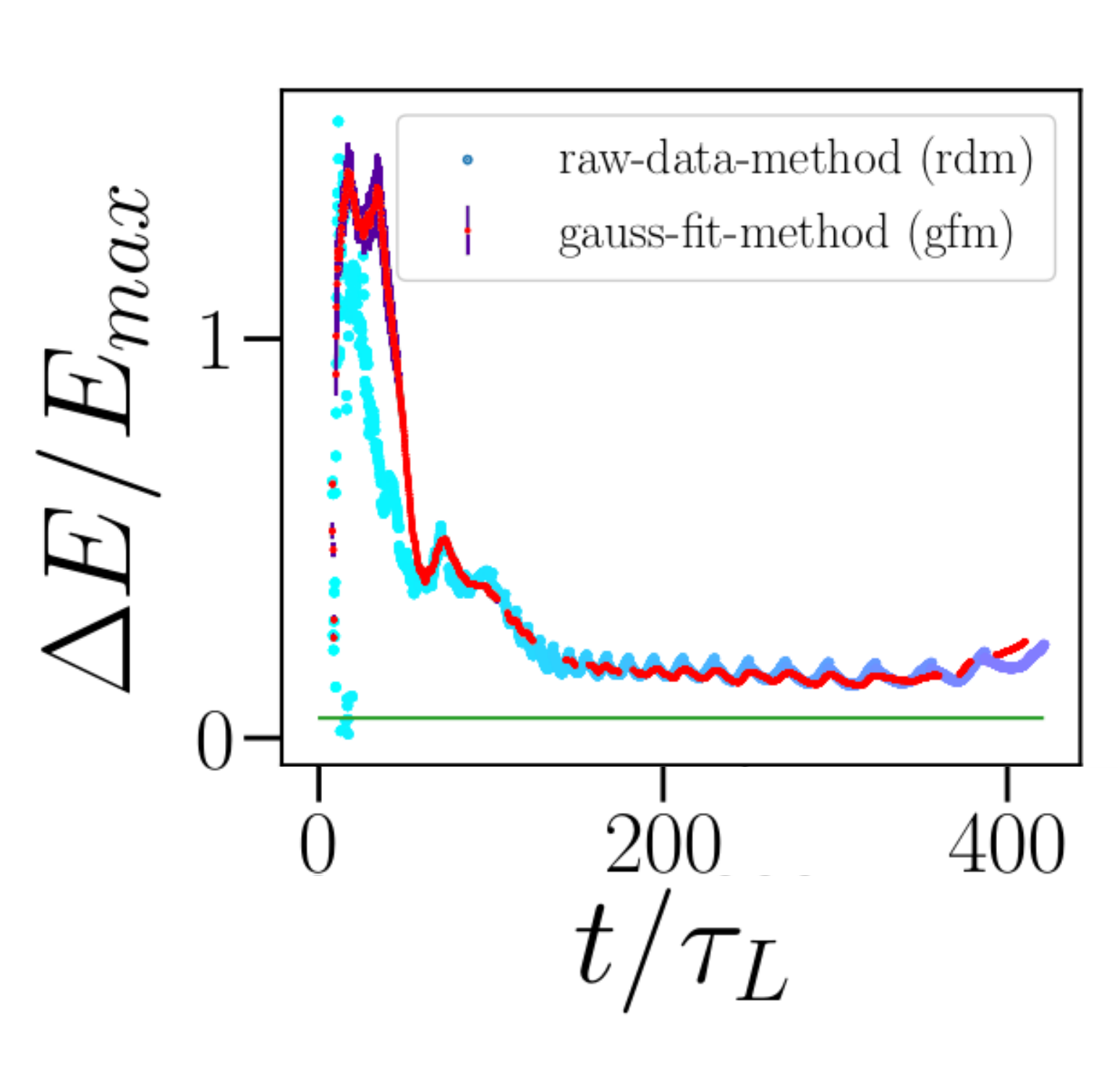}};
	\node[above right] (img) at (4.9cm,0.2cm) {\includegraphics[width=0.45\textwidth]{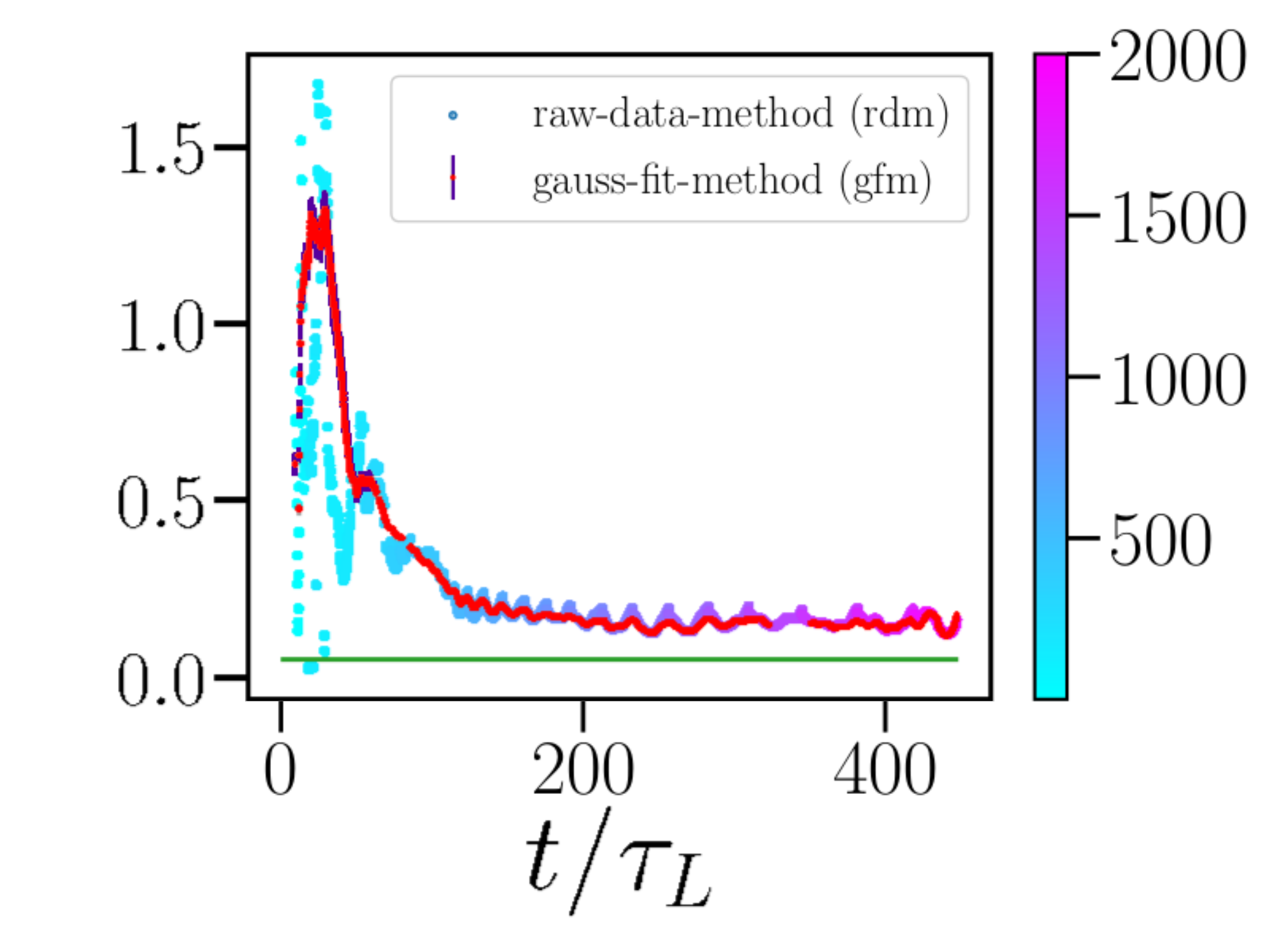}};	
	\node[above right] (img) at (0.cm,5.4cm) {\includegraphics[width=0.37\textwidth]{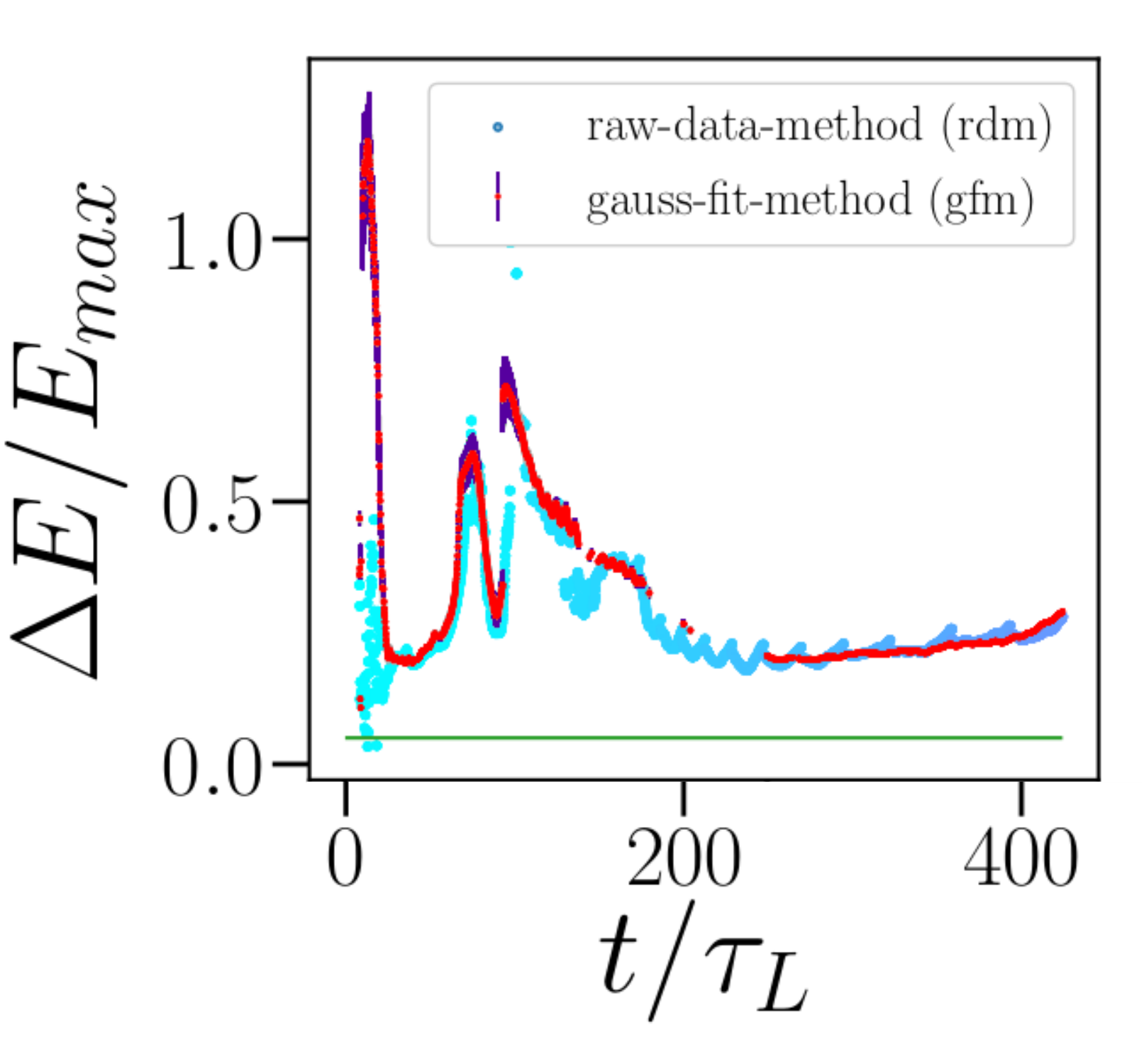}};
	\node[above right] (img) at (4.9cm,5.4cm) {\includegraphics[width=0.45\textwidth]{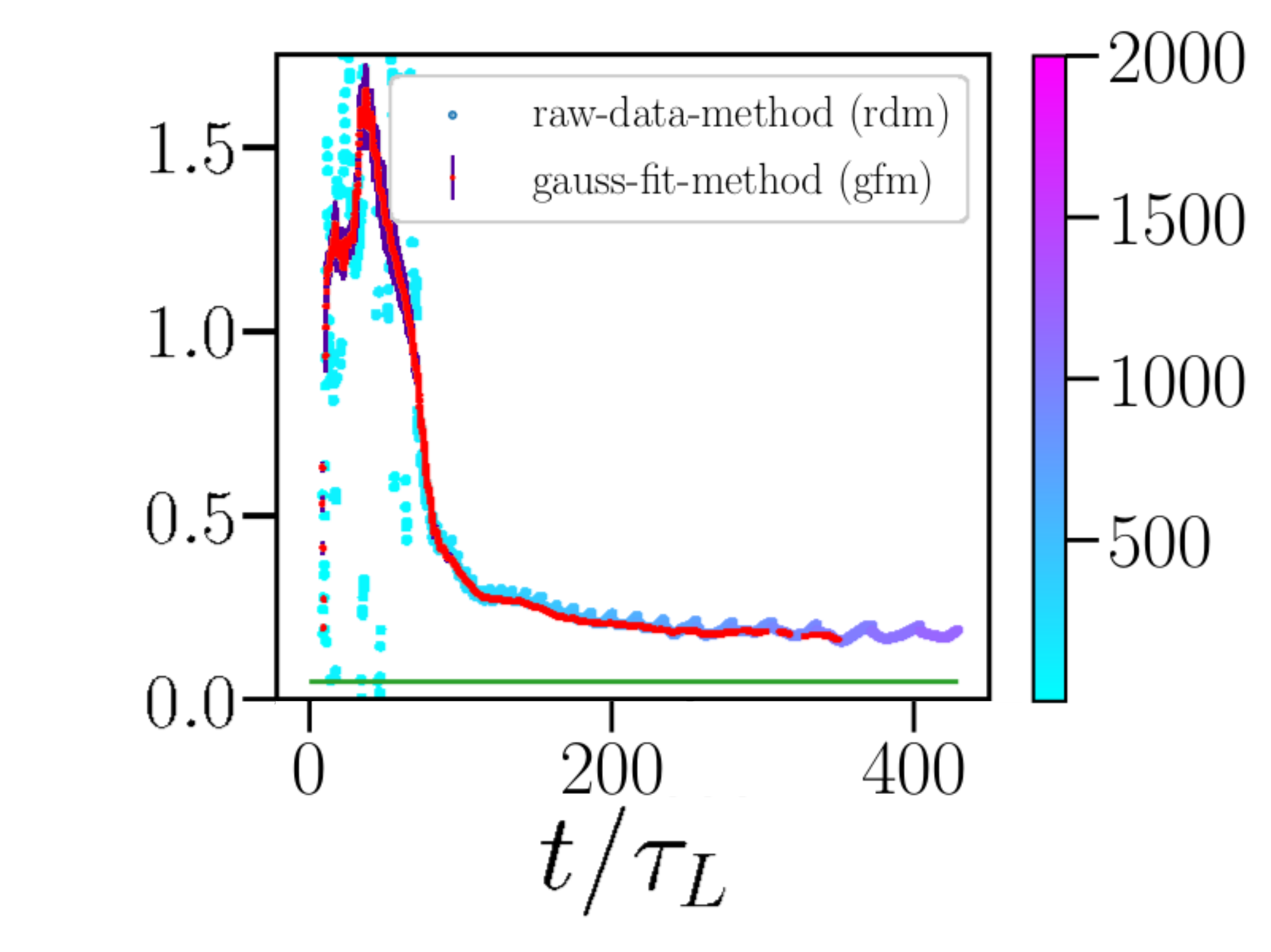}};	
	%\node[above right] (img) at (0.cm,0.0cm) {\includegraphics[width=0.295\textwidth]{figure4c}};
	%\node[above right] (img) at (3.9cm,0.2cm) {\includegraphics[width=0.35\textwidth]{figure4d}};	
	%\node[above right] (img) at (0.cm,3.6cm) {\includegraphics[width=0.285\textwidth]{figure4a}};
	%\node[above right] (img) at (3.8cm,3.6cm) {\includegraphics[width=0.35\textwidth]{figure4b}};
	\node at (9.15cm,8.65cm) {$\bm{(b)}$};
	\node at (4.3cm,3.35cm) {$\bm{(c)}$};
	\node at (4.3cm,8.7cm) {$\bm{(a)}$};
	\node at (3.1cm,10.0cm) {$\bm{flat}$};
	\node at (8.0cm,10.4cm) {$\bm{dm:}$};
	\node at (8.0cm,10.0cm) {$\bm{k_m=1;a_m=0.50}$};
	\node at (3.1cm,5.1cm) {$\bm{dm:}$};
	\node at (10.2cm,10.0cm) {\begin{scriptsize}$\bm{E_{max}}$\end{scriptsize}};
	\node at (3.1cm,4.7cm) {$\bm{k_m=2;a_m=0.50}$};
	\node at (8.0cm,5.1cm) {$\bm{dm:}$};
	\node at (8.0cm,4.7cm) {$\bm{k_m=2;a_m=0.25}$};
	\node at (10.2cm,4.8cm) {\begin{scriptsize}$\bm{E_{max}}$\end{scriptsize}};
	\node at (9.15cm,3.40cm) {$\bm{(d)}$};
	%\node at (7cm,4.65cm) {$\bm{(a)}$};
	%\node at (4cm,0cm) {\tiny $t/\tau_L$};
\end{tikzpicture}
\caption{\label{fig:DeltaE_E} \textcolor{black}{Evolution of the $\Delta E/E_{max}$ with time. The colorbar denotes the $E_{max}$ (in MeV)} in each case.  $\bm{(a)}$ the flat target, $\bm{(b)}$  density modulated (dm) with $k_m=1$, $a_m=0.50$, $\bm{(c)}$ $k_m=2$, $a_m=0.50$ and $\bm{(d)}$ $k_m=2$, $a_m=0.25$. The target width is {$d=1.0\lambda_L$} in each case. The green line is the sought limit of $\Delta E/E_{max} = 0.05\, (5\%)$. The other parameters are same as in Fig.\ref{fig:Ekin}.}
\end{figure}
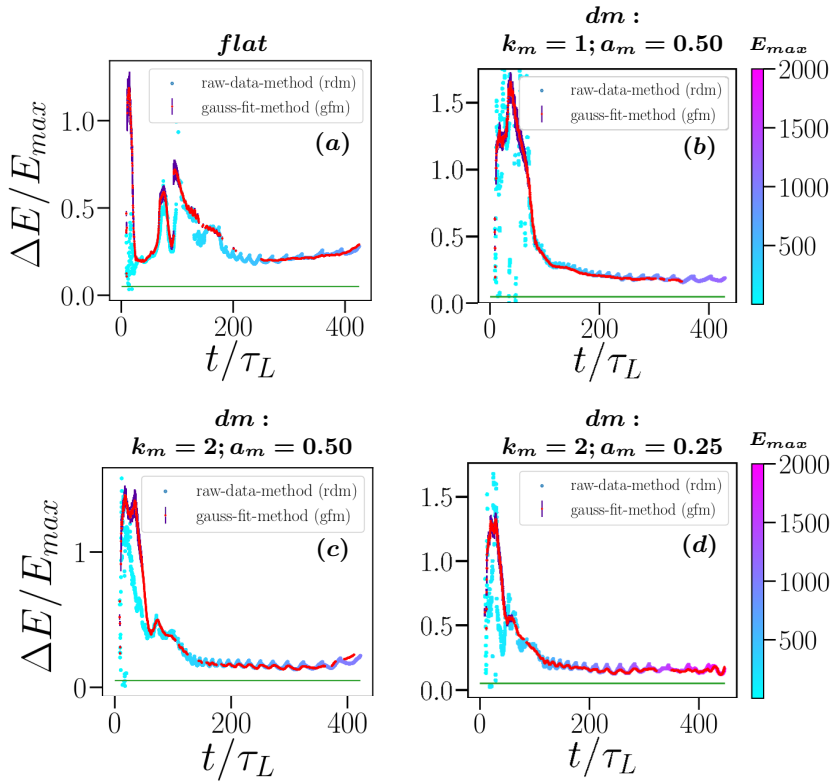

Fig.~\ref{fig:Interpoldm} shows the maps depicting the dependence of the proton energy and its spread on $a_m$ and $k_m$ for the density modulated targets with different thicknesses. The first and second rows show the FWHM and maximum ion energies for a density modulated target with the target thicknesses, {$d=1.0\lambda_L$ and $d=2.0\lambda_L$}, respectively. \textcolor{black}{These maps are generated from the 12 simulation data-points interpolated using a cubic interpolation scheme.} One can observe a few trends quickly. First, for the thinner target ({$d=1.0\lambda_L$}, top row), the optimum range for $a_m$ extends to $a_m \approx 0.35$ while range of $k_m$ shrinks to {$k_m \approx 3$}. For a thicker target ({$d=2.0\lambda_L$}, bottom row), the pre-imposed modulations have only beneficial effect for {$a_m  \lesssim 0.25$  and $k_m \in(2,5)$}. This can be understood based as follows: for a fixed $a_0$, the thinner target has lower target mass and consequently lower $\xi$, resulting into the dominance of the RPA mechanism and higher ion acceleration energies for {$d=1.0\lambda_L$} target [compare panels (b) and (d)]. Large $a_m$ and $k_m$ facilitate stronger absorption of the laser pulse, resulting in the stronger electron heating that can lower the RPA of ions and degrade the FWHM of the ions for thick target ({$d=2.0\lambda_L$}), \textcolor{black}{presumably due to the TNSA process playing a role}. \textcolor{black}{Further increasing the $a_m$ for {$d=1.0\lambda_L$} target, one again reaches the regime of stronger laser penetration and heating of the plasma electrons resulting into lower proton energy gain and degradation in the proton spectrum quality possibly due to the effect of the TNSA process~\cite{Andreev:2011aa,Ferri:2020aa,Zigler:2013aa}. }

 \begin{figure}
\centering
\begin{tikzpicture}
	\node[above right] (img) at (0.0cm,0.0cm) {\includegraphics[width=0.40\textwidth]{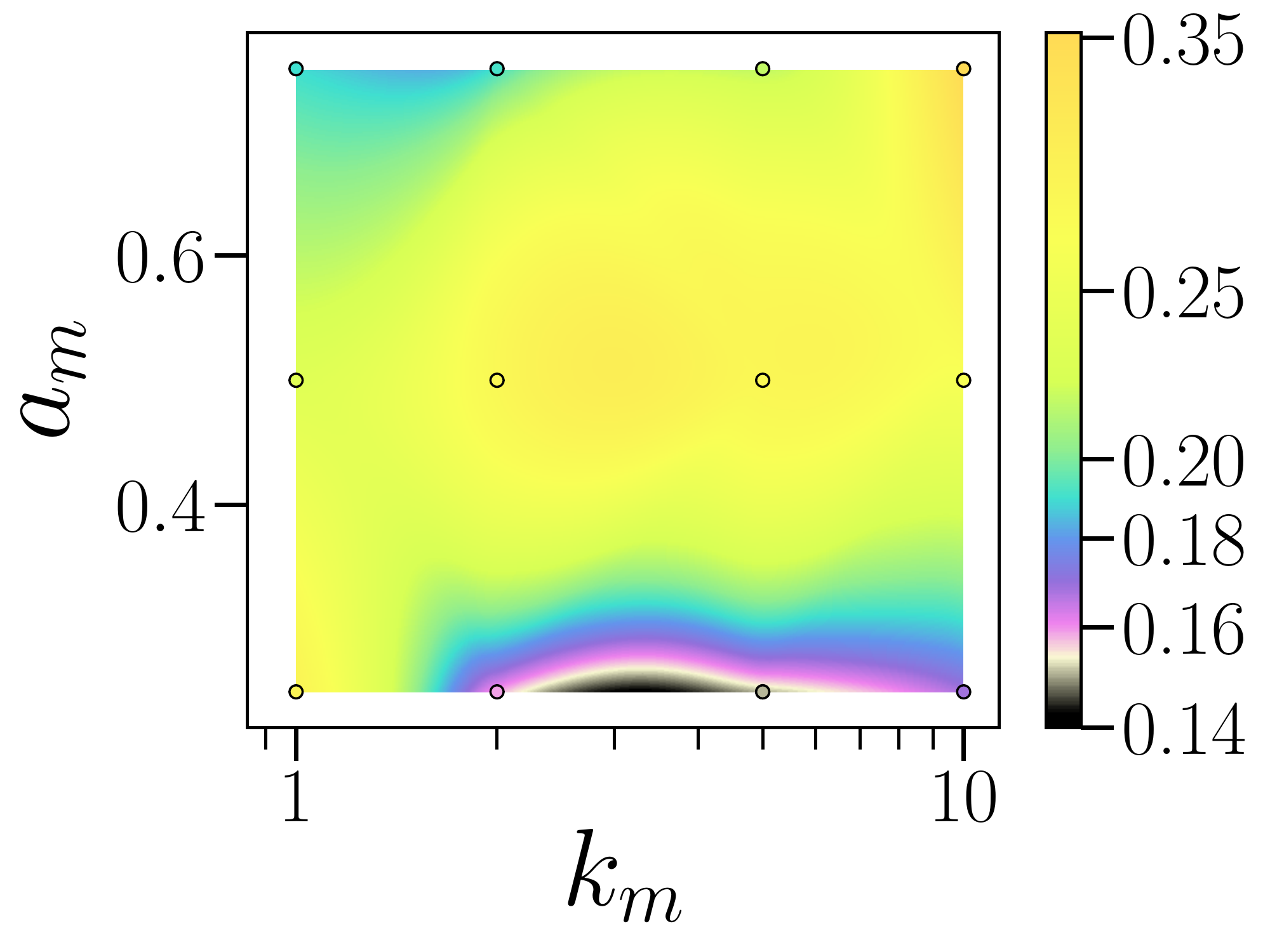}};
	\node[above right] (img) at (0cm,4.5cm) {\includegraphics[width=0.40\textwidth]{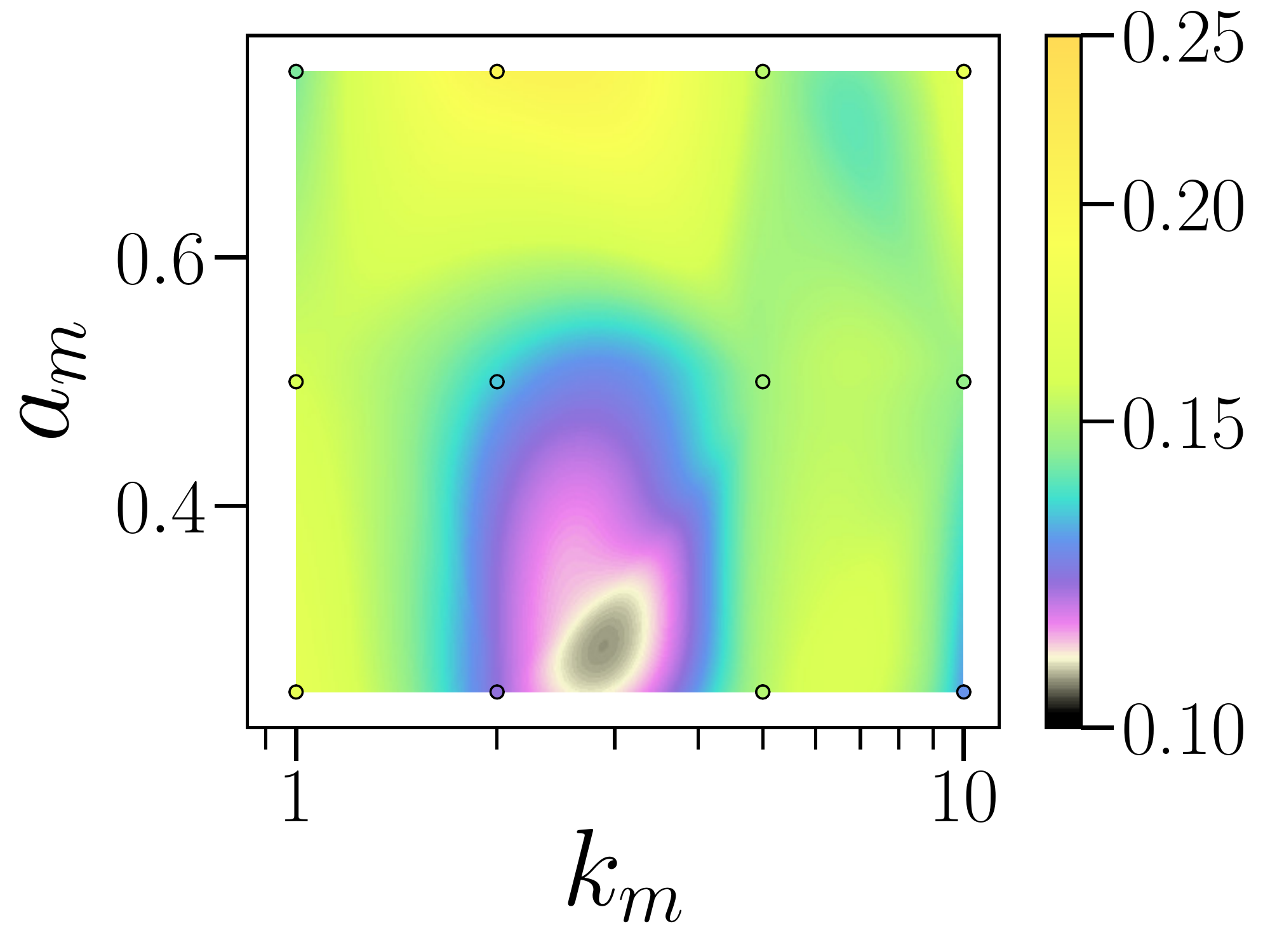}};
	\node[above right] (img) at (6.0cm,0.0cm) {\includegraphics[width=0.40\textwidth]{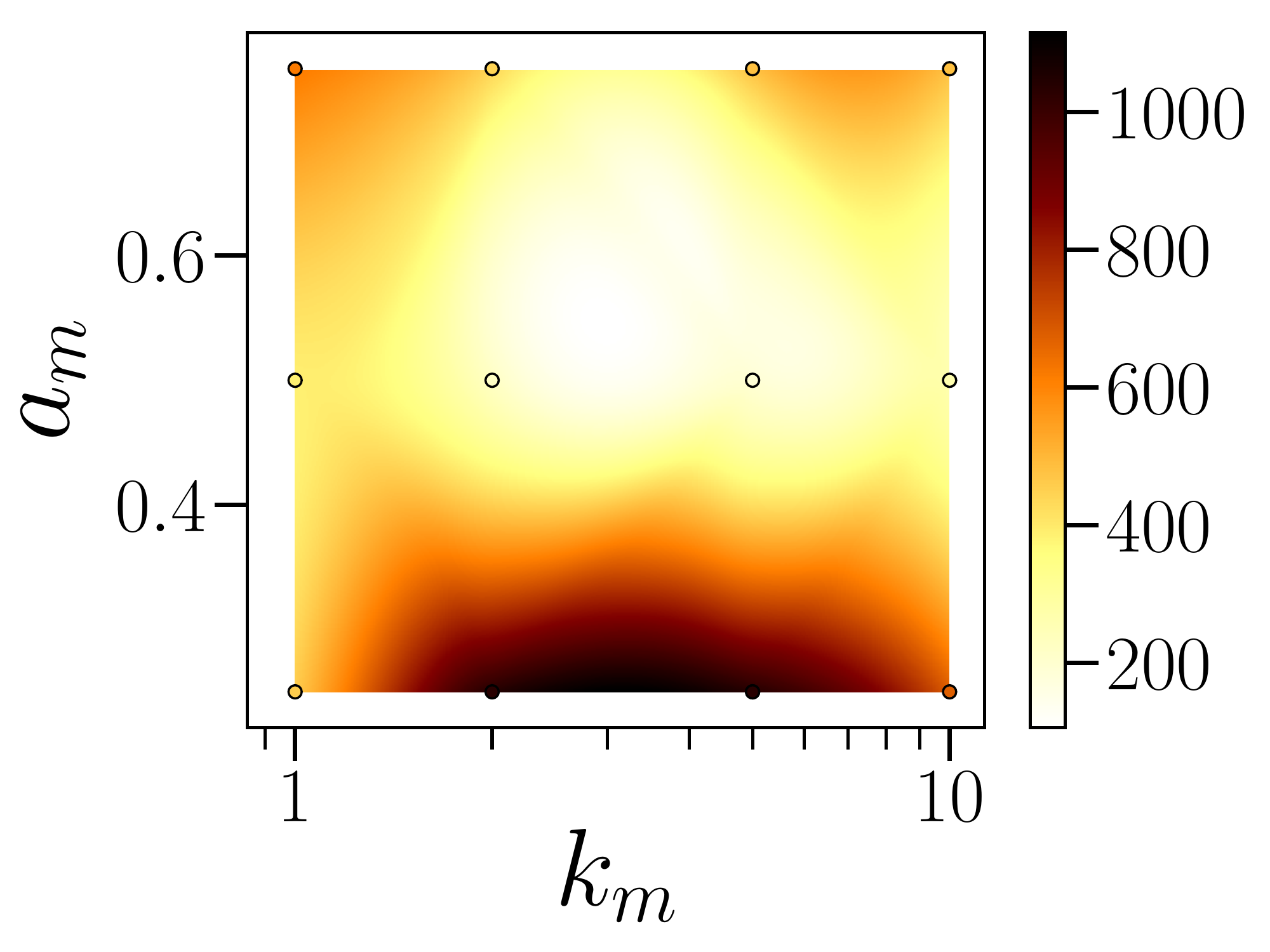}};
	\node[above right] (img) at (6.0cm,4.5cm) {\includegraphics[width=0.40\textwidth]{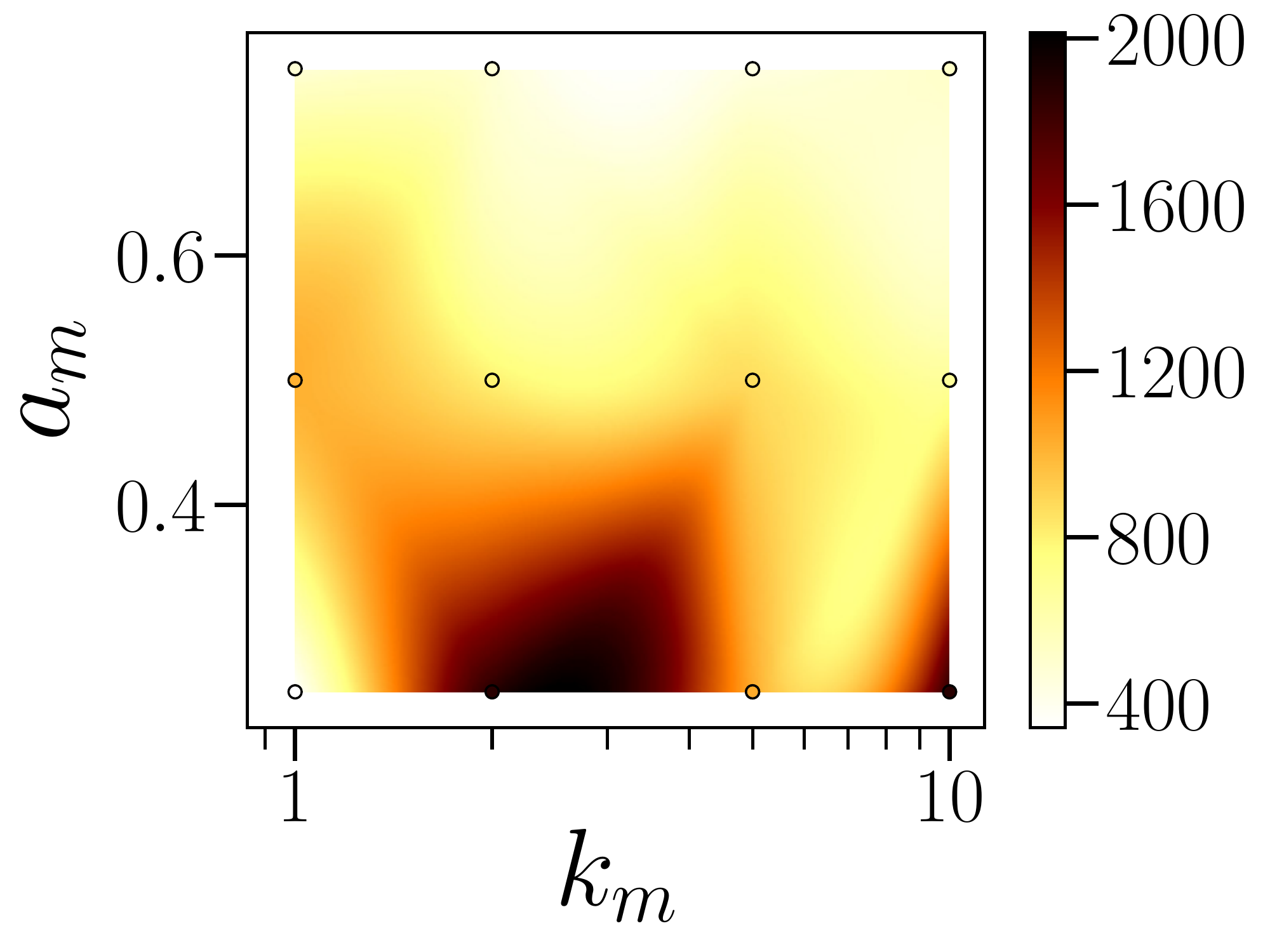}};
	\node at (3.8cm,3.5cm) {$\bm{(c)}$};
	\node at (3.8cm,8.0cm) {$\bm{(a)}$};
	\node at (9.75cm,3.5cm) {$\bm{(d)}$};
	\node at (9.75cm,8.0cm) {$\bm{(b)}$};
	\node at (8.75cm,8.75cm) {$\bm{dm, d=1.0\lambda_L}$};
	\node at (10.8cm,8.8cm) {\begin{scriptsize}$\bm{E_{max}}$\end{scriptsize}};
	\node at (2.8cm,8.75cm) {$\bm{dm, d=1.0\lambda_L}$};
	\node at (4.85cm,8.8cm) {\begin{scriptsize}$\bm{\Delta E/E_{max}}$\end{scriptsize}};
	\node at (2.8cm,4.25cm) {$\bm{dm, d=2.0\lambda_L}$};
	\node at (10.8cm,4.25cm) {\begin{scriptsize}$\bm{E_{max}}$\end{scriptsize}};
	\node at (8.75cm,4.25cm) {$\bm{dm, d=2.0\lambda_L}$};
	\node at (4.85cm,4.3cm) {\begin{scriptsize}$\bm{\Delta E/E_{max}}$\end{scriptsize}};
\end{tikzpicture}
\caption{\label{fig:Interpoldm} Parameter maps for the ion energy spectra [$\Delta E/E_{max}$, panels \textcolor{black}{$\bm{(a)}$} and \textcolor{black}{$\bm{(c)}$}] and ion acceleration energies $E_{max}$ [in MeV, panels \textcolor{black}{$\bm{(b)}$} and \textcolor{black}{$\bm{(d)}$}] with  $a_m$ and $k_m$ for the density modulated target. First and second rows correspond to the targets with {$d=1.0\lambda_L$ and $d=2.0\lambda_L$} widths, respectively. Please notice that here and afterwards, unless stated otherwise, $k_m$ is normalised with the laser wavevector $k_L$, while $a_m$ is a dimensionless number as mention before in Sec.\ref{sec:Param}. The small circles are datapoints used for the interpolation. The other parameters are same as in Fig.\ref{fig:Ekin}.}
\end{figure}
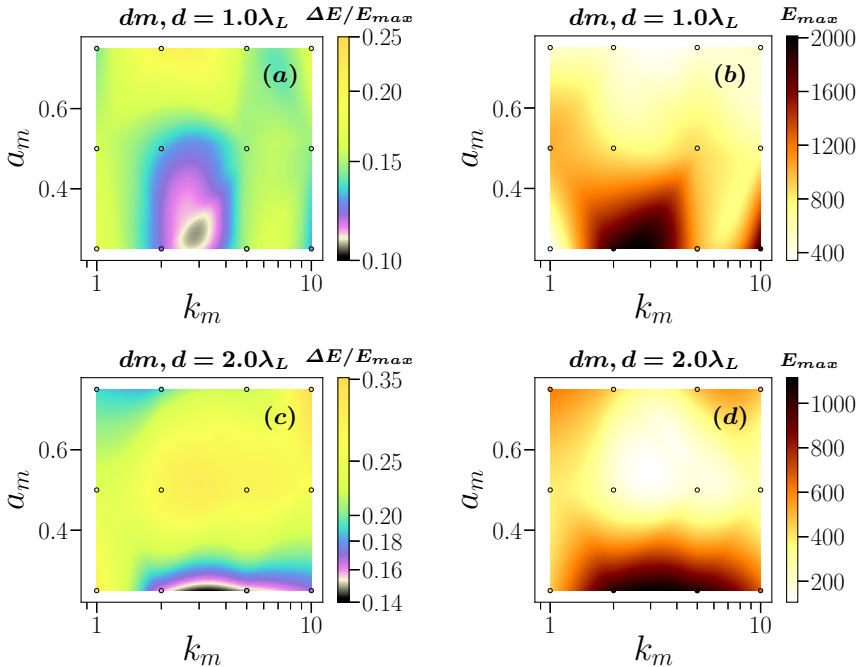

Fig.~\ref{fig:smt_int} shows the same parameter maps for other surface modulation shapes as in Fig.\ref{fig:shapes} {Here, the parameters maps are generated from 16 simulation data-points}. First, it can be seen, that all maximum energy peaks are roughly identical, \emph{i.e} all surface modulated targets have similar values of \(E_{{max}}\), which is smaller compared to the density modulated target as shown in Fig.~\ref{fig:Interpoldm}. The trends for optimum value of $k_m$ are similar in the cases of rectangular and rippled groovings, but a nonlinear behaviour for the rippled grooving with varying (rpg) density is observed. In general, larger value of $a_m$ \emph{e.g.} 
 $a_m \ge 0.2$ leads to smaller FWHM of the ion energy spectra with $k_m$ being largely centered between $2 < k_m < 5$ (except for the rpg shape). The corresponding values of $E_{{max}}$ are essentially following  the same pattern as for the corresponding FWMH of the energy spectra.  The higher acceleration energies for larger $a_m$ and $k_m \le 2$, can be explained by the locally enhanced electric field and higher absorption of the laser field in different targets. The different behaviour in three cases exemplify the different evolutions of the RTI-like interchange instabilities due to perturbations fed by different structured targets. To study this we carry out fast Fourier transforms (FFT) of the ion plasma densities and the results are discussed in Sec.\ref{Insta_ana}.

\begin{figure}
\centering
\begin{tikzpicture}
	\node[above right] (img) at (0cm,0.0cm) {\includegraphics[width=0.40\textwidth]{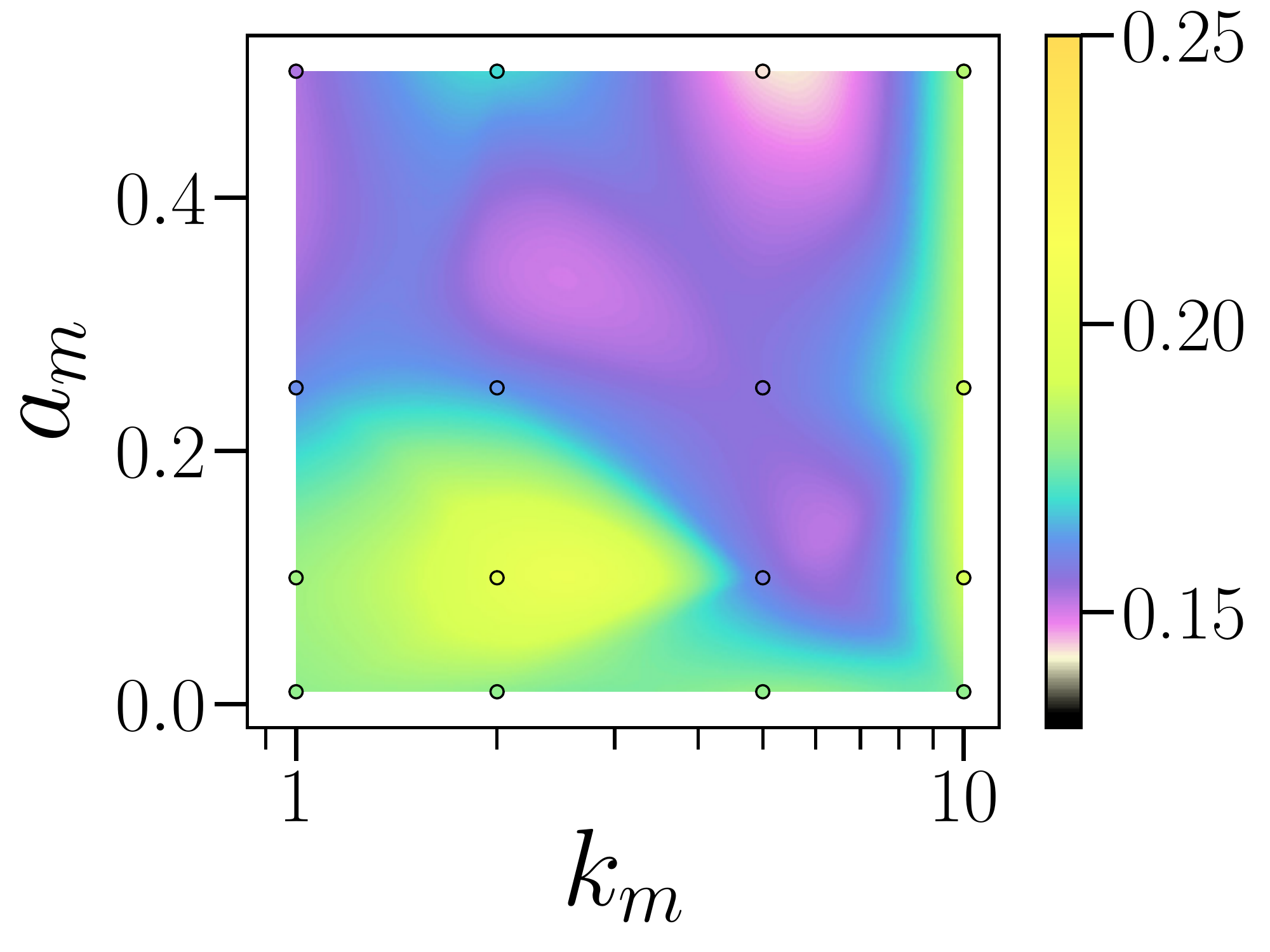}};
	\node[above right] (img) at (6.0cm,0.0cm) {\includegraphics[width=0.40\textwidth]{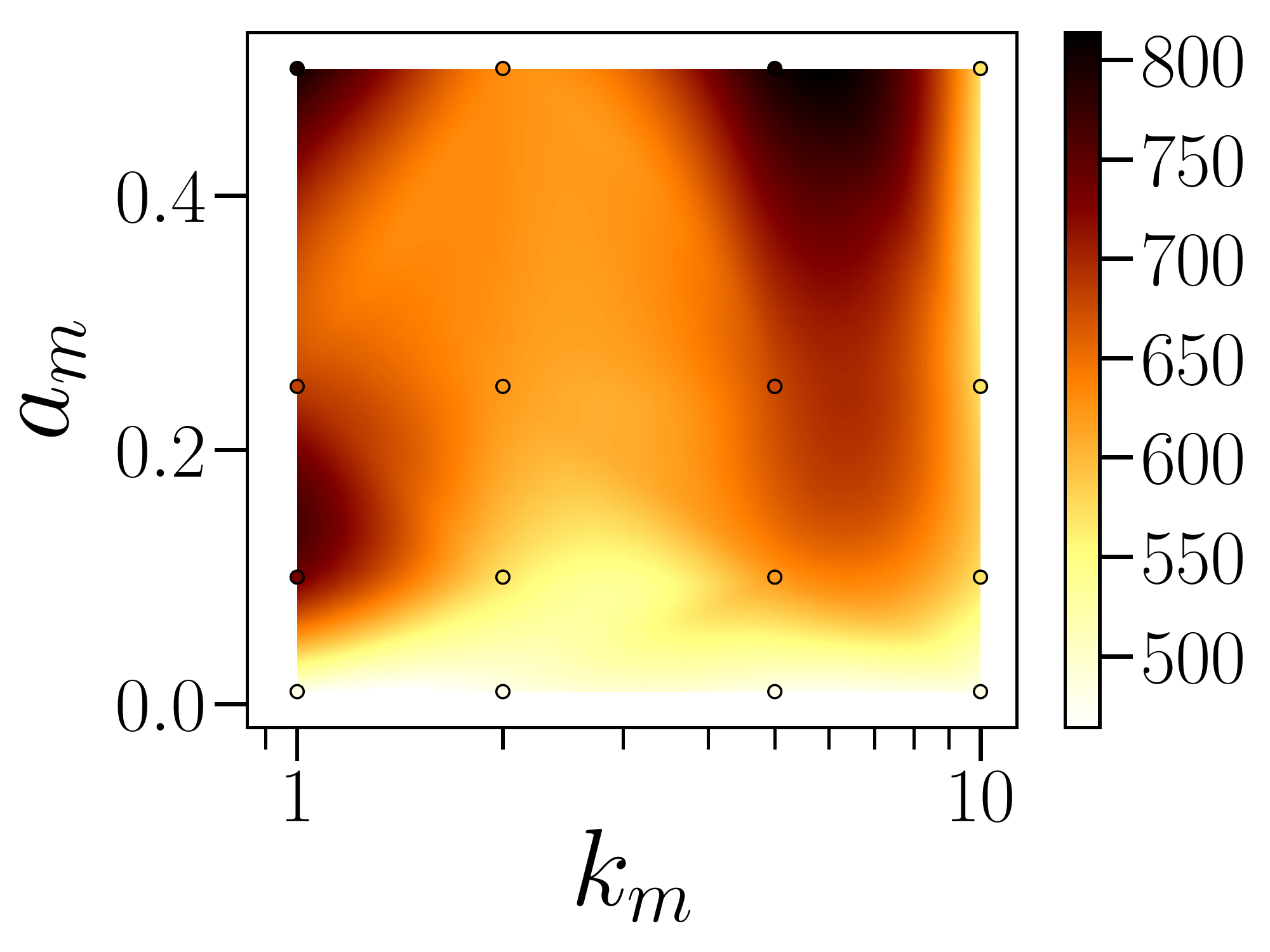}};
	\node[above right] (img) at (0cm,4.5cm) {\includegraphics[width=0.40\textwidth]{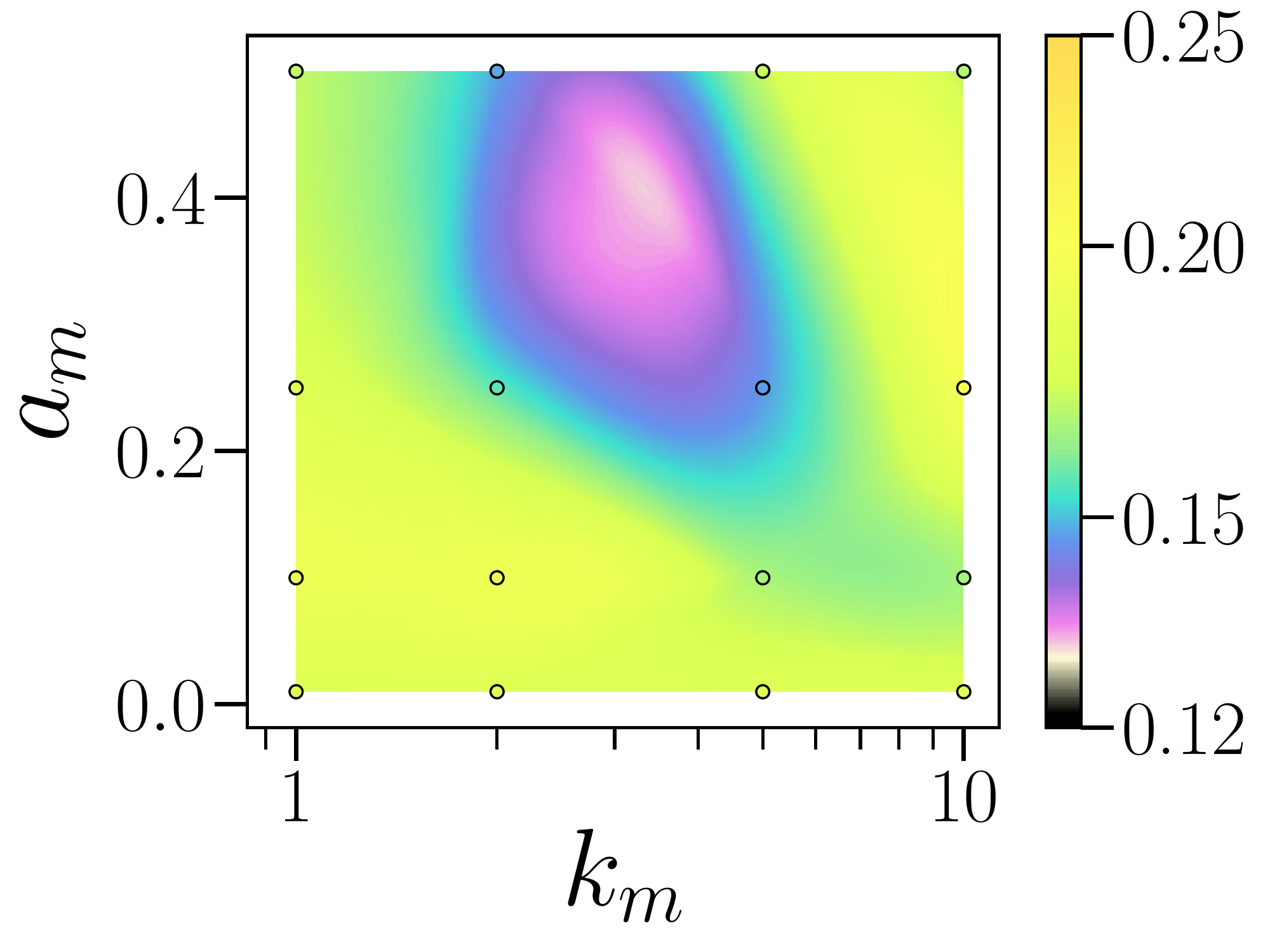}};
	\node[above right] (img) at (6.0cm,4.5cm) {\includegraphics[width=0.40\textwidth]{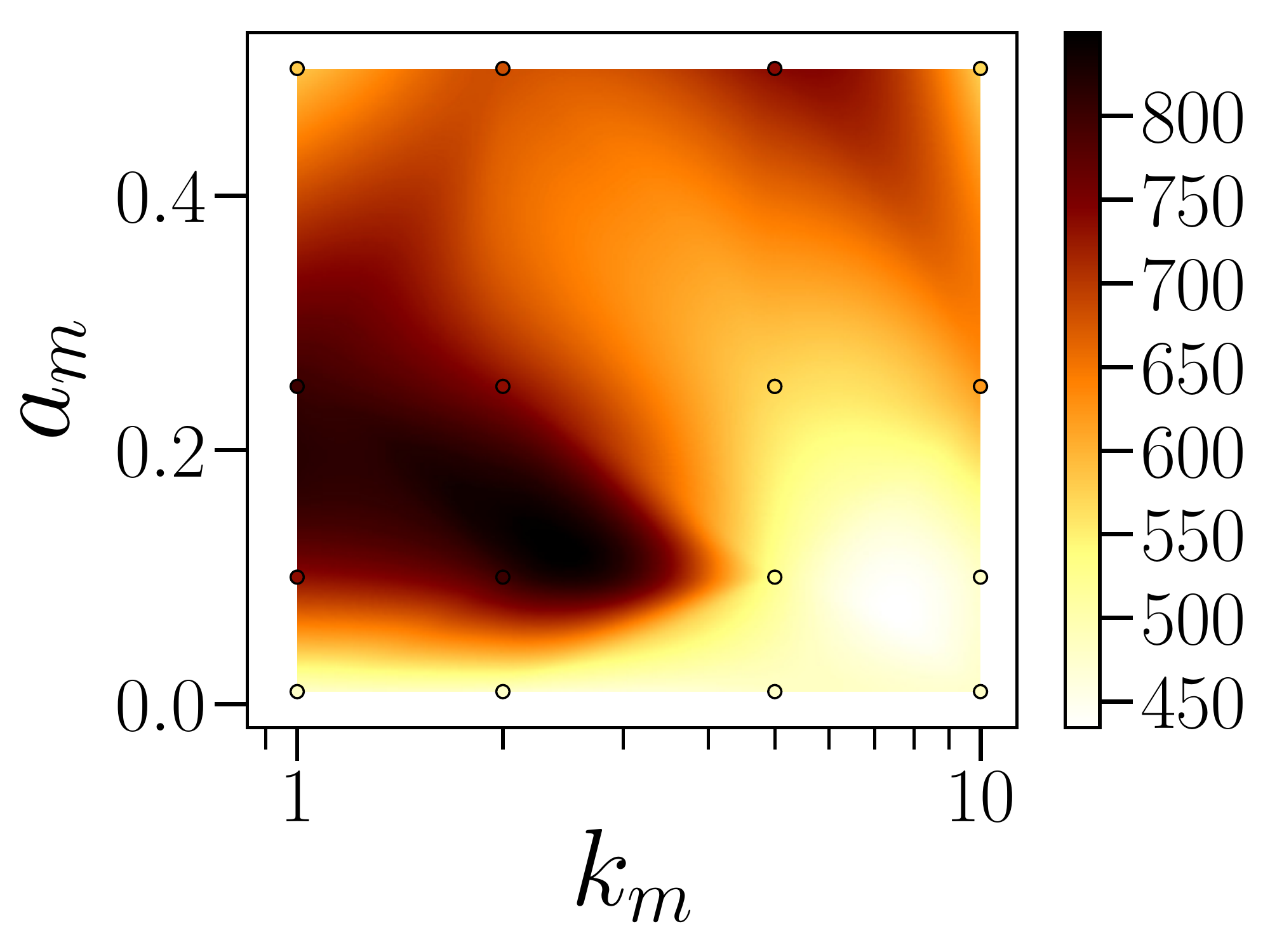}};
	\node[above right] (img) at (0cm,9.0cm) {\includegraphics[width=0.40\textwidth]{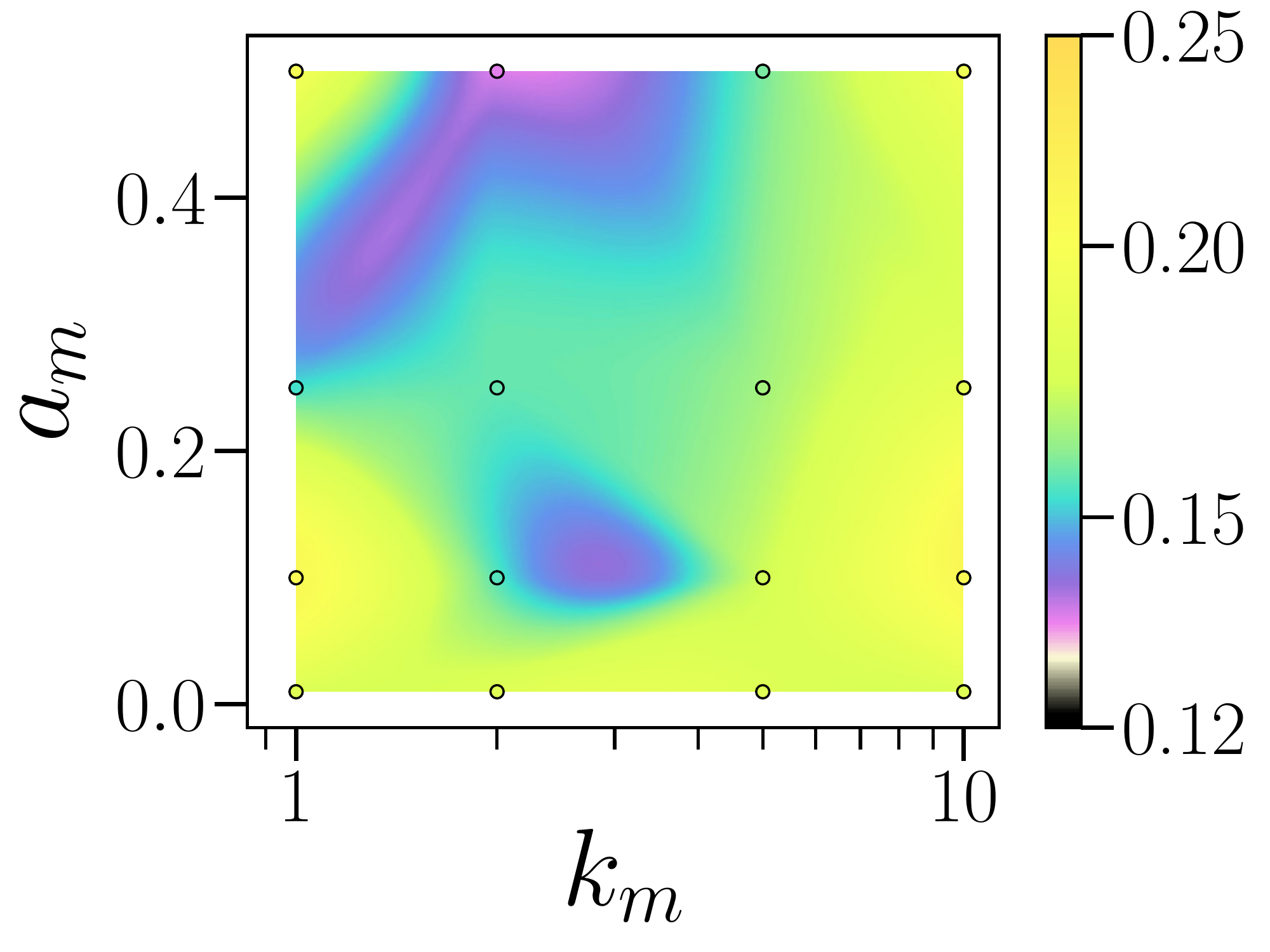}};
	\node[above right] (img) at (6.0cm,9.0cm) {\includegraphics[width=0.40\textwidth]{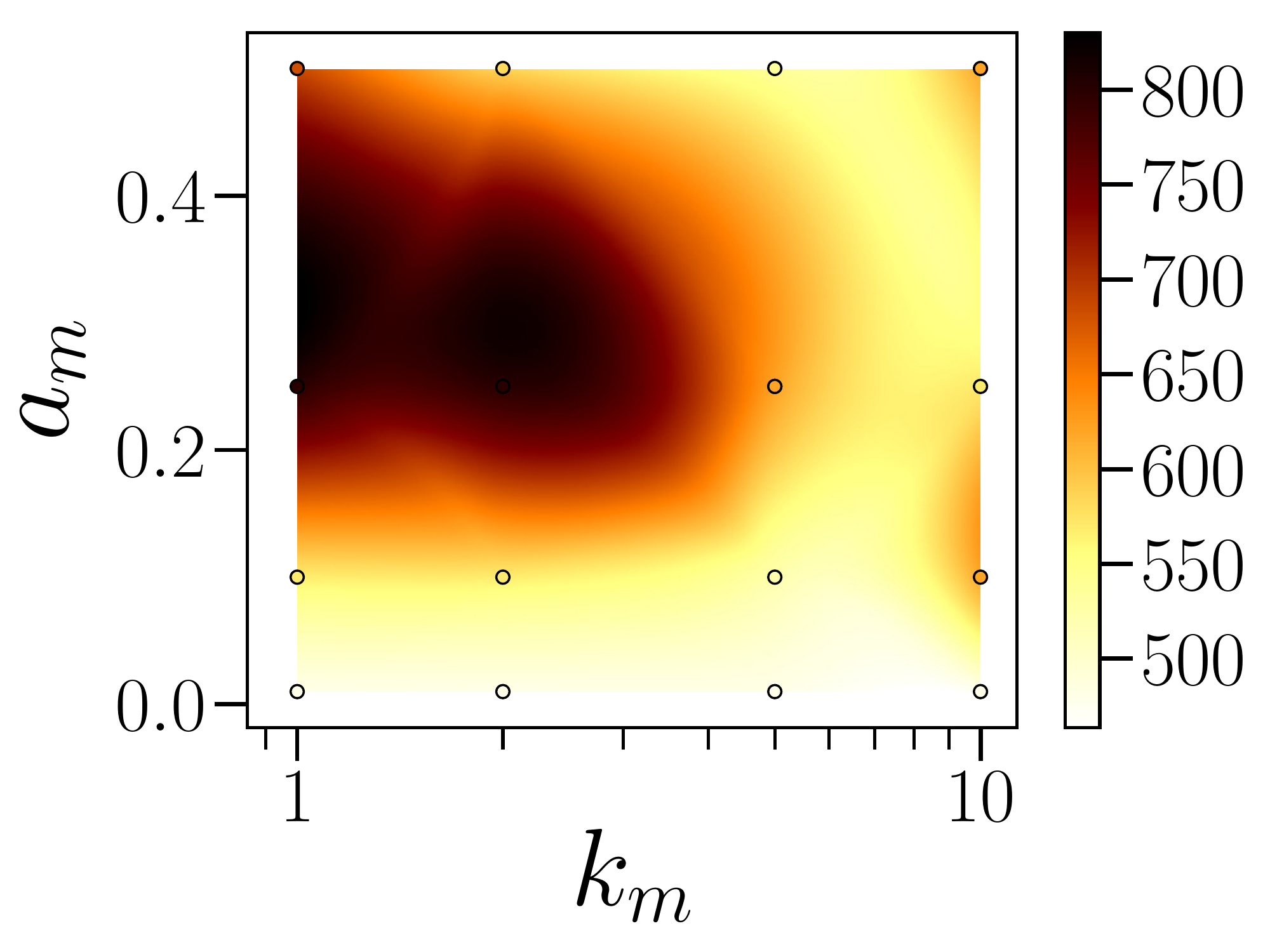}};
	\node at (3.8cm,8.0cm) {$\bm{(c)}$};
	\node at (3.8cm,12.5cm) {$\bm{(a)}$};
	\node at (9.75cm,8.0cm) {\textcolor{white}{$\bm{(d)}$}};
	\node at (9.75cm,12.5cm) {$\bm{(b)}$};
	\node at (8.75cm,8.75cm) {$\bm{rp}$};
	\node at (10.8cm,8.8cm) {\begin{scriptsize}$\bm{E_{max}}$\end{scriptsize}};
	\node at (2.8cm,8.75cm) {$\bm{rp}$};
	\node at (4.85cm,8.8cm) {\begin{scriptsize}$\bm{\Delta E/E_{max}}$\end{scriptsize}};
	\node at (2.8cm,4.25cm) {$\bm{rpg}$};
	\node at (10.8cm,4.25cm) {\begin{scriptsize}$\bm{E_{max}}$\end{scriptsize}};
	\node at (8.75cm,4.25cm) {$\bm{rpg}$};
	\node at (4.85cm,4.3cm) {\begin{scriptsize}$\bm{\Delta E/E_{max}}$\end{scriptsize}};
	\node at (2.8cm,13.25cm) {$\bm{rec}$};
	\node at (10.8cm,13.25cm) {\begin{scriptsize}$\bm{E_{max}}$\end{scriptsize}};
	\node at (8.75cm,13.25cm) {$\bm{rec}$};
	\node at (4.85cm,13.30cm) {\begin{scriptsize}$\bm{\Delta E/E_{max}}$\end{scriptsize}};	
	\node at (3.8cm,3.5cm) {$\bm{(e)}$};
	\node at (9.75cm,3.5cm) {\textcolor{white}{$\bm{(f)}$}};

\end{tikzpicture}
\caption{\label{fig:smt_int} Parameter maps for $a_m$ and $k_m$ for target thickness $d=1.0\lambda_L$. The colorbars denote $\Delta E/E_{max}$  for panels $\bm{(a)}$, $\bm{(c)}$ and $\bm{(e)}$ and $E_{max}$ (in MeV) for panels $\bm{(b)}$, $\bm{(d)}$ and $\bm{(f)}$. First, second and third rows show structured targets viz. rec [panels $\bm{(a)}$ and $\bm{(b)}$], rp [panels $\bm{(c)}$ and $\bm{(d)}$], rpg [panels $\bm{(e)}$ and $\bm{(f)}$], respectively. The other parameters are same as in Fig.\ref{fig:Ekin}.}
\end{figure}
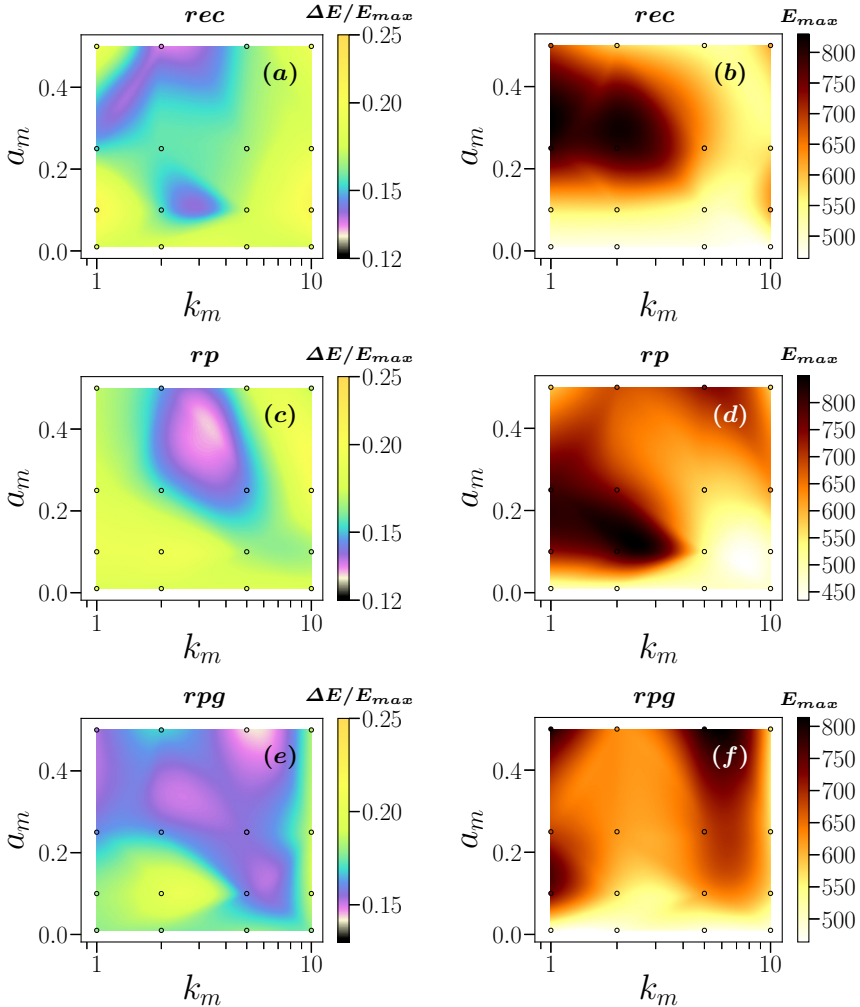

\subsection{\label{sec:rr}Radiation reaction effects on the RPA of ions from density modulated and structured targets}
 
\begin{figure}
\centering
\begin{tikzpicture}
	\node[above right] (img) at (0cm,0cm) 
	{\includegraphics[width=0.65\textwidth]{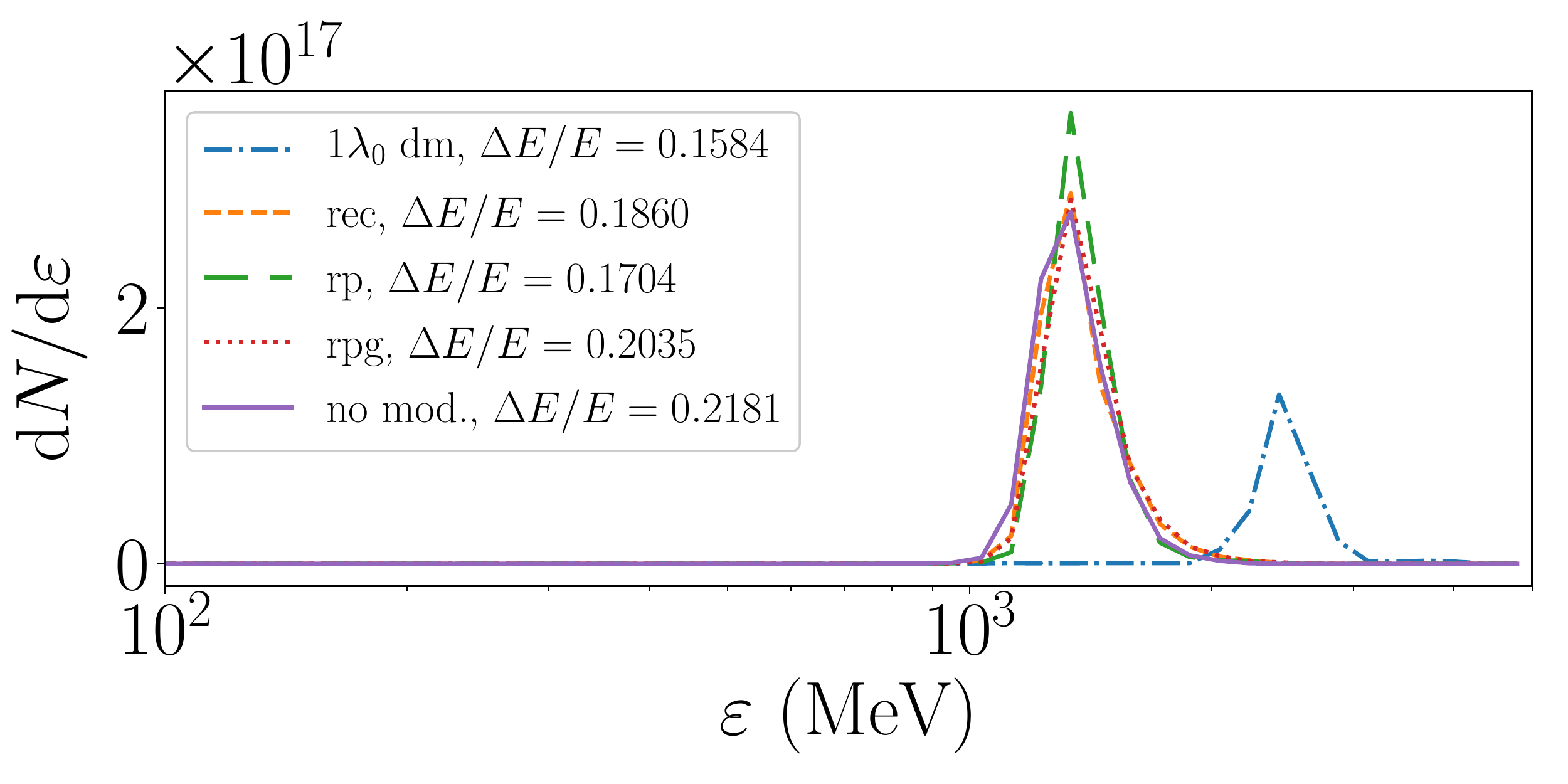}};
	%\node[above right] (img) at (0cm,3.1cm) {\includegraphics[width=0.45\textwidth]{Ion_energy1a}};
	%\node at (1cm,1.0cm) {$\bm{(b)}$};
	%\node at (1cm,3.85cm) {$\bm{(a)}$};
	%\node at (0.9cm,3.05cm) {\tiny $\times 10^{17}$};
	%\node at (0.9cm,5.9cm) {\tiny $\times 10^{16}$};
\end{tikzpicture}
\caption{\label{fig:Ekin_RR} Kinetic energy of ions for different targets with modulation parameters $k_m=2$, $a_m=0.25$, at $t/\tau_L = 314$ with radiation reaction ($a_0=250$). The target width is $d=1.0\lambda_L$ in each case. \textcolor{black}{Moving window velocities are $\upsilon_{\rm mov} = 0.84\, c$ for dm target, and $\upsilon_{\rm mov}=0.8\,c$ for surface modulated and flat targets.} }
\end{figure}

\begin{figure}
\centering
\begin{tikzpicture}
	\node[above right] (img) at (0cm,0cm) {\includegraphics[width=0.65\textwidth]{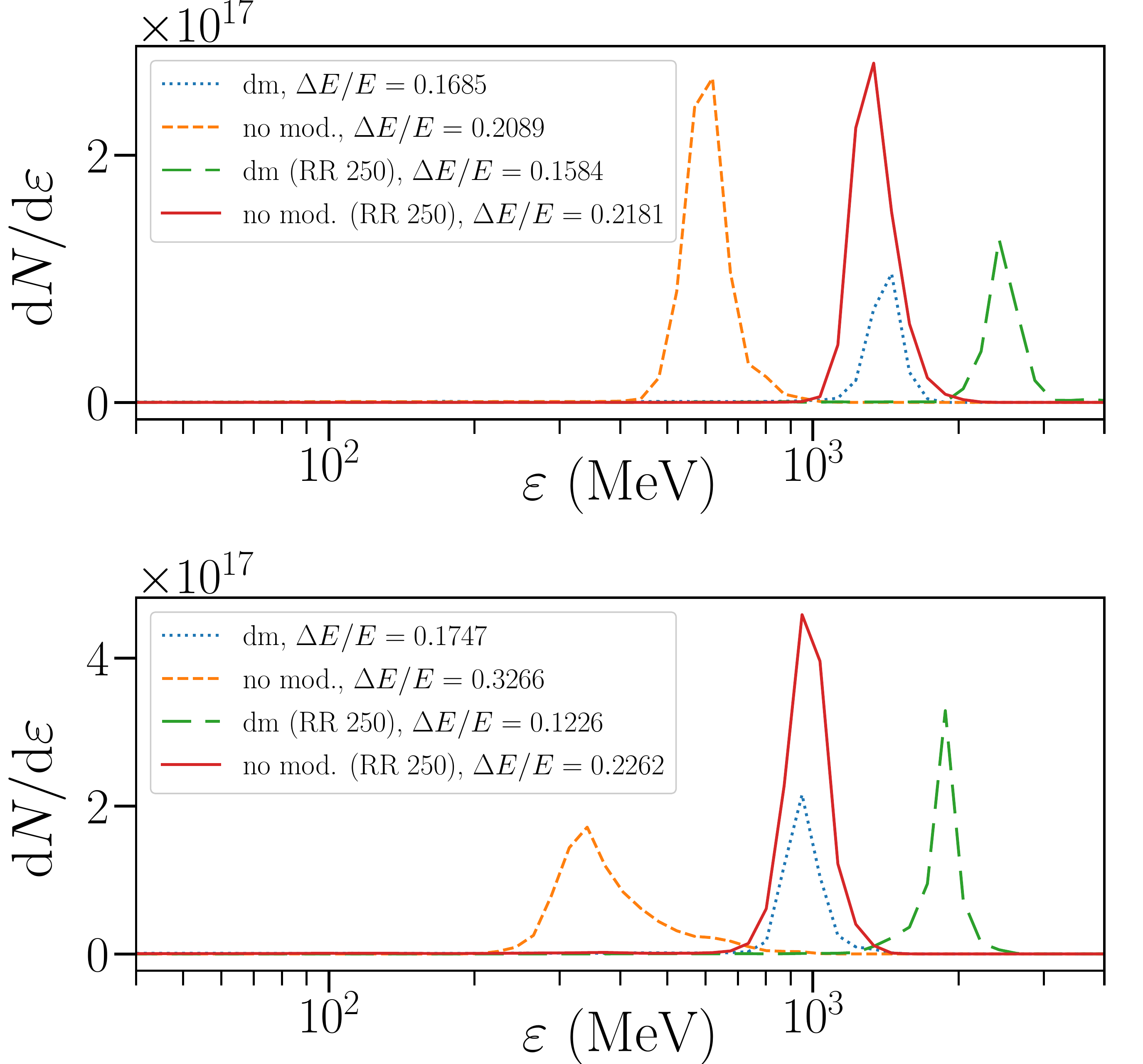}};
	%\node[above right] (img) at (0cm,3.1cm) {\includegraphics[width=0.45\textwidth]{Ion_energy1a}};
	\node at (8.25cm,3.3cm) {$\bm{(b)}$};
	\node at (8.25cm,7.6cm) {$\bm{(a)}$};
	%\node at (0.9cm,3.05cm) {\tiny $\times 10^{17}$};
	%\node at (0.9cm,5.9cm) {\tiny $\times 10^{16}$};
\end{tikzpicture}
\caption{\label{fig:Ekin_1l0_2l0_RR250} Kinetic energy of ions for different modulations for $k_m=2$, $a_m=0.25$, at $\bm{(a)}$ $t/\tau_L = 314$ ({$d=1.0\lambda_L$} target width) and $\bm{(b)}$ $t/\tau_L = 394$ ({$d=2.0\lambda_L$} target width) with radiation reaction ($a_0=250$). The dm (blue dotted) and no modulation ({orange} dashed) lines are for $a_0=150$.
\textcolor{black}{Panel (a), moving window velocities are  $\upsilon_{\rm mov}= 0.80\,c\, (0.75\,c)$ for flat, and $\upsilon_{\rm mov} = 0.84\,c\, (0.80\,c)$ for dm targets at $a_0=250$\, ($a_0=150$). 
Panel (b), moving window velocities are $\upsilon_{\rm mov} = 0.75\,c\, (0.75\,c)$ for flat, and $\upsilon_{\rm mov} = 0.75\,c \,(0.67\,c)$  for dm targets at $a_0=250$\, ($a_0=150$). }}
\end{figure}

\begin{figure}
\centering
	\begin{tikzpicture}
	\node[above right] (img) at (0cm,0cm) {\includegraphics[width=0.65\textwidth]{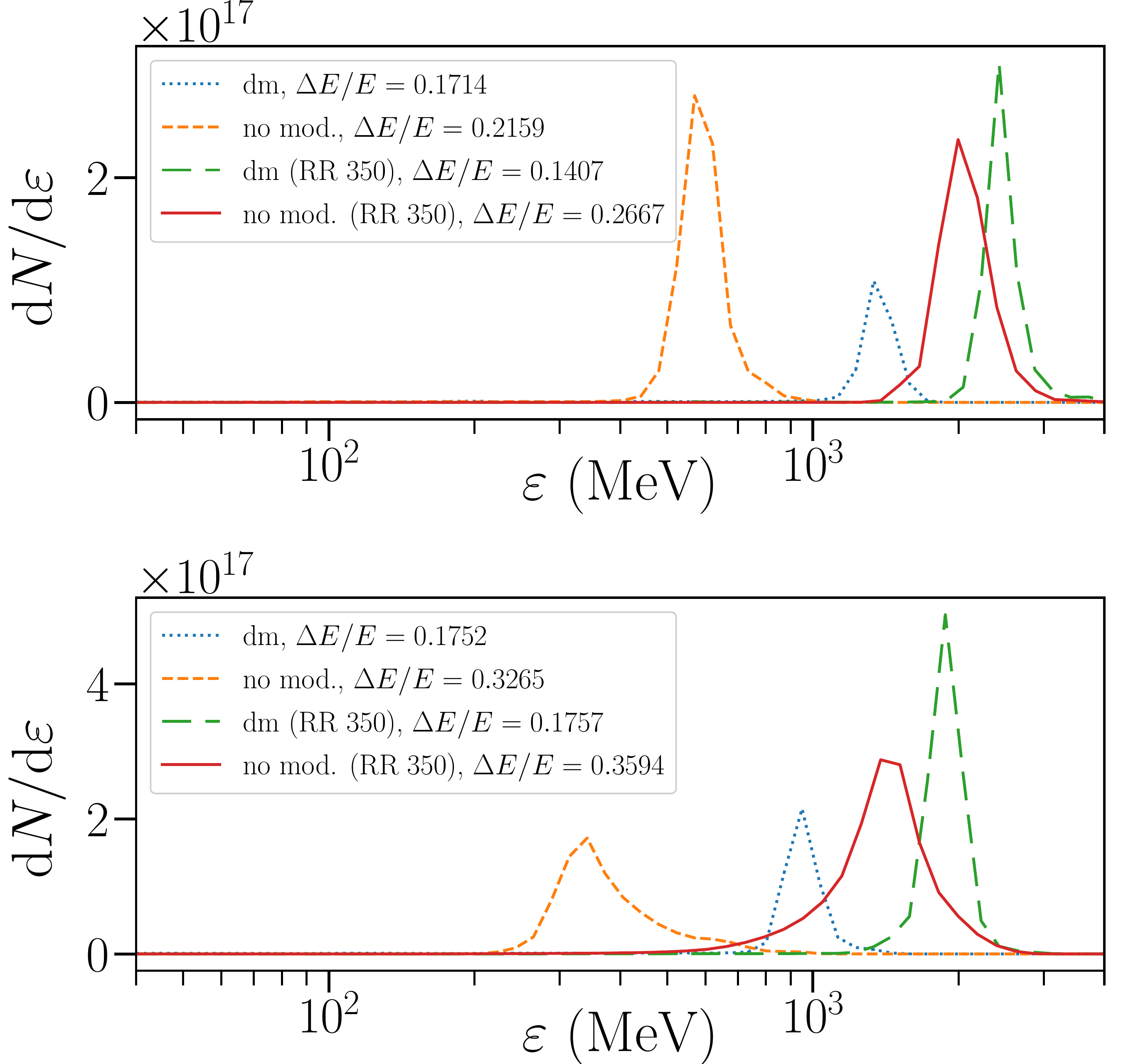}};
	%\node[above right] (img) at (0cm,3.1cm) {\includegraphics[width=0.45\textwidth]{Ion_energy1a}};
	\node at (8.25cm,3.3cm) {$\bm{(b)}$};
	\node at (8.25cm,7.6cm) {$\bm{(a)}$};
	%\node at (0.9cm,3.05cm) {\tiny $\times 10^{17}$};
	%\node at (0.9cm,5.9cm) {\tiny $\times 10^{16}$};
	\end{tikzpicture}
	\caption{\label{fig:Ekin_1l0_2l0_RR350} Kinetic energy of ions for different modulations, $k_m=2$, $a_m=0.50$, at $\bm{(a)}$ $t/\tau_L = 307$ for {$d=1.0\lambda_L$} target width. $\bm{(b)}$ $t/\tau_L = 394$  for {$d=2.0\lambda_L$} target width with radiation reaction ($a_0=350$). The dm (blue dotted) and no modulation ({orange} dashed) lines are for $a_0=150$. \textcolor{black}{For $a_0=350$, we have $\upsilon_{\rm mov} = 0.80\,c$ for flat targets (in both panels);  for dm targets, $\upsilon_{\rm mov} = 0.84\,c$ and $\upsilon_{\rm mov} = 0.75\,c$ in panels (a) and (b), respectively. Moving window velocity is same for $a_0=150$ as in Fig.\ref{fig:Ekin_1l0_2l0_RR250}. } }
\end{figure}

We also performed simulations including the effect of the radiation reaction on the RPA of ions from density modulated and structured targets. \textcolor{black}{SMILEI includes both Landau-Lifshitz and quantum description of the RR force~\cite{derouillat2018smilei}}. For this we used the same structured and density modulated targets as in Sec.~\ref{sec:Param}. Fig.\ref{fig:Ekin_RR} shows the results on the ion energy spectra from the structured and density modulated targets for $a_0=250$.  One can see that use of the density modulated and structured targets results in the lower FWHM of ions compared to a flat target. However, compared to the other structured targets, the biggest reduction  occurs in the case of density modulated target which shows the FWHM of ion energy spectrum to be $\sim 15\%$. For this value of $a_0$, one may begin to see the influence of the radiation reaction (RR) force in laser-plasma interaction. In order to further examine the role of the RR force and the target widths on ion acceleration from the density modulated target, we show in Fig.\ref{fig:Ekin_1l0_2l0_RR250} and Fig.~\ref{fig:Ekin_1l0_2l0_RR350} the best results with radiation reaction (RR) ($a_0=250$ for Fig.~\ref{fig:Ekin_1l0_2l0_RR250} and $a_0=350$ for Fig.~\ref{fig:Ekin_1l0_2l0_RR350}) for the two target widths (upper and lower panels). In order to compare the results, we kept $a_m$ and $k_m$ same in Fig.~\ref{fig:Ekin_1l0_2l0_RR250}(a) and ~\ref{fig:Ekin_1l0_2l0_RR250}(b) ($a_m=0.25$ and $k_m=2$) and also in Figs.~\ref{fig:Ekin_1l0_2l0_RR350}(a) and (b) ($a_m=0.5$ and $k_m=2$). Moreover, we also show the results for $a_0=150$ for comparison in each respective case, which facilitate the comparison with the respective no radiation reaction force limiting case since radiation reaction effects are significantly weaker at $a_0=150$. For target with $d=1.0\lambda_L$ width (upper panels) in each case, the ion energy gain is higher compared to the thicker target  $d=2.0\lambda_L$ (lower panels).  \textcolor{black}{Also ions gain larger energies for dm (dashed green line) target compared to the flat target (solid red line)  in Figs.\ref{fig:Ekin_1l0_2l0_RR250} and  \ref{fig:Ekin_1l0_2l0_RR350}. These trends can be explained on the basis of the lower target mass in respective cases, since the lower target mass is expected to result in ions acquiring higher energies in accordance with the scalings of RPA of ions; also discussed before in Sec.\ref{sec:Eprofile}~\cite{Macchi:2009tx}.} Also with the inclusion of the radiation reaction force, the density modulated target continues to show the higher ion energy gain and lower FWHM, compared to the flat target case. However, the trend with respect to FWHM of the ion energy spectra shows interesting features. The ion energy spread is lower for the the thicker target; the FWHM $\sim 12\%$ for {$d=2.0\lambda_L$} width (lower panel) at $a_0=250$  in Fig.\ref{fig:Ekin_1l0_2l0_RR250}(b). But it is smaller for the thinner target ({$d=1.0\lambda_L$}) case when the radiation reaction force is strong, see Fig.\ref{fig:Ekin_1l0_2l0_RR350}. Thus,   
the thinner target (upper panel in Fig.\ref{fig:Ekin_1l0_2l0_RR350}) shows not only higher ion energy ($\sim 2.5$ GeV) but also lower energy spread (FWHM$=14\%$), yielding the best results in the radiation dominated regime. Moreover, the density of the accelerated ion bunch is also higher at higher $a_0$ in Fig.\ref{fig:Ekin_1l0_2l0_RR350} compared to Fig.\ref{fig:Ekin_1l0_2l0_RR250}.
This highlights the nonlinear role of radiation reaction force for the density modulated target.

\section{\label{sec:interp} Interpretation of the PIC simulation results}

First we briefly discuss the theoretical analysis of the RTI-like transverse instability from the surface modulated targets. Afterwards, we carry out the Fourier analysis of the ion density oscillation and discuss the development of the RTI-like transverse instability for density modulated and structured targets including the effect of the RR force.

\subsection{\label{sec:theory} Theoretical analysis of the transverse instability from surface modulated targets}

To understand the behaviour of the transverse instability development, we calculate the growth rate of the RTI-like transverse instability in the RPA regime of ions for surface modulated targets. We wish to stress out that although we follow the analysis of Ref.\cite{Pegoraro:2007aa}, we analytically also consider the effect of the pre-imposed density modulation on the developement of the RTI-like instabilities, which was not considered before in Refs.\cite{Pegoraro:2007aa,Bulanov:2009aa}. Ref.\cite{Bulanov:2009aa} considered the effect of modulating the laser field in PIC simulations and theoretically allowing for temporal variation of the target mass density in the transverse direction. This is different from our set-up since we impose modulations which have only spatial dependence. The evolution of these modulations, in feeding different modes of the RTI, is self-consistently simulated in PIC simulations. Most importantly, our emphasis on explaining the competitive feeding of different modes of the RTI was not done in earlier works. Notwithstanding with the fact that the analysis is carried out for the surface modulated targets, one can also gain valuable physical insights for the density modulated target. \textcolor{black}{We also do not take into account the variation in the radiation pressure (included in PIC simulations) due to the surface density modulations as studied before in lower $a_0$ and $n_0$ limits~\cite{Sgattoni:2015aa,Eliasson:2015aa}. These studies suggest that pre-imposed surface modulations can lower the growth rate of short-wavelength perturbations of the RTI-like instability, and  importantly the growth rate of this instability becomes higher around the laser wavelength due to the plasmonic effects~\cite{Sgattoni:2015aa,Eliasson:2015aa}. In our simulations (shown later in Fourier transforms), we do not observe these trends. It appears that for our parameters (higher $a_0$ and $n_e$), plasmonic effects discussed before are not dominant and consequently we can ignore them in the theoretical analysis.} We briefly recall here the key points involved in the development of the analytical model to describe the interchange instability development for surface modulated targets. The Eq. of motion for a thin-foil target driven by the radiation pressure  is written as
 \begin{equation}\label{eq2}
\frac{d p_{i}}{d t}=\frac{\mathcal{N}}{\sigma_{0}}\,\epsilon_{i j k}\,\partial_{\zeta}x_{j}\,\partial_{\eta}x_{k},
\end{equation}
where $\mathcal{N}=(E^{2}/2\pi)(1-{\beta})/(1+{\beta}),\,{\beta}={\upsilon}/c$ is the relativistically invariant pressure, $E$ is the electric field of the laser pulse, $\epsilon_{i j k}$ is the Levi-Civita tensor, $\sigma_{0} = n_{0} l_{0}$ ($n_{0}$ and $l_{0}$ are the foil density and thickness respectively) is the initial surface mass density, $p_{x,y}=m_{i}\, c\, \beta_{x,y}\gamma, \gamma = (1-\beta^{2})^{-1/2}$, $\varphi,\, \zeta,\,\eta$  is a set of  curvilinear coordinate system  to describe the evolution of a differential element of the thin-foil. So motion of any point $\bm{r}$ on the surface of the thin-foil is defined as $\bm{r}[x(\xi,\,\zeta,\,\eta),\,y(\xi,\,\zeta,\,\eta),\,z(\xi,\,\zeta,\,\eta)]$. The $x$ and $y$ components of the \ref{eq2} reads as~\cite{Pegoraro:2007aa,Bulanov:2009aa}
\begin{eqnarray}
\frac{dp_{x}}{dt}=\frac{E^{2}}{2\pi \sigma_{0}} \frac{(m_{i}\,c\,\gamma_{0}-p_{x}^{0})}{(m_{i}\,c\,\gamma_{0}+p_{x}^{0})}\left[\frac{\partial y}{\partial \zeta}\frac{\partial z}{\partial \eta}-\frac{\partial z}{\partial \zeta}\frac{\partial y}{\partial \eta}\right],\\
\frac{dp_{y}}{dt}=\frac{E^{2}}{2\pi \sigma_{0}} \frac{(m_{i}\,c\,\gamma_{0}-p_{x}^{0})}{(m_{i}\,c\,\gamma_{0}+p_{x}^{0})}\left[\frac{\partial z}{\partial \zeta}\frac{\partial x}{\partial \eta}-\frac{\partial x}{\partial \zeta}\frac{\partial z}{\partial \eta}\right].
\end{eqnarray}
We investigate the stability of the thin foil in the long-wavelength limit (wavelength of perturbation higher than the thickness of the foil) by extending the approach of Ref.\cite{Pegoraro:2007aa} for surface modulated targets. The stability of the thin-foil target against the long-wavelength perturbation is important as long-wavelength perturbations are detrimental and lead to the breaking of the target.  We define $\varphi=\omega_{0}(t-x_{0}(t)/c)$ as a new variable. The initial conditions are
\begin{eqnarray}\label{initialConditions}
 \gamma=\gamma_{0},\,
  p_{x}=p_{x}^{0},\,
  p_{y}=0,\,
  x_{0}=0,\,
  y_{0}=\zeta+a_{m}\exp( \,i k_{m}\zeta),
\end{eqnarray} 
where $a_{m}$ denotes the depth of the modulation\footnote{It has to be of the order of or less than the non-relativistic plasma skin-depth $\sim c/\omega_{p}$ in order to avoid strong plasma electron heating.}, while $k_{m}$ is the modulation wave-vector. We again wish to stress out that Refs.\cite{Pegoraro:2007aa,Bulanov:2009aa} take $y_{0}=\zeta$, thus not including the effect of pre-imposed density modulations on the growth of the RTI-like instabilities. The $x$-component of the momentum on solving gives
\begin{gather}
 p_{x}^{0} = L(\varphi)\frac{[1+2 L(\varphi)]}{(2[1+L(\varphi)])}, \nonumber \\
 L(\varphi) = R(\varphi)(1+i\,k_{m}l_{m}),\quad l_{m}=a_{m}\exp( \, ik_{m}\zeta),\nonumber \\ R(\varphi)=\int_{0}^{\varphi}\Delta(\varphi^{'})\,d\varphi^{'}/\lambda_{0},\quad \Delta(\varphi)=E^{2}(\varphi)/m_{i}\omega_{0}^{2}\sigma_{0}. 
\end{gather}  

\begin{figure}
	\centering
	\begin{tikzpicture}
	\node[above right] (img) at (0cm,0cm) {\includegraphics[width=0.60\textwidth]{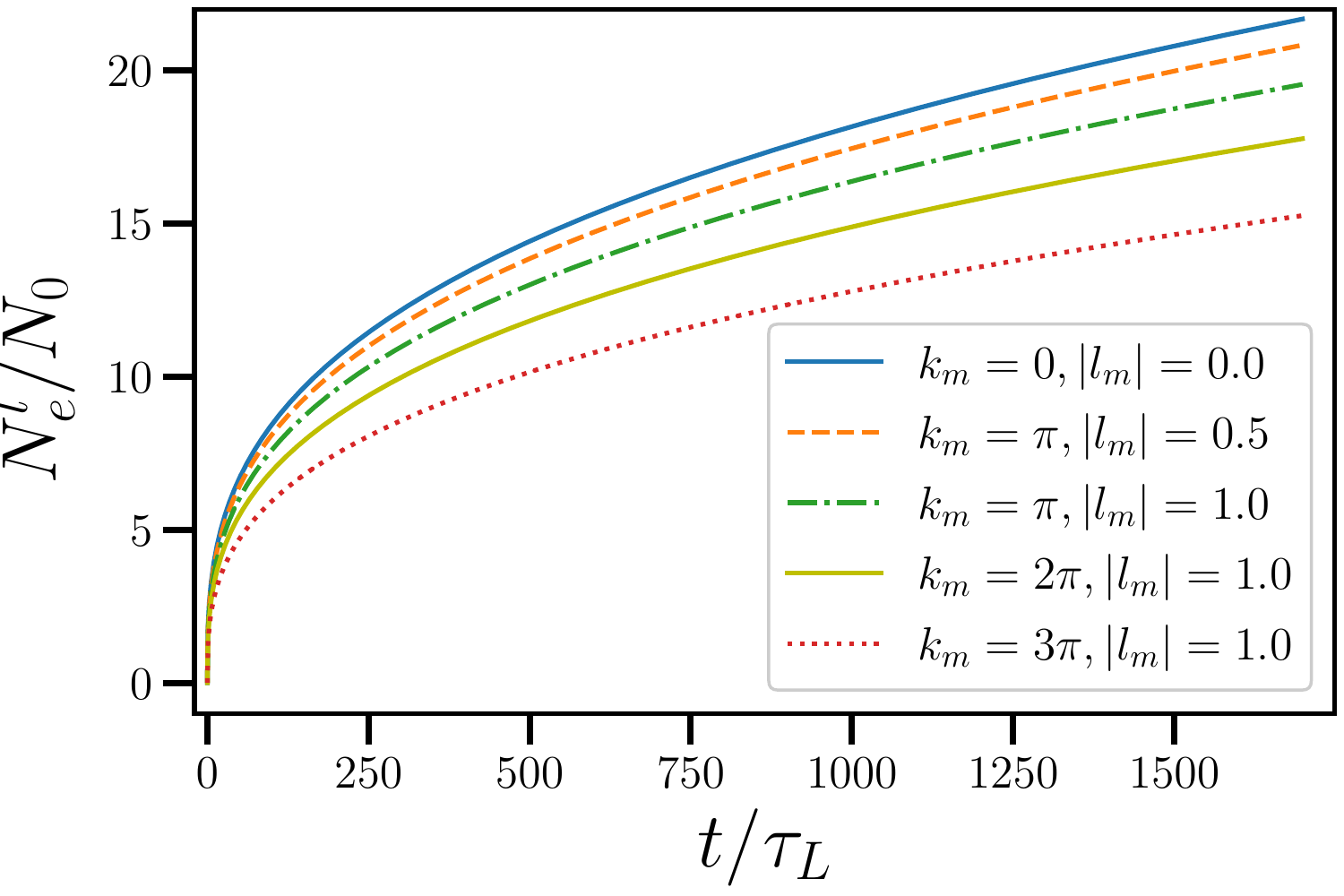}};
	%\node[above right] (img) at (0cm,3.1cm) {\includegraphics[width=0.45\textwidth]{Ion_energy1a}};
	%\node at (6.4cm,5.3cm) {$\bm{(b)}$};
	%\node at (6.4cm,11.0cm) {$\bm{(a)}$};
	%\node at (0.9cm,3.05cm) {\tiny $\times 10^{17}$};
	%\node at (0.9cm,5.9cm) {\tiny $\times 10^{16}$};
	\end{tikzpicture}
	\caption{\label{fig:growth_rate} Number of e-foldings for the late-time growth of the perturbations; see Eq.~\eqref{eq:5}. The reduction in the e-folding is apparent in  the case of surface modulations.}
\end{figure}
\noindent
For a constant amplitude pulse $E(\varphi)=E_{0},\, \Delta(\varphi)=\Delta_{0}$, for $\omega_{0}\,t\ll (\lambda_{0}/\Delta_{0})$ (early-time), we have $\varphi\approx \omega_{0} t$, and for $\omega_{0}\,t\gg (\lambda_{0}/\Delta_{0})$ (late-time), we have $\varphi^{3}=(\omega_{0}\,t) \, 6 \lambda_{0}^{2}/\Delta_{0}^{2}(1+ k_{m}^2l_{m}^2/4)$. On perturbing the equilibrium as
\begin{eqnarray*}
x=\delta x,\quad y=\zeta+l_m+\delta y, \quad l_m=a_{m}\exp( \,i k_{m}\zeta),
\end{eqnarray*} 
we get the following Eqs. for the $x$ and $y$ components as in Refs.\cite{Pegoraro:2007aa,Bulanov:2009aa},
\begin{eqnarray}
\frac{\partial}{\partial \varphi} \left[\frac{p_{x}^{0}}{m_{i}c} \frac{\partial \delta x}{\partial \varphi}  \right] = \frac{\Delta(\varphi)}{2\pi} \frac{\partial \delta y}{\partial \zeta} , \\ 
\frac{\partial}{\partial \varphi} \left[\frac{m_{i}c}{p_{x}^{0}} \frac{\partial \delta y}{\partial \varphi}  \right] = -\frac{\Delta(\varphi)}{2\pi} \frac{\partial \delta x}{\partial \zeta}.
\end{eqnarray}
Assuming the perturbation of the form $\delta x,\,\delta y \sim \exp( \,\int_{0}^{\varphi}\Gamma (\varphi^{'})d\varphi^{'}-i q \zeta)$, and $\partial \Gamma/\partial \varphi \ll \Gamma^{2}$ and $\Gamma \gg 1$, we get the growth rate of long wavelength perturbations as $\Gamma= (\Delta(\varphi) q/2\pi)^{1/2}$. However, the actual growth of the perturbation for a constant amplitude pulse is written as $\delta x,\delta y \sim e^{\Gamma \varphi -i q \zeta}$. On using the relation between the variables $\varphi$ and $t$ for early and late times, we get the number of e-foldings for early and late-times growths of the perturbation (see also Fig.\ref{fig:growth_rate}) as

\begin{gather}
N^{e}_{e} =  \left(\frac{q \Delta_{0}}{2\pi}\right)^{1/2} \omega_{0} t,\nonumber \\
N^{l}_{e} =   \left(\frac{q \Delta_{0}}{2\pi}\right)^{1/2}
\left(\frac{\lambda_{0}}{\Delta_{0}}\right)^{2/3}
\frac{(6\omega_{0} t)^{1/3}}{(1+k_{m}^2l_{m}^2/4)^{1/12}} =  N_0 \,\frac{(6 \,t/\tau_{L})^{1/3}}{(1+k_{m}^2l_{m}^2/4)^{1/12}}.\label{eq:5}
\end{gather}
\noindent
\textcolor{black}{One may note that the early time asymptote of the instability shows no dependence on the pre-imposed modulations. This justifies the assumption of taking the equilibrium solution for a flat target and imposing the modulations in the initial conditions as done in Eq.\eqref{initialConditions} }.
{From here the role of pre-imposed density modulations in reducing the growth rate of the RTI-like instabilities is apparent (see Fig.\ref{fig:growth_rate}).} In the case of no modulation $(k_{m}=0)$, we recover the same growth of the perturbation as in Refs.~\cite{Pegoraro:2007aa,Bulanov:2009aa}. One may observe that during the early stage, surface modulations do not play any role in the growth  of the perturbation.  However, for late-time  of the instability development, modulations tend to lower the growth of the instability. In fact, for $|k_{m}l_{m}|\gg 4$ (short wavelength modulation), the growth of the long-wavelength modes of the instability reads as $N_{e}^{l} \propto  t^{1/3}/ (k_{m}l_{m}/2)^{1/6}$. \textcolor{black}{ This clearly shows reduction in the growth of the short-wavelength perturbation, consistent with the results presented before~\cite{Sgattoni:2015aa,Eliasson:2015aa}.} In the opposite limit $|k_{m}l_{m}|\ll 4$ (long wavelength modulation), there is no reduction in the growth rate of the long-wavelength perturbation. \textcolor{black}{However, the introduction of the short-wavelength modulation amounts to selectively feeding the short-wavelength modes of the instability. This selective feeding can suppress the generation of the long-wavelength modes of the instability, which are detrimental for the stability of the target. In the opposite case of the pre-imposed long wavelength density modulations, the long-wavelength modes of the instability grow faster to break the target. Consequently, one can expect to get lower energy spread in the ion energy spectra for the pre-imposed short-wavelength density modulations.} Fig.\ref{fig:smt_int} qualitatively agrees with this scaling and shows better agreement for the rippled structured target [Fig.\ref{fig:shapes}(d)] for which the theoretical analysis is most suited. Other structured targets, except the density modulated target [Fig.\ref{fig:Interpoldm}], also show similar trends with the theoretical analysis.

\subsection{\label{Insta_ana} Fourier analysis of the ion density}

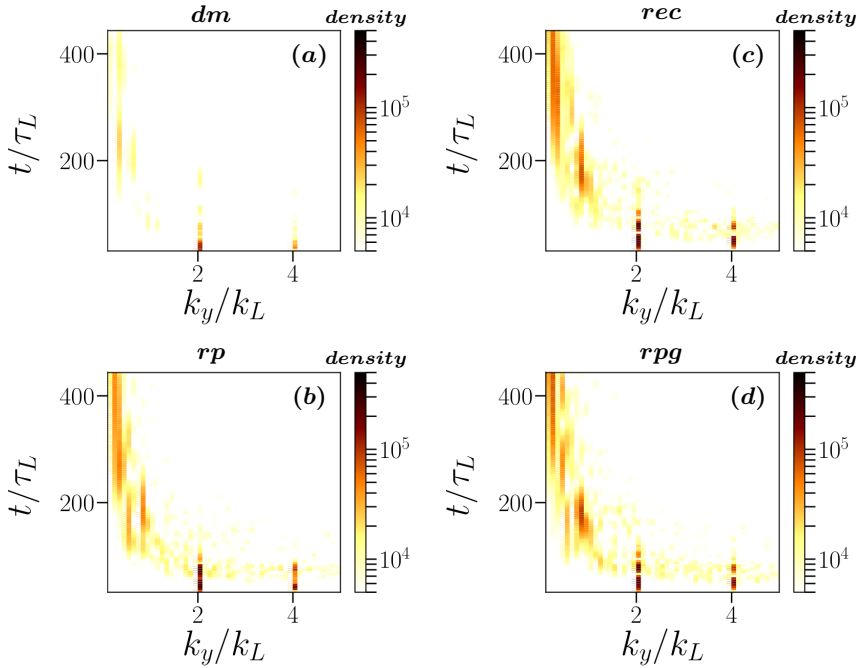
\begin{figure}
\centering
\begin{tikzpicture}
	\node[above right] (img) at (0cm,4.5cm) {\includegraphics[width=0.40\textwidth]{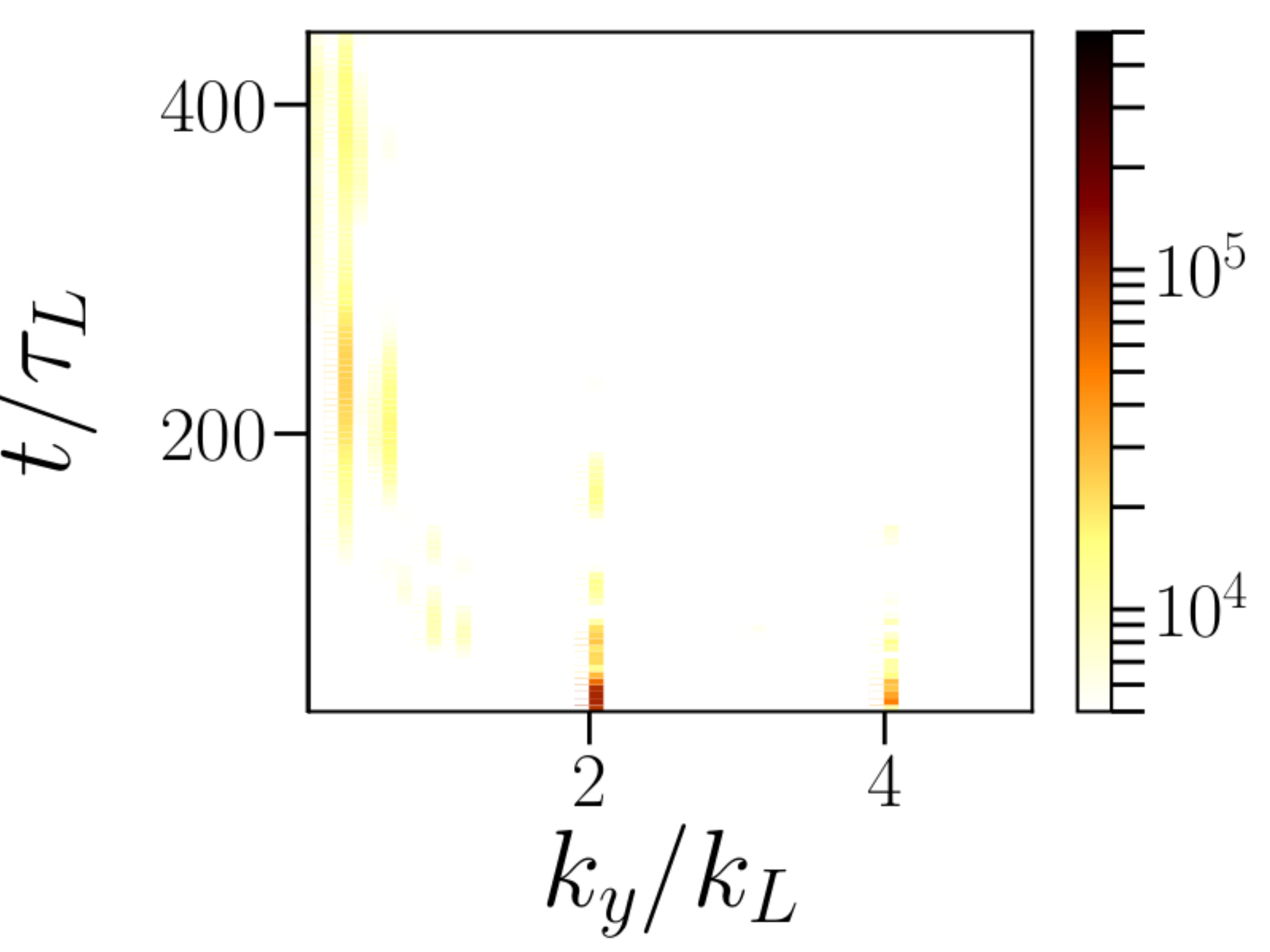}};
	\node[above right] (img) at (5.8cm,4.5cm) {\includegraphics[width=0.40\textwidth]{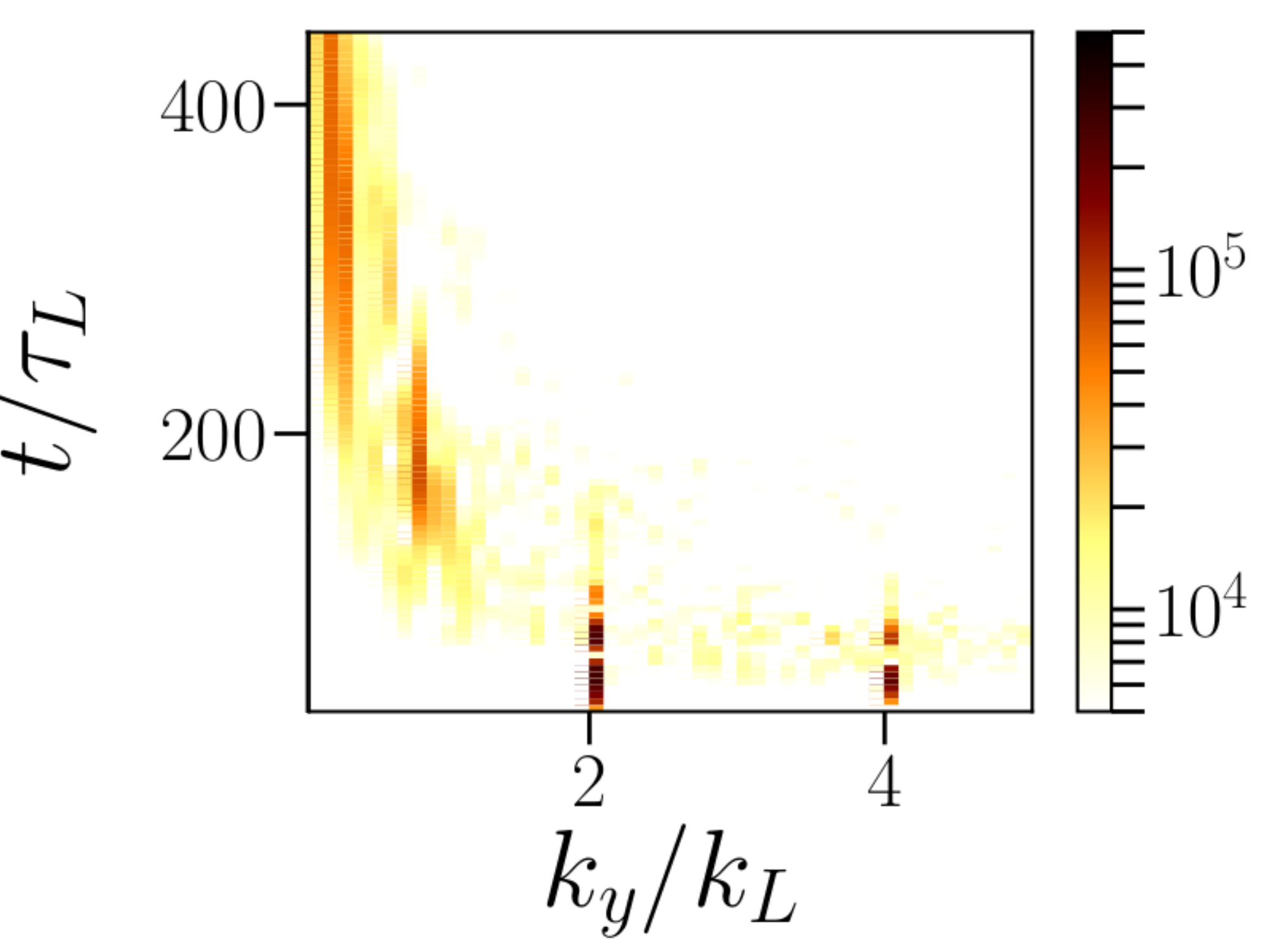}};	
	\node[above right] (img) at (0cm,0.0cm) {\includegraphics[width=0.40\textwidth]{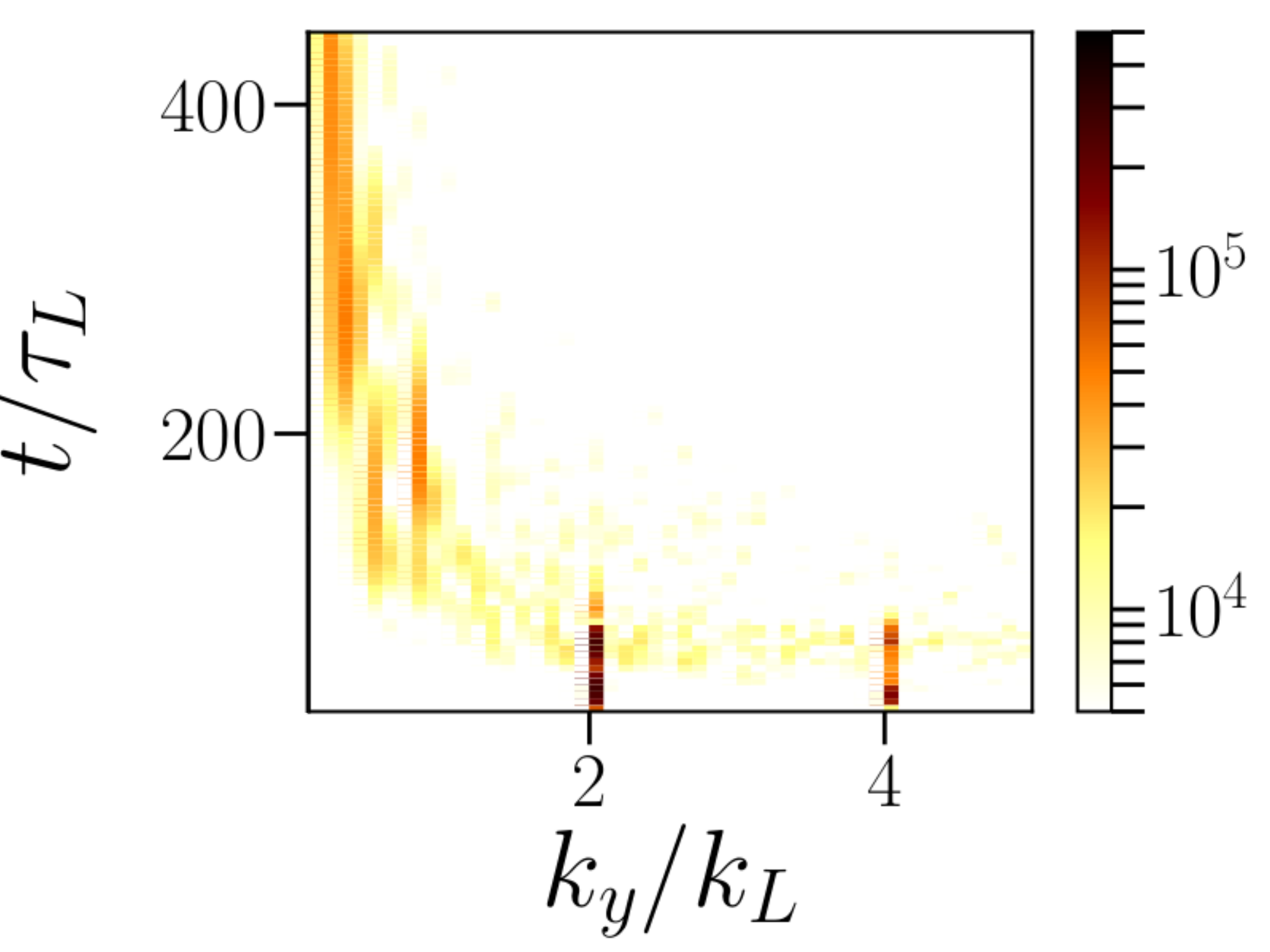}};
	\node[above right] (img) at (5.8cm,0.0cm) {\includegraphics[width=0.40\textwidth]{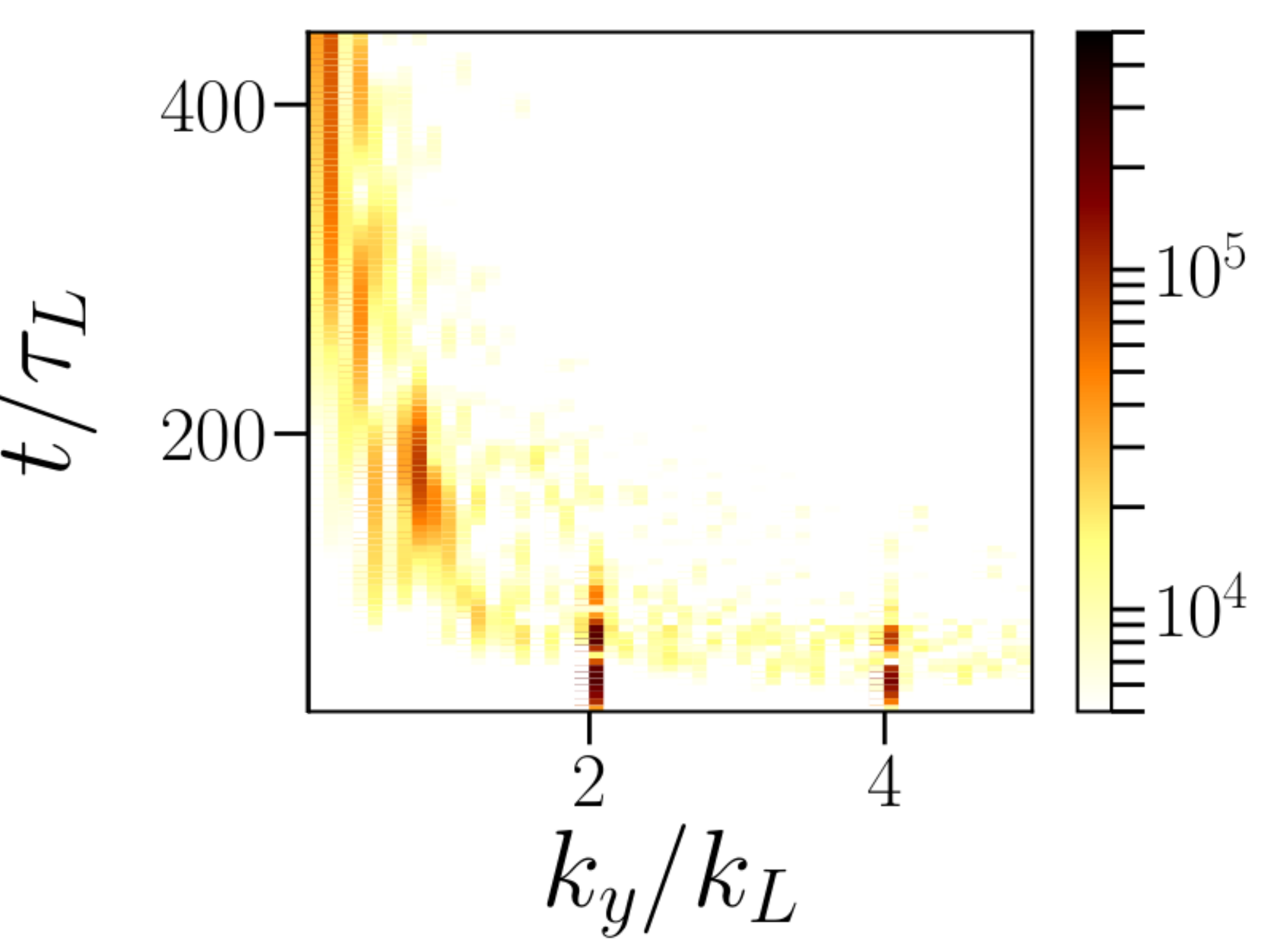}};
	\node at (4.1cm,3.7cm) {$\bm{(b)}$};
	\node at (4.1cm,8.2cm) {$\bm{(a)}$};
	\node at (9.9cm,3.7cm) {$\bm{(d)}$};
	\node at (9.9cm,8.2cm) {$\bm{(c)}$};
	
	\node at (8.75cm,8.75cm) {$\bm{rec}$};
	\node at (10.8cm,8.7cm) {\begin{scriptsize}$\bm{density}$\end{scriptsize}};
	\node at (2.8cm,8.75cm) {$\bm{dm}$};
	\node at (4.85cm,8.7cm) {\begin{scriptsize}$\bm{density}$\end{scriptsize}};
	
	\node at (2.8cm,4.25cm) {$\bm{rp}$};
	\node at (10.8cm,4.2cm) {\begin{scriptsize}$\bm{density}$\end{scriptsize}};
	\node at (8.75cm,4.25cm) {$\bm{rpg}$};
	\node at (4.85cm,4.2cm) {\begin{scriptsize}$\bm{density}$\end{scriptsize}};
\end{tikzpicture}
\caption{\label{fig:fft_mods} Evolution of the FFT of the ion density oscillations with ($k_y/k_L$) for $\bm{(a)}$ the density modulated target, $\bm{(b)}$ the rp structured, $\bm{(c)}$ the rec structured and $\bm{(d)}$ the rpg structured targets. The modulation parameters are $a_m = 0.25$, $k_m = 2$ and the target width is {$d=1.0\lambda_L$} in each case. \textcolor{black}{The FFT spectra for the flat target is shown in the upper row of Fig.\ref{fig:fft_no_mod} }}
\end{figure}

\begin{figure}
\centering
\begin{tikzpicture}
	\node[above right] (img) at (0cm,4.5cm) {\includegraphics[width=0.40\textwidth]{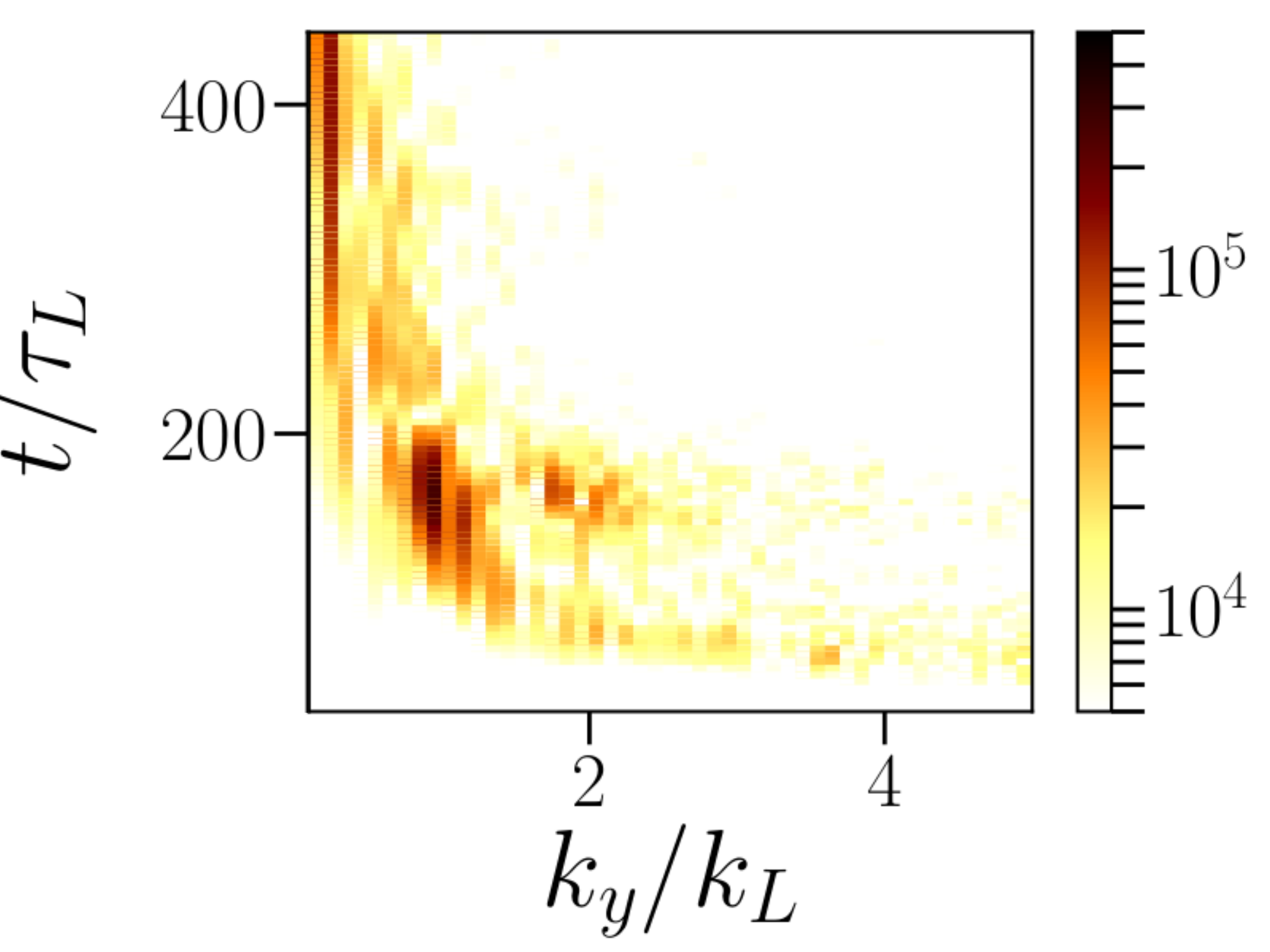}};
	\node[above right] (img) at (5.8cm,4.5cm) {\includegraphics[width=0.40\textwidth]{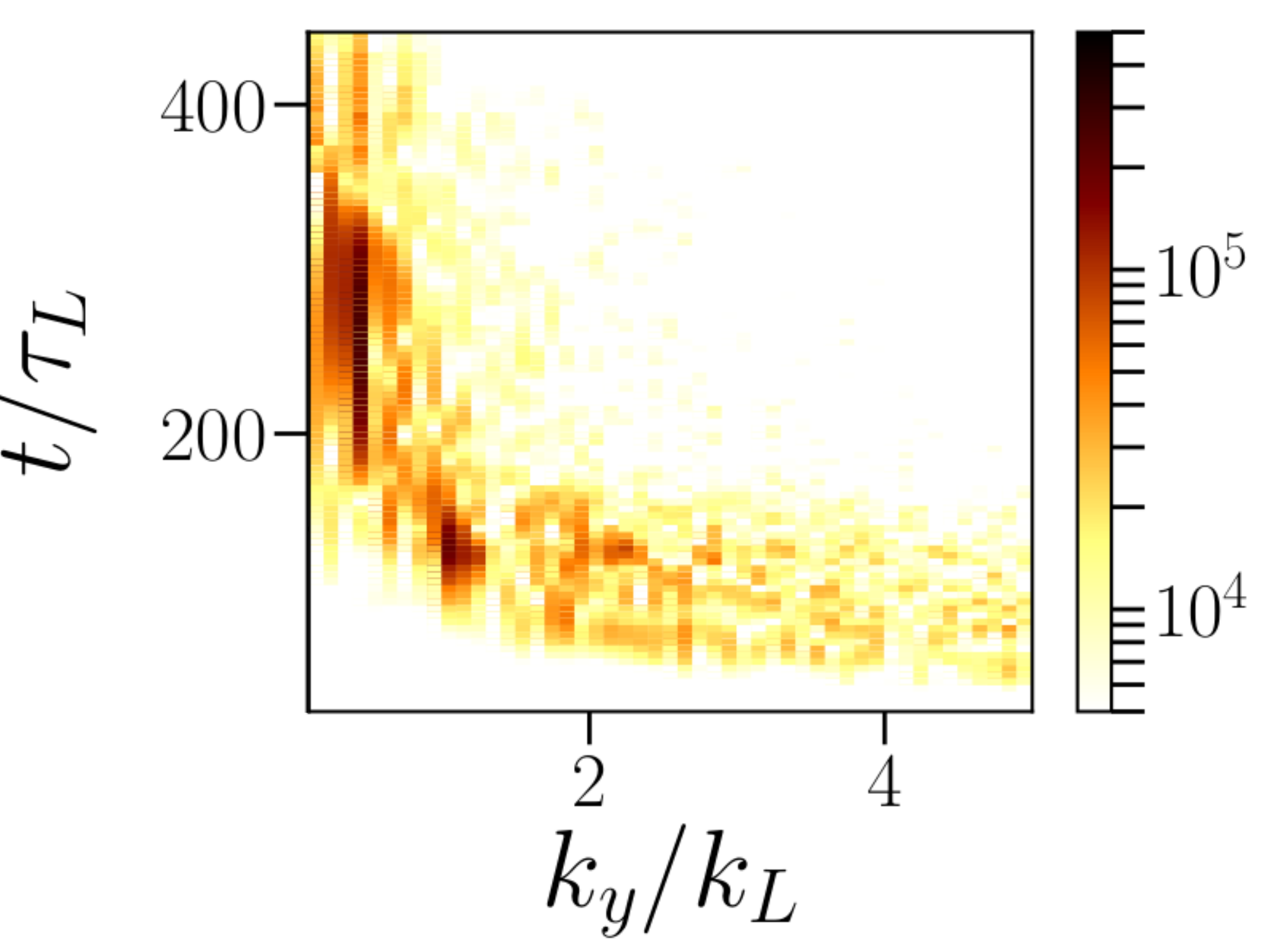}};
	\node[above right] (img) at (0cm,0.0cm) {\includegraphics[width=0.40\textwidth]{figure11a12c}};
	\node[above right] (img) at (5.8cm,0.0cm) {\includegraphics[width=0.40\textwidth]{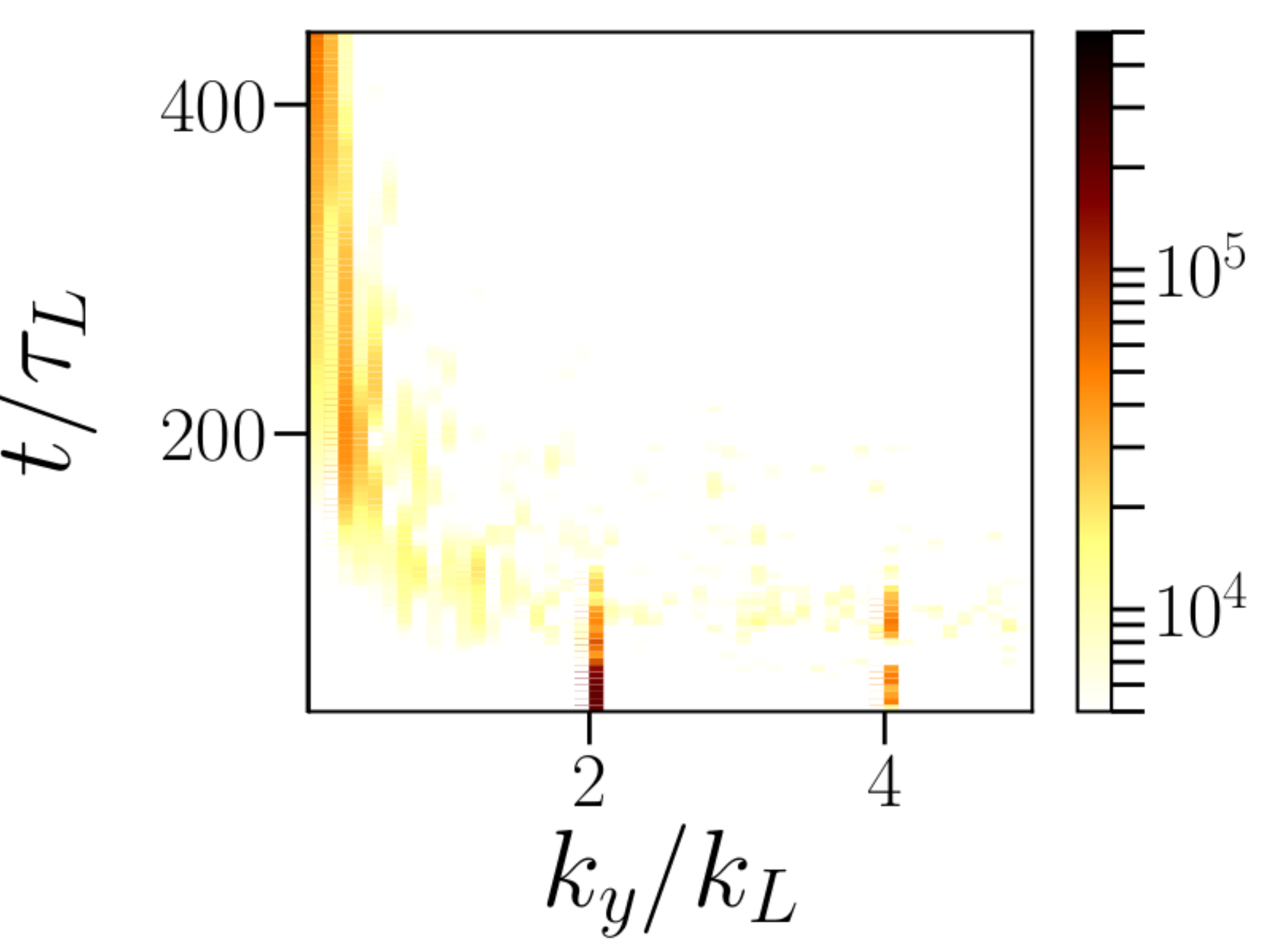}};
	\node at (4.1cm,3.7cm) {$\bm{(c)}$};
	\node at (4.1cm,8.2cm) {$\bm{(a)}$};
	\node at (9.9cm,3.7cm) {$\bm{(d)}$};
	\node at (9.9cm,8.2cm) {$\bm{(b)}$};
	
	\node at (8.75cm,8.75cm) {$\bm{flat, d=2.0\lambda_L}$};
	\node at (10.8cm,8.7cm) {\begin{scriptsize}$\bm{density}$\end{scriptsize}};
	\node at (2.8cm,8.75cm) {$\bm{flat, d=1.0\lambda_L}$};
	\node at (4.85cm,8.7cm) {\begin{scriptsize}$\bm{density}$\end{scriptsize}};
	
	\node at (2.8cm,4.25cm) {$\bm{dm, d=1.0\lambda_L}$};
	\node at (10.8cm,4.2cm) {\begin{scriptsize}$\bm{density}$\end{scriptsize}};
	\node at (8.75cm,4.25cm) {$\bm{dm, d=2.0\lambda_L}$};
	\node at (4.85cm,4.2cm) {\begin{scriptsize}$\bm{density}$\end{scriptsize}};
\end{tikzpicture}
\caption{\label{fig:fft_no_mod} Time development $(t/\tau_L)$ of the FFT of the ion density of a flat target (upper row), and for a density modulated target (bottom row) with the normalized wave vector \(k_y/k_L\). The first and second columns are for {$d=1.0\lambda_L$}  and {$d=2.0\lambda_L$} target widths in respective cases. \textcolor{black}{The modulation parameters are $a_m = 0.25$, $k_m = 2$.} }
\end{figure}

In order to understand the instability growth and development in the nonlinear stage, we look into the spatial Fourier spectra of the protons. It is obtained by taking the Fast Fourier transform (FFT) of the ion density distribution:

\begin{equation}
n(k,t) = \int_{0}^{L_y} \int_{0}^{L_x} n(x,y,t) e^{iky} dx\, dy,
\label{eq:12}
\end{equation}
\noindent where \(n(x,y,t)\) is averaged in $x$-direction and the FFT is taken along the $y$-direction. Fig.\ref{fig:fft_mods} shows temporal evolution of the FFT of proton density oscillations for the density modulated and structured targets. On comparing the figures, one can see the presence of modes at $k/k_L=1, 2,4$, signifying the role of density and surface modulations in the RPA of ions.  The ion density oscillations with $k/k_L \le 1$ are detrimental for the stability of the target as they tend to break the target at later times. On comparing with Fig.\ref{fig:fft_no_mod}(a) one can see that there is a significant suppression of the modes at $k/k_L \le 1$ and instead the modes at $k/k_L=1, 2,4$ are stronger. This is selective feeding of the modes as discussed before in Sec.\ref{sec:theory}. \textcolor{black}{Since for a target of thickness $\sim \lambda_L$, any transverse instability modes with $k_y/k_L \le 1$ can break the target easily. The modes at $k/k_L=1, 2,4$ (shorter wavelengths) are not detrimental for the stability of $\lambda_L$ thickness target, and one can expect a better RPA of ions for density and surface modulated targets.} One can see from Fig.\ref{fig:fft_mods}(a) that the density modulated target is most effective at suppressing the long wavelength modes ($k/k_L \le 1$) compared to other structured targets. Since this reduction is pronounced in the case of density modulated target, we compare the temporal evolution of FFT of ion density oscillations of a density modulated target with a flat target in 
Fig.\ref{fig:fft_no_mod}. Upper row of Fig.\ref{fig:fft_no_mod} shows the temporal evolution of FFT of ion density oscillation for a flat target of widths {$d=1.0\lambda_L$} [panel (a)] and {$d=2.0\lambda_L$} [panel (b)] while lower row shows the corresponding cases for the density modulated targets. One can clearly see that for the flat target (upper row) the ion density oscillations have wavelengths extending up to $\lambda/\lambda_L \ge 0.25$. For the thin target ({$d=1.0\lambda_L$}, first column), the dominant mode of the RTI-like transverse instabilities is concentrated around $\lambda/\lambda_L \approx 1$, while for the thicker target ({$d=2.0\lambda_L$}, second column) the dominant mode of the ion density oscillations is located around $\lambda \le \lambda_L$. At later times, the ion density oscillations exhibit oscillations at wavelengths $\lambda \ge 0.5\lambda_L$. These longer wavelengths modes are responsible for breaking the target and hence are detrimental for the stable RPA of ions. While for the density modulated target, appearance of these longer wavelength modes has considerably suppressed, though the thicker target [panel (d)] appears to show the excitation of weaker longer wavelength modes at later times. This further confirms that thinner targets are optimum for RPA of ions. For thicker targets ($d > \lambda_L$), it is difficult to suppress the long-wavelength modes of RTI-like transverse instabilities. 

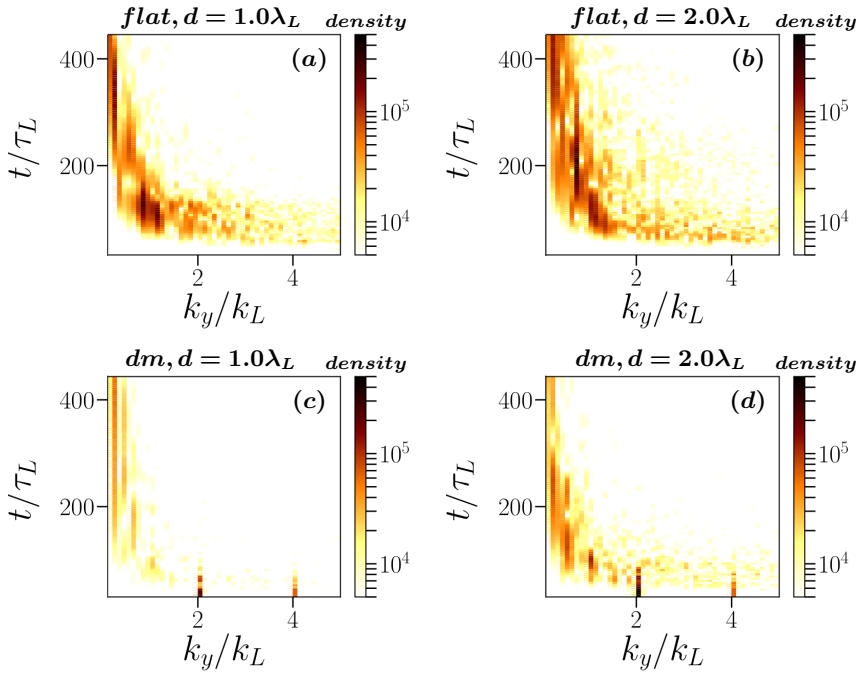
\begin{figure}
\centering
\begin{tikzpicture}
	\node[above right] (img) at (0cm,4.5cm) {\includegraphics[width=0.40\textwidth]{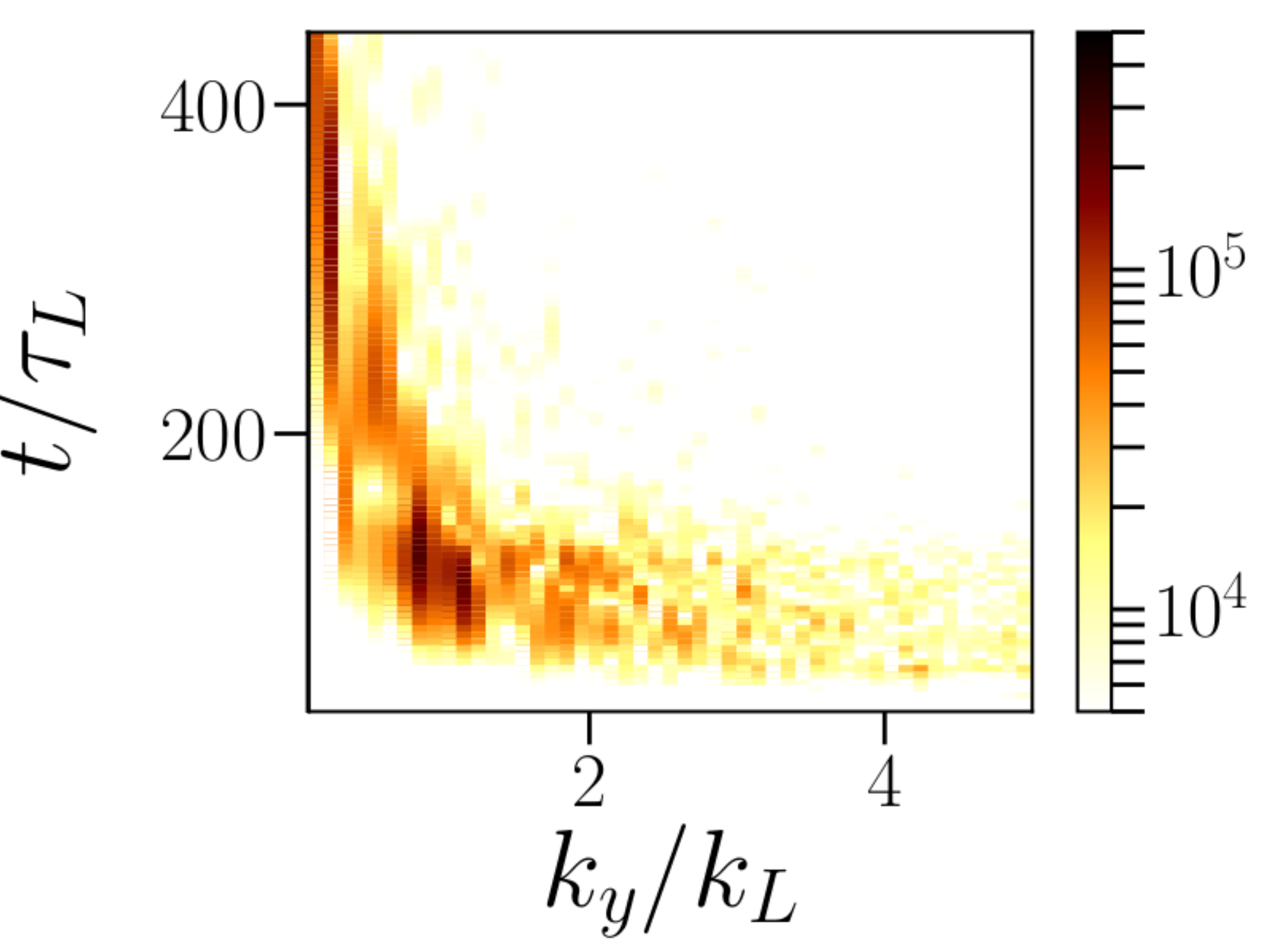}};
	\node[above right] (img) at (5.8cm,4.5cm){\includegraphics[width=0.40\textwidth]{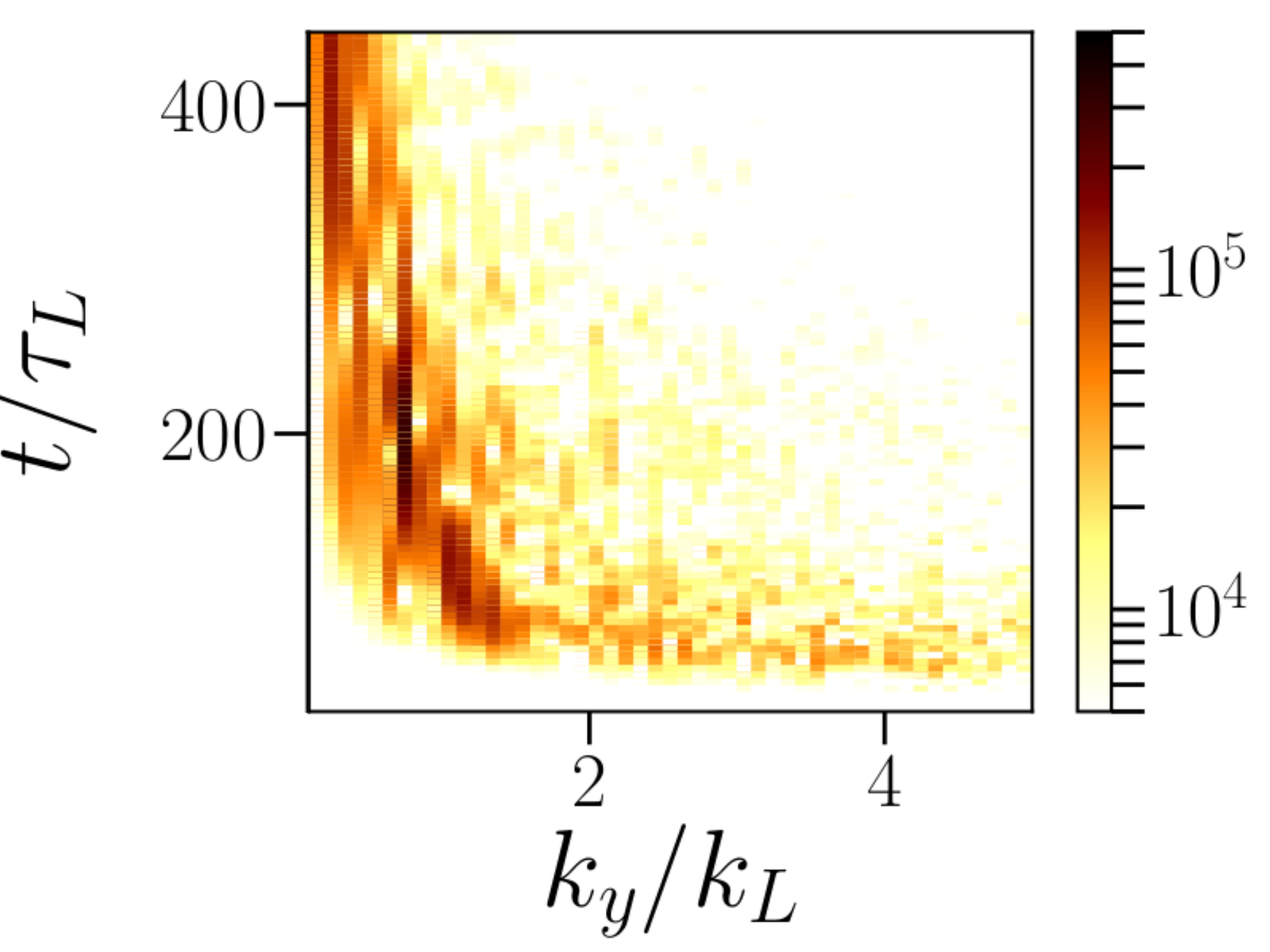}};
	\node[above right] (img) at (0cm,0.0cm){\includegraphics[width=0.40\textwidth]{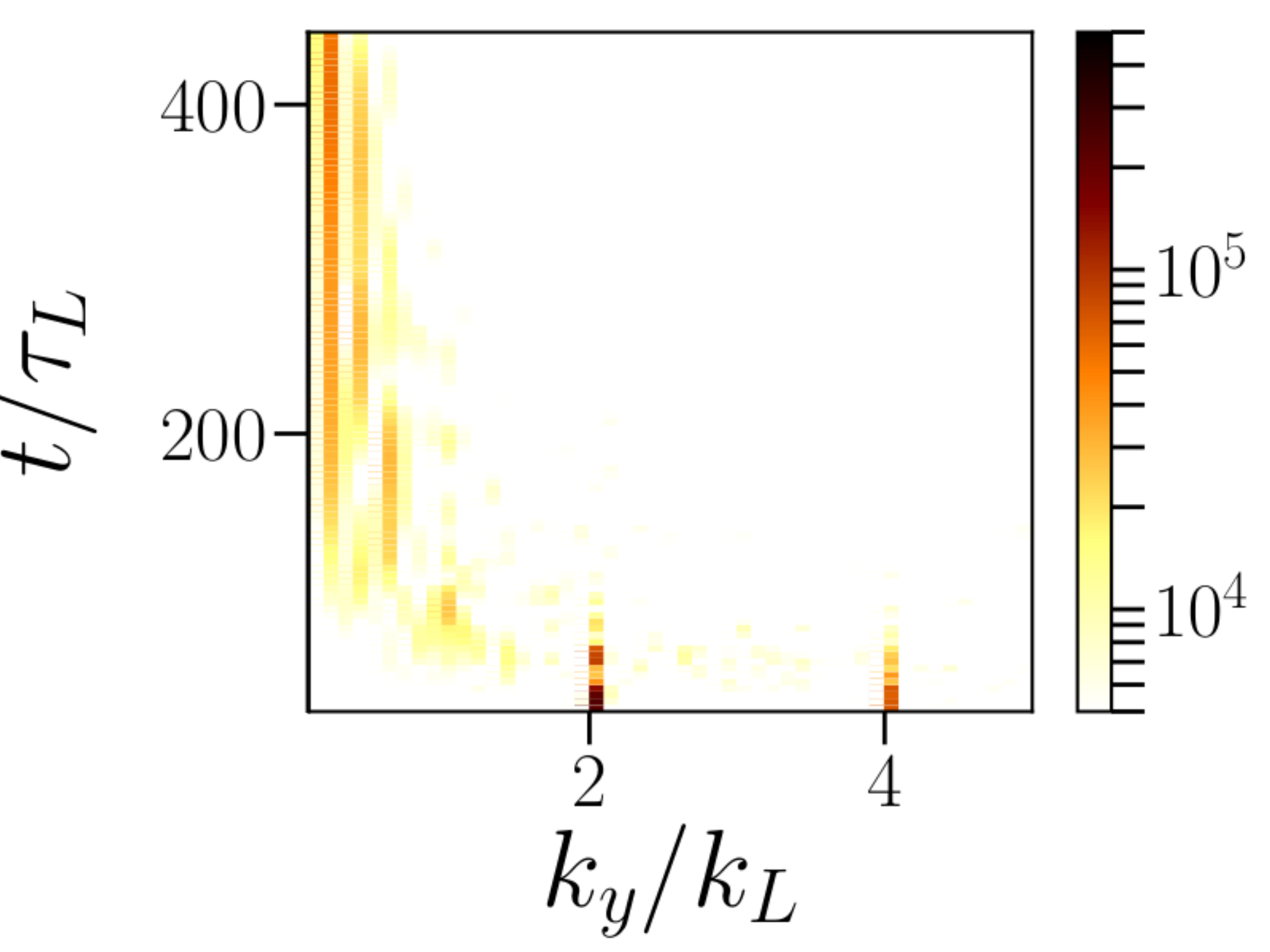}};
	\node[above right] (img) at (5.8cm,0.0cm){\includegraphics[width=0.40\textwidth]{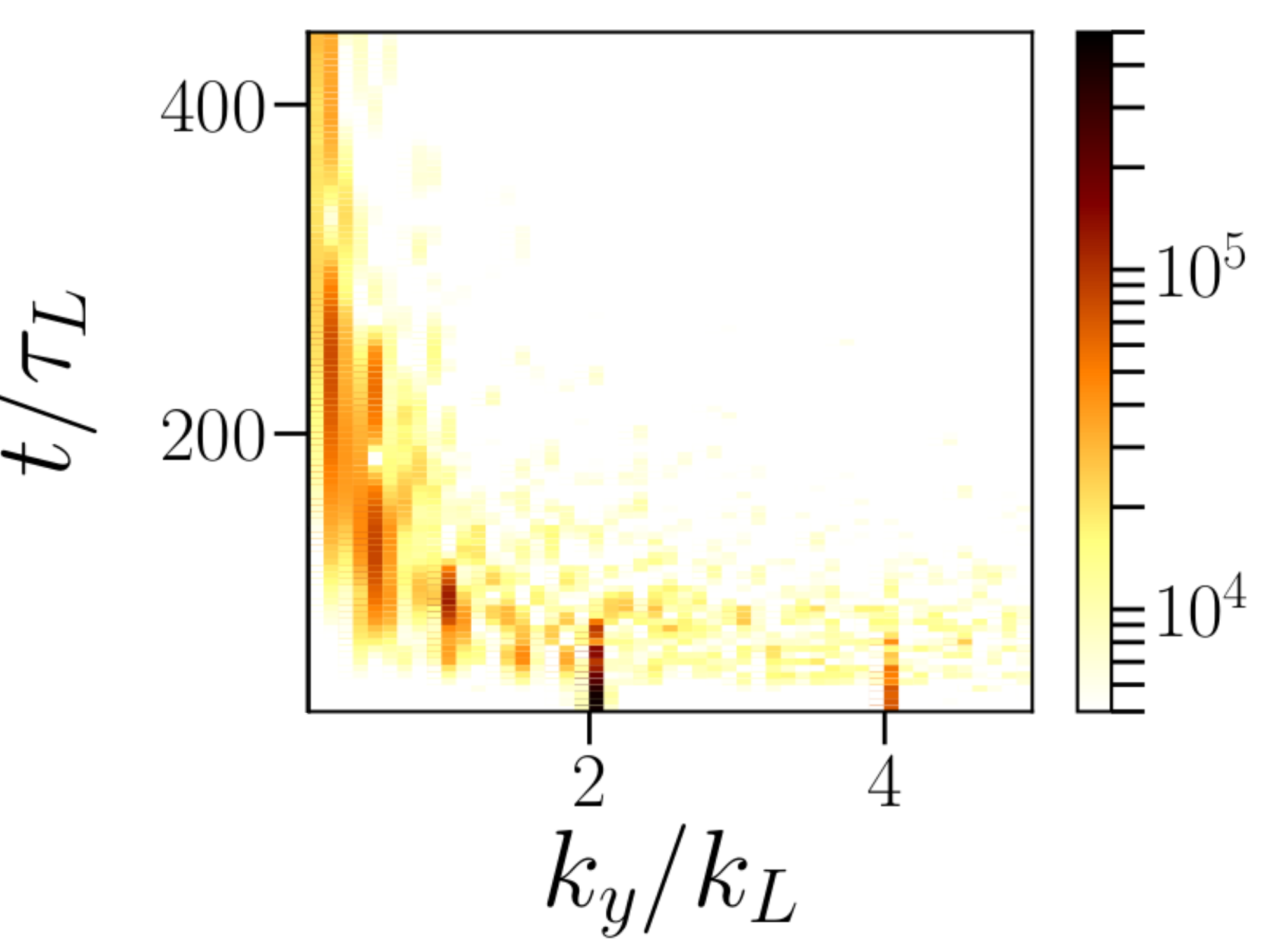}};	
	\node at (4.1cm,3.7cm) {$\bm{(c)}$};
	\node at (4.1cm,8.2cm) {$\bm{(a)}$};
	\node at (9.9cm,3.7cm) {$\bm{(d)}$};
	\node at (9.9cm,8.2cm) {$\bm{(b)}$};
	
	\node at (8.75cm,8.75cm) {$\bm{flat, d=2.0\lambda_L}$};
	\node at (10.8cm,8.7cm) {\begin{scriptsize}$\bm{density}$\end{scriptsize}};
	\node at (2.8cm,8.75cm) {$\bm{flat, d=1.0\lambda_L}$};
	\node at (4.85cm,8.7cm) {\begin{scriptsize}$\bm{density}$\end{scriptsize}};
	
	\node at (2.8cm,4.25cm) {$\bm{dm, d=1.0\lambda_L}$};
	\node at (10.8cm,4.2cm) {\begin{scriptsize}$\bm{density}$\end{scriptsize}};
	\node at (8.75cm,4.25cm) {$\bm{dm, d=2.0\lambda_L}$};
	\node at (4.85cm,4.2cm) {\begin{scriptsize}$\bm{density}$\end{scriptsize}};
\end{tikzpicture}
\caption{\label{fig:fft_mods_RR250} \textcolor{black}{Time development of the FFT of ion density oscillations including radiation reaction force for the flat target (upper row) and the density modulated target (bottom row) at $a_0=250$. First and second columns correspond to the target widths of {$d=1.0\lambda_L$} and {$d=2.0\lambda_L$}, respectively. The modulation parameters are $a_m = 0.50, k_m = 2$. }}
\end{figure}

\begin{figure}
\centering
\begin{tikzpicture}
	\node[above right] (img) at (0cm,4.5cm){\includegraphics[width=0.40\textwidth]{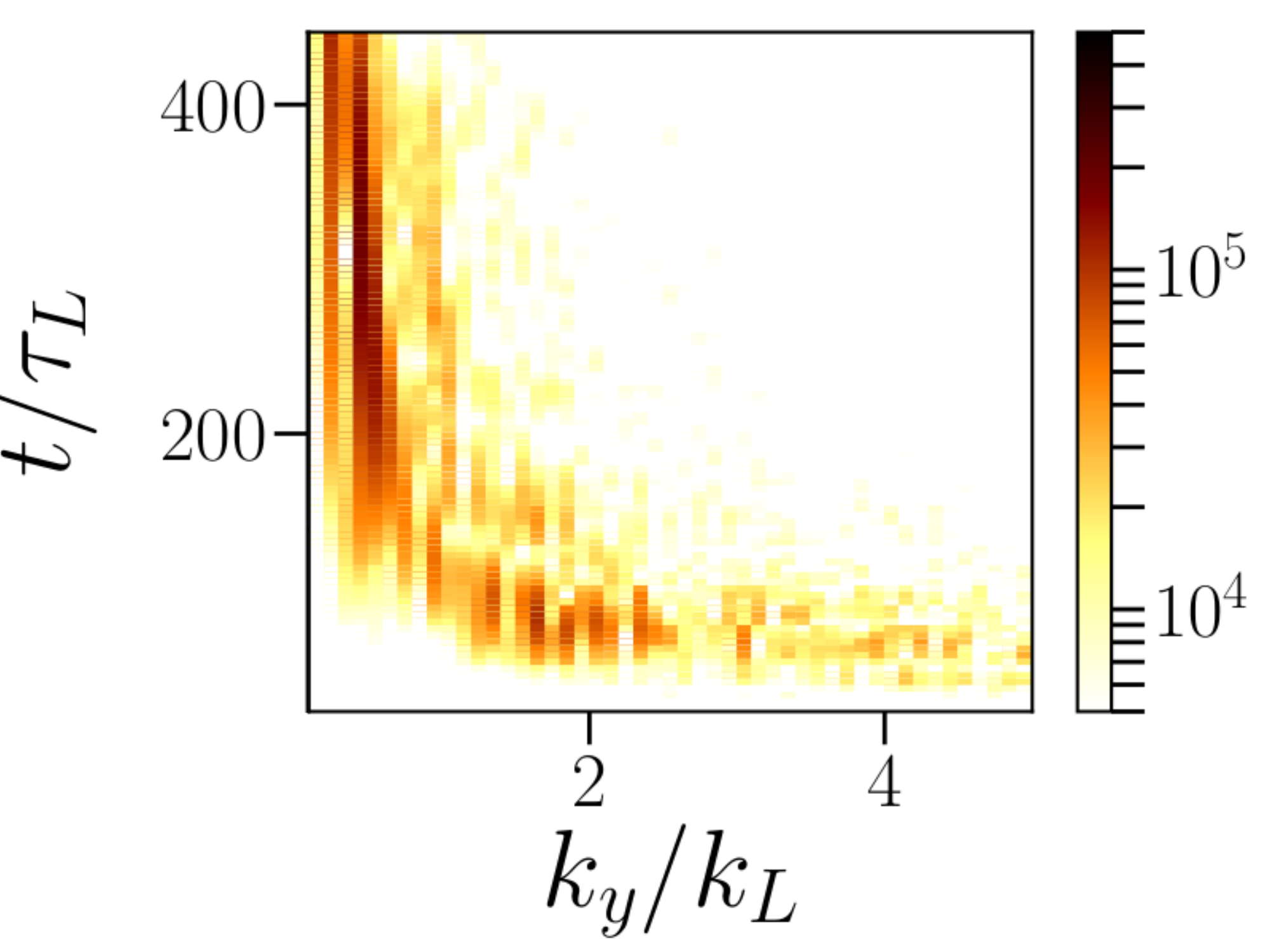}};
	\node[above right] (img) at (5.8cm,4.5cm){\includegraphics[width=0.40\textwidth]{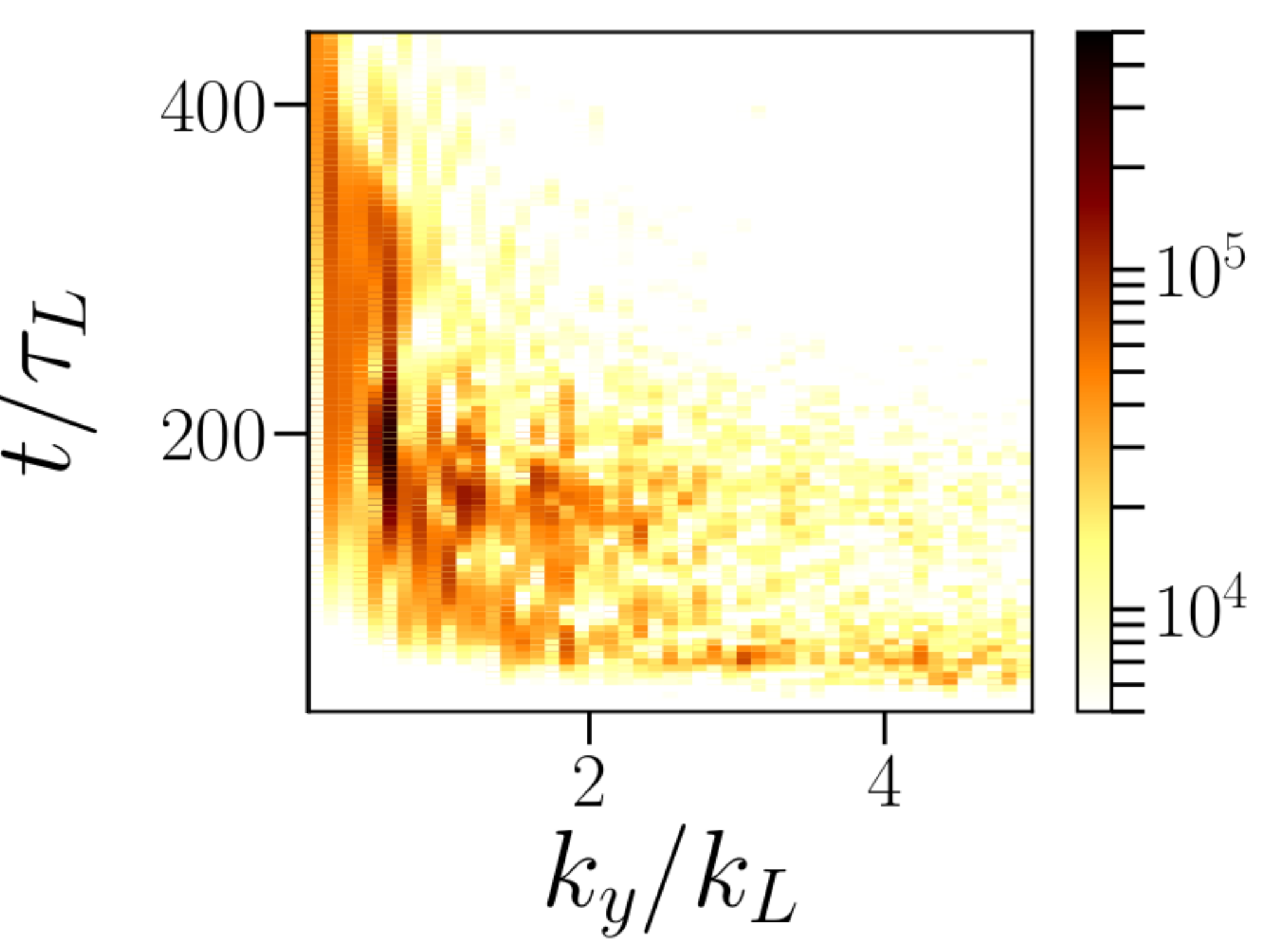}};
	\node[above right] (img) at (0cm,0.0cm) {\includegraphics[width=0.40\textwidth]{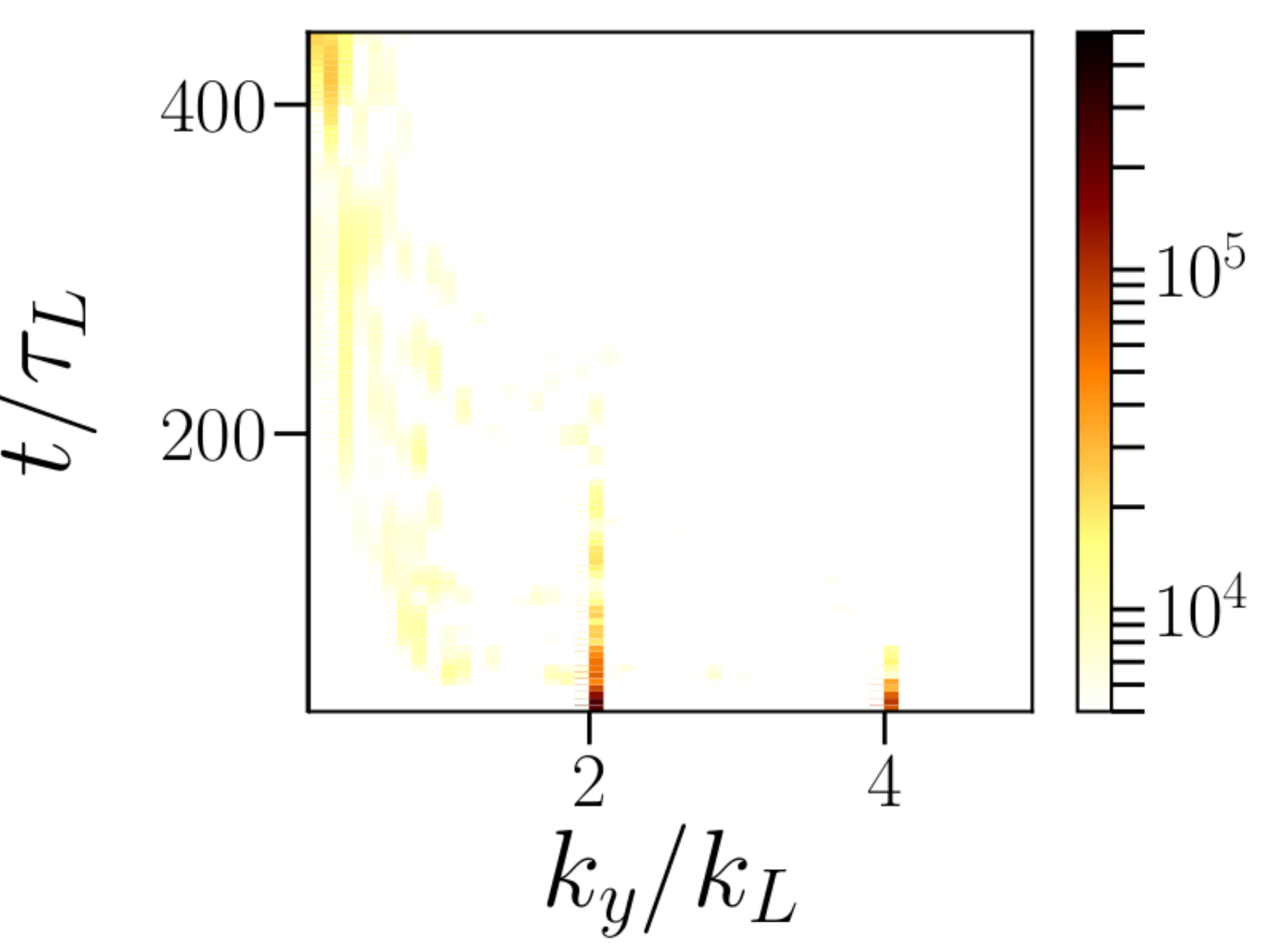}};
	\node[above right] (img) at (5.8cm,0.0cm) {\includegraphics[width=0.40\textwidth]{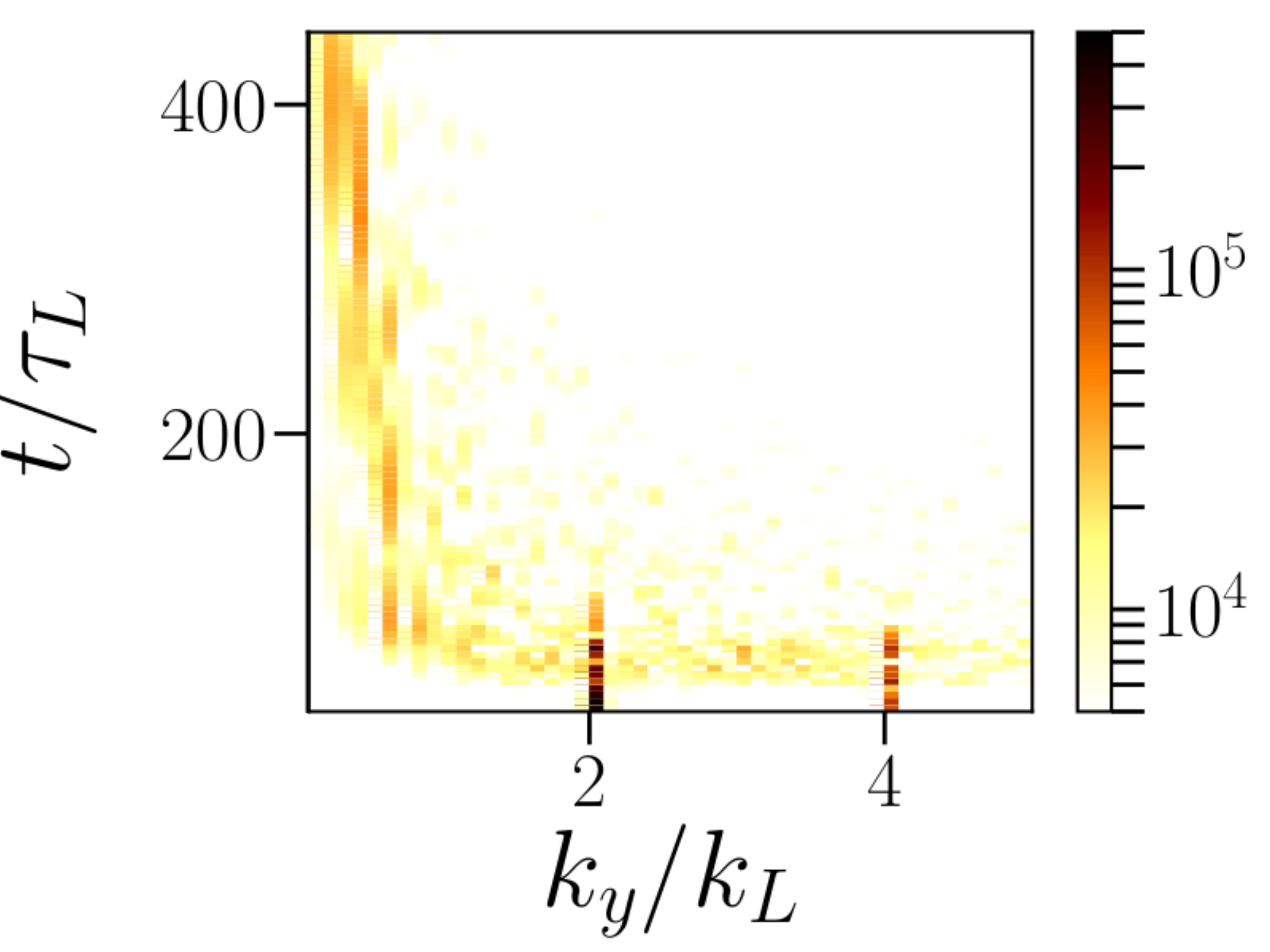}};
	\node at (4.1cm,3.7cm) {$\bm{(c)}$};
	\node at (4.1cm,8.2cm) {$\bm{(a)}$};
	\node at (9.9cm,3.7cm) {$\bm{(d)}$};
	\node at (9.9cm,8.2cm) {$\bm{(b)}$};
	
	\node at (8.75cm,8.75cm) {$\bm{flat, d=2.0\lambda_L}$};
	\node at (10.8cm,8.7cm) {\begin{scriptsize}$\bm{density}$\end{scriptsize}};
	\node at (2.8cm,8.75cm) {$\bm{flat, d=1.0\lambda_L}$};
	\node at (4.85cm,8.7cm) {\begin{scriptsize}$\bm{density}$\end{scriptsize}};
	
	\node at (2.8cm,4.25cm) {$\bm{dm, d=1.0\lambda_L}$};
	\node at (10.8cm,4.2cm) {\begin{scriptsize}$\bm{density}$\end{scriptsize}};
	\node at (8.75cm,4.25cm) {$\bm{dm, d=2.0\lambda_L}$};
	\node at (4.85cm,4.2cm) {\begin{scriptsize}$\bm{density}$\end{scriptsize}};
	\end{tikzpicture}
	\caption{\label{fig:fft_mods_RR350} \textcolor{black}{Time development of the FFT of ion density oscillations including radiation reaction force for the flat target (upper row) and the density modulated target (bottom row) at $a_0=350$. First and second columns corresponds to the target widths of {$d=1.0\lambda_L$ and $d=2.0\lambda_L$}, respectively. The modulation parameters are $a_m = 0.50, k_m = 2$. }}
\end{figure}

We follow the same procedure and study the temporal evolution of the FFT of ion density oscillations for density modulated and structured targets for higher $a_0$ cases to see the influence of the RR force on the RPA of ions. \textcolor{black}{In the case of radiation reaction force, a significant fraction of the laser energy gets converted into high-energy photons. Consequently the instability that breaks the target becomes only stronger at late times. Additionally, due to radiation reaction force  bunching of plasma ions is also possible.} Since the density modulated targets show better results on the ion acceleration spectra, we compare the cases of a flat target with a density modulated target (corresponding to the best modulation parameters) for $a_0=250$ and $a_0=350$. For former case, the RR force effects begin to appear in the ion energy spectra. While for the later case ($a_0=350$), the RR force effects are stronger, but still not requiring to include the quantum recoil  and pair-production  in PIC simulations. \textcolor{black}{Fig.\ref{fig:fft_mods_RR250} depicts the expected trend as observed before in Fig.\ref{fig:fft_no_mod}. The thinner target ({$d=1.0\lambda_L$}) shows significant suppression of the long-wavelength mode of the ion density oscillations, while for thicker target ({$d=2.0\lambda_L$}) there is indeed an appearance, albeit weaker in magnitude, of the long-wavelength mode. This suggest that eventually the RR force wash out the pre-imposed modulations in the plasma density and this use of density modulated targets may not be effective for higher $a_0$. Indeed this is further confirmed in Fig.\ref{fig:fft_mods_RR350} which shows appearance of strong long-wavelength modes of ion density oscillation being generated at later times. This limits the improvements in the FWHM of ions for the density modulated targets. Though, not shown here, we see similar trends for the other structured targets.}

Finally, we also carried out 2D PIC simulations with a laser pulse with Gaussian spatial profile and we recover the same trends as shown before. We show here one simulation run for a target with both density and surface modulations (dm-rpg) and Gaussian shape at the rear end (see Fig.\ref{fig:gaussian}). This target has a spatial density profile, $n(x,y) = n_e a_m [3+\cos(k_m y)]/2$, and is located between
\begin{equation}
2\pi a_c\exp\left(-\left(\frac{y-10\pi}{2\pi\cdot b_c\cdot 0.6}\right)^2\right)+ 4\pi \ge x \ge 2\pi - a_m\cos\left(k_m\frac{y-10\pi}{2\pi}\right),
\label{eq:14}
\end{equation}
where $a_c=1.0$ and $b_c=5.0$ are different dimensionless parameters. The  laser pulse has a waist of ${w}=7.0\lambda_L$, \textcolor{black}{and $x$ and $y$ coordinates of the focus-points, $f_x=1.0\lambda_L, f_y=5.0\lambda_L$ in the simulation box.}
\noindent
\begin{figure}
\centering
\begin{tikzpicture}
	\node[above right] (img) at (0cm,0cm){\includegraphics[width=0.6\textwidth]{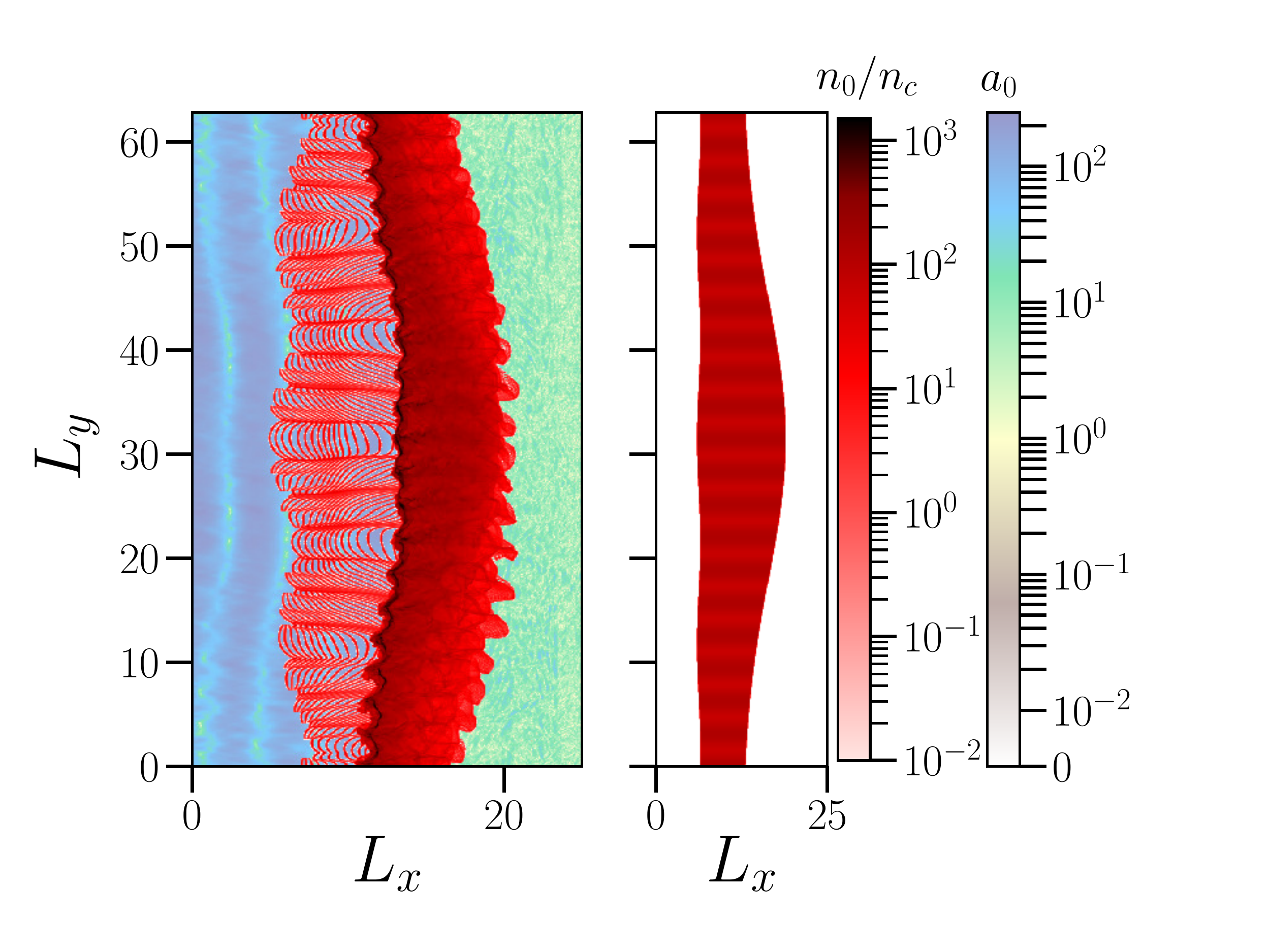}};
	\node[above right] (img) at (7.5cm,0.6cm){\includegraphics[width=0.40\textwidth]{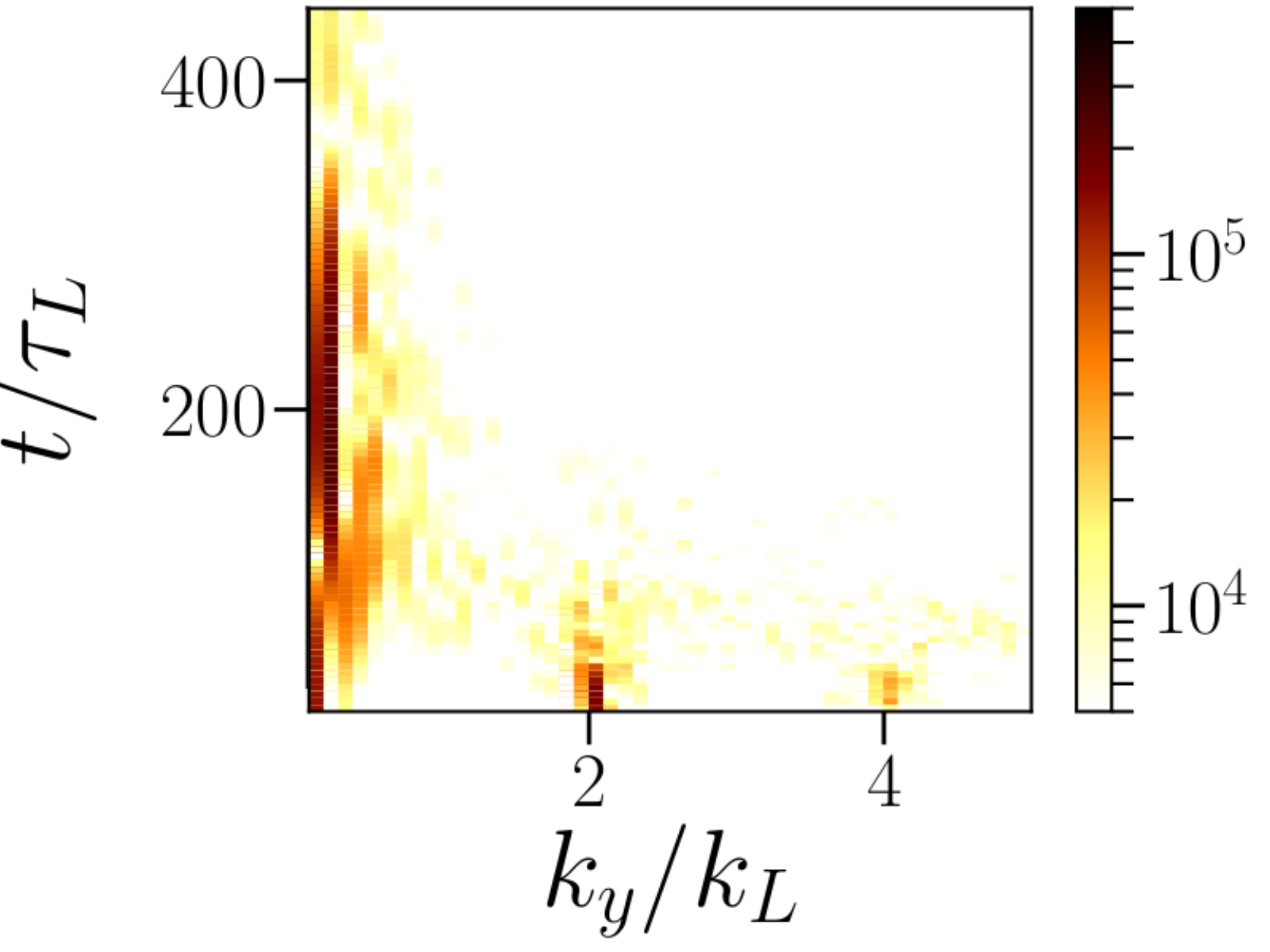}};
	\node at (5.15cm,5.15cm) {$\bm{(b)}$};
	\node at (3.5cm,5.15cm) {$\bm{(a)}$};
	%\node at (2.5cm,5.7cm) {$\bm{t/\tau_L=151}$};
	%\node at (4.6cm,5.7cm) {\begin{scriptsize}$\bm{t/\tau_L=0}$\end{scriptsize}};
	\node at (11.6cm,4.4cm) {$\bm{(c)}$};
	\node at (12.3cm,4.9cm) {\begin{scriptsize}$\bm{density}$\end{scriptsize}};
	\end{tikzpicture}
	\caption{\label{fig:gaussian} \textcolor{black}{Proton density of a target having both density and surface modulations (dm-rpg). The laser pulse has a spatial Gaussian profile with $a_0=150$. Panel $\bm{(a)}$ is at $t/\tau_L=40$, while the initial profile at  $t/\tau_L=0$ is shown in $\bm{(b)}$. $\bm{(c)}$ Time development of the FFT of the proton density. Modulation parameters are $a_m = 0.25, k_m = 2$, and the target widths varies between $d=2.0\lambda_L$ and $d=1.0\lambda_L$ [see Eq.\eqref{eq:14}].} \textcolor{black}{Moving window velocity is $\upsilon_{\rm mov} = 0.8\, c$.}	}
\end{figure}
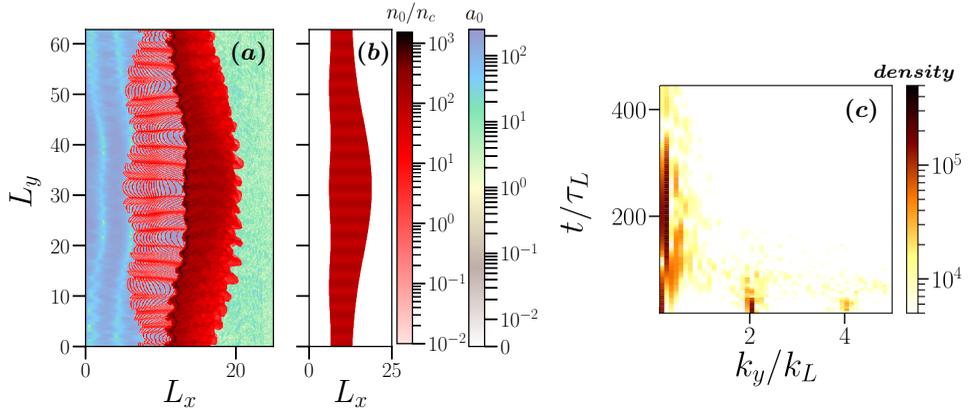
\noindent
This combination of density and surface modulations help the target to remain stable in time. The laser pulse can wash out the surface modulations after a while, but as the main target density is also modulated, the laser pulse can not wash out the density modulations in the early stages [Fig.\ref{fig:gaussian}(a)]. The suppression of long wavelength modes by competitive feeding is therefore most effective for density modulated targets [Fig.\ref{fig:gaussian}(c)]. The Gaussian shape at the rear end of the target in Fig.\ref{fig:gaussian}(b) is necessary\footnote{The dm-rpg target without the Gaussian profile at the rear end for a plane wave laser pulse does not improve the ion energy spread. It is only beneficial for a laser pulse with Gaussian spatial profile.} in order to counter the target breaking caused by the laser pulse with a spatial Gaussian profile~\cite{Chen:2009aa}. For this simulation, one gets $\Delta E/E_{max} = 17.78\%$ at $t/\tau_L=316$ with $E_{max}=1.13$ GeV for a density modulated (dm) target with Gaussian shape at the rear end and $\Delta E/E_{max} = 14.51\%$ at $t/\tau_L=280$ with $E_{max}=872.33$ MeV,  for an additional surface modulation on top (dm-rpg), incorporated by the cosine-term in Eq.\eqref{eq:14}; see Fig.\ref{fig:gaussian_parameters}(a). We carried other simulations for different $b_c$ and $a_c$, and the results on the FWHM of the ion energy spectra are depicted in Fig.\ref{fig:gaussian_parameters}(b). These results confirm the trends shown before.

\begin{figure}
\centering
\begin{tikzpicture}
	\node[above right] (img) at (0cm,0.7cm){\includegraphics[width=0.45\textwidth]{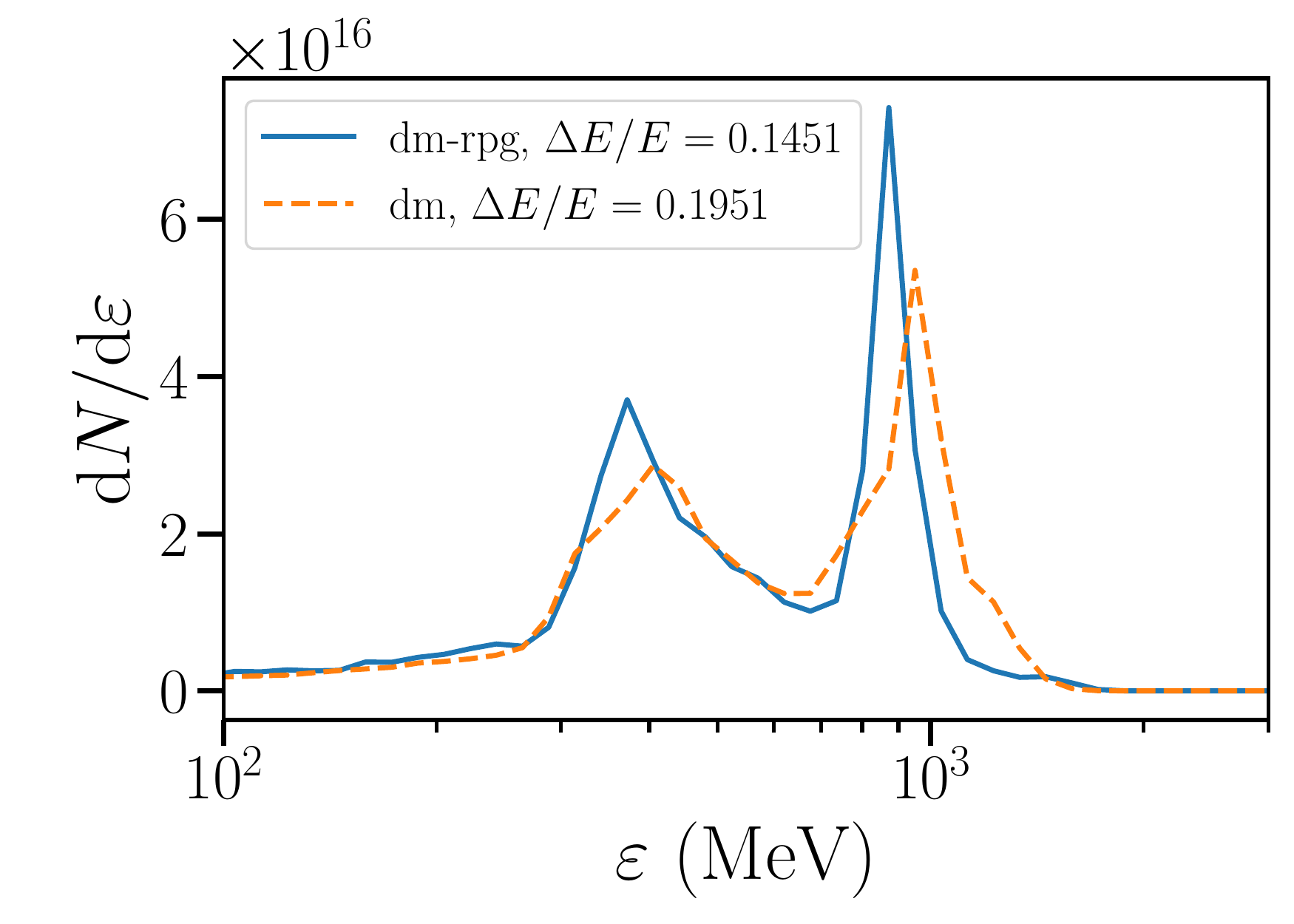}};
	\node[above right] (img) at (6cm,0cm){\includegraphics[width=0.5\textwidth]{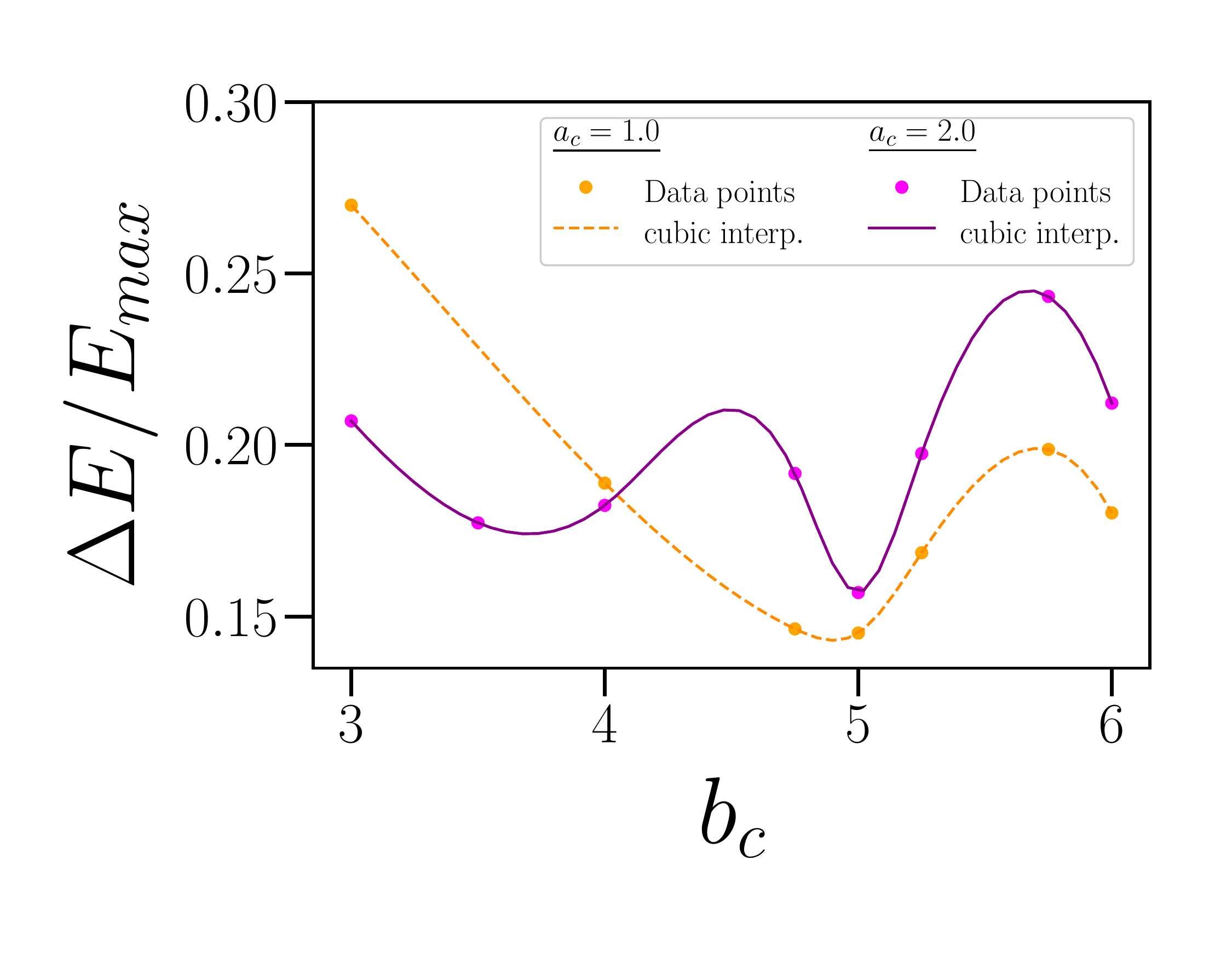}};
	\node at (5.65cm,2.15cm) {$\bm{(a)}$};
	\node at (12.0cm,2.15cm) {$\bm{(b)}$};
	\end{tikzpicture}
	\caption{\label{fig:gaussian_parameters} $\bm{(a)}$ Proton energy spectra corresponding to Fig.\ref{fig:gaussian}(a) at $t/\tau_L =280$ for $a_c=1.0, b_c=5.0$. $\bm{(b)}$ Cubic interpolation of $\Delta E/E_{max}$ with $b_c$ and with $a_c=1.0$ and $a_c=2.0$. The other parameters are $a_m = 0.25, k_m = 2, a_0=150$.}
\end{figure}

\section{\label{sec:conclusion}Conclusions and Discussions}

    \textcolor{black}{Our 2D simulation results, especially for the density modulated targets, showing the minimum energy spread ($\Delta E/E_{max} \approx 12\%$) for 1 GeV protons, are encouraging}. } Pushing the boundaries of RPA of ions in the radiation dominated regime, one can gain higher proton energies ($E_k \ge 2.0$ GeV) albeit the  FWHM remains stagnate at $\Delta E/E_{max} \sim (12-15)\%$. At lower $a_0=250$, thicker target ($d=2.0\lambda_L$) yields better results, while at higher $a_0=350$, this trend is reversed with the thinner target ($d=1.0\lambda_L$) showing improved results. Also, the density of accelerated protons is considerably higher at higher $a_0=350$. {These 2D simulation results \textcolor{black}{(also with spatial Gaussian laser profile in Fig.\ref{fig:gaussian_parameters})} from the density modulated target show \emph{substantial enhancements} over the flat target case, especially for the FWHM of the proton spectra. This enhancement continues in the radiation dominated regime of proton acceleration. Thus, the improvement in the FWHM for density modulated target, compared to the flat targets, is robust for the large range of the laser-plasma interaction parameters.} The density modulated as well as structured targets can be manufactured with newer technological advancements~\cite{Fischer:2013aa,Klimo:2011aa,Cantono:2021td}. {In particular, it has been shown experimentally that by adjusting the spatial profile of the laser pre-pulse and  introducing a variable delay with the main laser pulse, one can create transient plasma gratings on the surface of the thin-foil target with controllable precisions~\cite{Monchoce:2014tg}.  A realistic laser pulse has a long pre-pulse which ionises the target and creates a pre-formed plasma in front of the main target. The interaction of pre-pulse and the rising part of the main laser pulse can also excite, by parametric instabilities, long-wavelength modes of Brillouin instability, which may also have transverse wavevector associated in a 2D geometry~\cite{Giacone:1995wb}.} This may help in creating the conditions of the density modulated target.  For the sake of computational efficiency, we take protons instead of high-Z ions. The results presented here for protons can be recovered for high-$Z$ ions albeit requiring longer laser pulse durations.

Even though our 2D simulation results are \textcolor{black}{encouraging}, in real experiments 3D and other physical effects are likely to play a strong role, potentially limiting the ion energy gain and the FWHM of ion energy spectra~\cite{Dollar:2012ws}. Some of the geometrical effects related to target bending arising due to the finite laser spot-size in a 3D geometry can be overcome by using the target shapes as studied in Fig.\ref{fig:gaussian}. Further optimisation of the geometry and parameters, for higher ion energy gain in 3D geometry, can also be undertaken as shown recently~\cite{Wang:2021aa}. \textcolor{black}{Nevertheless, results presented here for density modulated target need to be further explored by taking into account realistic spatio-temporal laser profiles and high-$Z$ targets in 3D geometry.}

This work presented here encompasses the bachelor thesis of Tim Arniko Meinhold submitted to the Physics Department of the Heidelberg University.

All authors declare no competing interests.

\bibliographystyle{jpp}

%\bibliography{RPA}

\end{document}